\documentclass[aps,prl,reprint,superscriptaddress,floats]{revtex4-2}
\usepackage[T1]{fontenc}
\usepackage[latin9]{inputenc}
\usepackage{amsmath}
\usepackage{amssymb}
\usepackage{stmaryrd}
\usepackage{graphicx}
\usepackage{esint}
\usepackage{subfigure}
\usepackage{multirow}
\usepackage[dvipsnames]{xcolor}
\usepackage{accents}
\usepackage{bibunits}
\usepackage{xcolor}

\makeatletter

\newcommand{\appcontentssection}[2]{%
    \noindent
    \hyperref[#1]{#2}%
    \hfill \pageref*{#1}\par
    \medskip
}

\newcommand{\beq}{\begin{equation}}
	\newcommand{\eeq}{\end{equation}}
\newcommand{\bea}{\begin{eqnarray}}
	\newcommand{\eea}{\end{eqnarray}}
\newcommand{\bwt}{\begin{widetext}}
	\newcommand{\ewt}{\end{widetext}}
\@ifundefined{textcolor}{}
{%
	\definecolor{BLACK}{gray}{0}
	\definecolor{WHITE}{gray}{1}
	\definecolor{RED}{rgb}{1,0,0}
	\definecolor{GREEN}{rgb}{0,1,0}
	\definecolor{BLUE}{rgb}{0,0,1}
	\definecolor{CYAN}{cmyk}{1,0,0,0}
	\definecolor{MAGENTA}{cmyk}{0,1,0,0}
	\definecolor{YELLOW}{cmyk}{0,0,1,0}
}

\newcommand{\mbf}[1]{\mathbf{#1}}

\providecommand{\U}{\ensuremath{\mathrm{U}}}
\providecommand{\ie}{i.e.}

\makeatletter
\newcommand{\doublewidetilde}[1]{{%
  \mathpalette\double@widetilde{#1}%
}}
\newcommand{\double@widetilde}[2]{%
  \sbox\z@{$\m@th#1\widetilde{#2}$}%
  \ht\z@=.9\ht\z@
  \widetilde{\box\z@}%
}
\makeatother

\DeclareMathOperator{\Tr}{Tr}

\usepackage{hyperref}

\newcommand{\eqa}[1]{\begin{align}\begin{split} #1 \end{split}\end{align}}
\newcommand{\bsl}[1]{{\boldsymbol{#1}}}
\usepackage{cleveref}
\crefname{appendix}{Supp.}{Supps.}
\Crefname{appendix}{Supplemental}{Supplementals}
\crefname{equation}{Eq.}{Eqs.}
\Crefname{equation}{Equation}{Equations}
\Crefname{figure}{Figure}{Figures}
\crefname{figure}{Fig.}{Figs.}
\crefname{theorem}{Thm.}{Thms.}
\Crefname{theorem}{Theorem}{Theorems}
\Crefname{table}{Table}{Tables}
\crefname{table}{Tab.}{Tabs.}
\Crefname{section}{Section}{Sections}
\crefname{section}{Sec.}{Secs.}
\creflabelformat{appendix}{#2#1#3}

\newcommand{\refcite}[1]{Ref.\cite{#1}}

\let\oldAA\AA
\renewcommand{\AA}{\text{\normalfont\oldAA}}

\newcommand{\eg}{{\emph{e.g.}}}

\newcommand{\G}{\mathcal{G}}

\newcommand{\BZ}{\ensuremath{\mathrm{BZ}}}

\newcommand{\eq}[1]{\begin{align}\begin{split} #1 \end{split}\end{align}}

\hypersetup{colorlinks=true, citecolor=Blue,linkcolor=red, breaklinks=true, pdfduplex=DuplexFlipLongEdge}

\newcommand{\llangle}{\mathopen{\langle\mkern-4mu\langle}}
\newcommand{\rrangle}{\mathclose{\rangle\mkern-4mu\rangle}}
\usepackage{newtxtext}

\usepackage{placeins} 
\usepackage{array}
\renewcommand{\refcite}[1]{Ref.~\onlinecite{#1}}

\makeatletter
\renewcommand{\NAT@bibsetnum}[1]{%
  \setlength{\topsep}{0pt}%
  \NATx@bibsetnum{#1}%
}
\makeatother

\usepackage{xspace}
\DeclareRobustCommand{\ie}{\emph{i.e.}\@\xspace}
\DeclareRobustCommand{\eg}{\emph{e.g.}\@\xspace}
\usepackage{babel}

\makeatother

\usepackage{babel}

\begin{document}
\title{Narrow-Band Magnetism Allowed by Quantum Geometry: Altermagnetism and Beyond}

\author{Minxuan Wang}
\affiliation{Department of Physics, Princeton University, Princeton, New Jersey 08544, USA}

\author{Pei-Han Lin}
\affiliation{Max-Planck-Institut f\"ur Physik komplexer Systeme, 01187 Dresden, Germany}

\author{Yang Ge}
\affiliation{Department of Physics and Quantum Theory Project, University of Florida, Gainesville, Florida 32611, USA}

\author{Jiabin Yu}
\email{yujiabin@ufl.edu}
\affiliation{Department of Physics and Quantum Theory Project, University of Florida, Gainesville, Florida 32611, USA}

\begin{abstract}
Quantum geometry plays a central role in shaping correlated phases in narrow bands, yet its influence on the competition between distinct magnetic orders, especially the emergence of altermagnetism, remains largely unexplored.
To address this question, we study the quadrupole insulator model with strictly onsite generalized Hubbard--Kanamori interactions, in which both N\'eel antiferromagnetism and altermagnetism are symmetry-permitted.
Depending on the parameters, the isolated narrow lowest band of the model can either be obstructed atomic with strong quantum geometry or trivial atomic with weak quantum geometry, while the dispersion is identical in the two cases.
By partially filling the lowest band, our Hartree--Fock calculations show only conventional ferromagnetic phases for the weak quantum geometric case, while we find two collinear altermagnetic phases, as well as nematic phases, for the strong quantum geometric case.
Such a dramatic difference can be understood by the distinct characters of the projected interaction: the projected interactions acquire sizable longer-range channels only for the strong quantum geometric case, while they are dominated by the on-site Hubbard channel in the weak quantum geometric case.
These results establish the essential role of quantum geometry for altermagnetism to emerge from interacting narrow bands where crystalline symmetries cannot sufficiently constrain the magnetic order.
\end{abstract}

\maketitle

\section{Introduction}

Quantum geometry describes how Bloch wave functions vary in momentum space, revealing electronic properties beyond those encoded in the band dispersion \cite{Rossi2021CurrentOpinion,torma2023essay,liu2024quantgeo,yu2025quantum,gao2025quantgeo,kitamura2026QGcorrelated,jiang2025QGnonlinear,Peotta2025QGSuperfluidity,yang2025QGMetrology,cano2026QGfraction,verma2026QGscale}.
The profound role of quantum geometry has been established in diverse aspects of quantum many-body physics, such as superfluid weight~\cite{peotta2015superfluidity,julku2016SFlieb,Paivi2017bandgeometrySFweight,PhysRevLett.123.237002,PhysRevB.101.060505,Xie2020TopologyBoundSCTBG,Yu2022EOCPTBG,torma2022superconductivity,herzog2022superfluid,tian2023evidence,hu2025sweight,sun2025geometricsuperfluidweightquasicrystals,zeng2025superfluidstiffnessboundstimereversal}, fractional Chern insulators~\cite{neupert2011fqh,sheng2011fqh,roy2014band,claassen2015position,Jie2021IdealBands,Parker2023IdealBands,Valentin2023IdealBands,Dong_2023_ideal_Higher_Chern,Roy2024FCINonLL,Queiroz_2024_TMD_Chiral,liu2025generallandau,liu2025ideal,Sarkar2025IdealChern,Emanuel2025unify,Yu_2025_WL_ideal}, electron-phonon coupling~\cite{yu2024nontrivial,Alexandradinata2024ShiftCurrent,Wang2025Optical_QG_EPC,hu2025quantumgeometricoriginstraininduced,Pellitteri_2025_Phonon_Geometry_Goldstone, jiang2026quantumgeometrycriticaltemperature}, 
and entanglement~\cite{paul_area-law_2024,tam2024corner,Wu2024QGCornerChargeFluctuationsManyBody,kruchkov2025bounds,balut2025entanglement,tam2026fermisurfacegeometrycharge}.
Recently, there have been studies on how quantum geometry affects spin stiffness~\cite{kitamura2026quantum}, the dynamics of magnetic textures~\cite{deng2026collective}, the magnon gap~\cite{chen_2026_QGdipole,goncalves_2026_SyMA}, and the competition between ferromagnetism (FM) and N\'eel  or stripe antiferromagnetism (AFM)~\cite{hu2026ferromagnetism,zhang2026identifying,kitamura2026QGFM,oh_2026_magnetic_transition,shimizu2026magneticfluctuations}, as well as different forms of multipole magnetism~\cite{kudo_2026_odd-parity}. 
Despite those studies, it is still unclear how quantum geometry affects the emergence of the spontaneous altermagnetic (AM) state, \ie, a collinear
\footnote{While noncollinear AM phases have recently been proposed~\cite{cheong2024noncollinearAM,lee2024noncollinearAM}, we follow the classification in Refs.~\onlinecite{Song_2024_Spin_Group,Jiang_2024_Spin_Group} and define AM to be collinear unless specified otherwise.}
and compensated magnetic order distinct from conventional AFM~\cite{Song_2024_Spin_Group,Jiang_2024_Spin_Group}, which has attracted extensive research efforts in the past several years~\cite{Smejkal2022LandscapeAM,Smejkal2022SpinMomentumAM,Jungwirth2026AMReview}.

Based on the studies on the emergence of AM~\cite{Mazin_2023_AM,Das2024FermiHubbardAM,Leeb2024OrbitalOrdering,Ferrari2024ShastrySutherland,BlancoPeces2025KagomeAM,bai2024altermagnetism,He2025FermiHubbardAM,Antonenko_2025_AM,Li2025HoleyGraphyneAM,Kaushal2025LiebAM,GiuliPRB2025,Peng_2025_Multicomponent_Altermagnet,Heinsdorf2025QGInstabilityAM,*heinsdorf2026erratum,Dong2026ExtendedHubbardAM,Yu2026MetalFreeAM,Equbal2026FrustratedSquareAM,Franz_2026_SpontaneousalterAM,yang2026reevaluatingquantumgeometriccriteria,song2026crystalsymmetry,devaraj_2026_altermagnetism,lange2026surface,li2026MirrorEngineer,Fernandes_unpublished}, the answer to any question related to the emergence of AM must depend on the symmetries.
A major established route to realizing AM is to lower the crystalline symmetry sufficiently to suppress the conventional N\'eel antiferromagnetic order protected by translation/inversion combined with time-reversal symmetry, such that the collinear compensated magnetism enforced by the remaining crystalline symmetries is instead altermagnetic.
For example, one can realize an AM by exploiting the crystalline structure to lower the symmetry of a N\'eel-like compensated spin pattern generated by strong electronic repulsion, such as an onsite Hubbard interaction~\cite{Jungwirth2026AMReview}.
In this scenario, quantum geometry is not necessary.

Therefore, we address the question of the role of quantum geometry in the scenario where the crystalline symmetries are high enough to allow both N\'eel AFM and AM states.
We focus on a partially filled isolated narrow band of a multi-orbital model with strictly onsite interactions and without spin-orbit coupling.
If the narrow band has very weak quantum geometry, its Wannier state has to be mainly localized on one atomic site, and the projected interaction has to be dominated by the conventional onsite Hubbard interaction.
Consequently, regardless of the form of the onsite interaction before the projection, the resulting phases are largely restricted to those familiar from the Hubbard model on the corresponding lattices.

However, when the narrow band has very strong quantum geometry, each of its Wannier states would have a large spread over multiple atomic sites and/or multiple unit cells, and the projected interaction would have multiple longer-range channels.
Thus, varying the form of the onsite interaction before projection can give rise to unconventional magnetic phases even when the symmetries remain unchanged, allowing AM to outcompete the conventional states in Hubbard models.
This intuition suggests that for isolated narrow bands near the Fermi energy, quantum geometry would play an essential role in the emergence of spontaneous AM when the symmetries alone are not enough to constrain the magnetic order, in addition to the obviously important role played by the form of the onsite interaction before the projection.
Our claim is different from that of the previous work~\cite{Heinsdorf2025QGInstabilityAM} which suggested that quantum geometry generically favors AM instabilities, irrespective of the crystalline symmetry, under the complete decoupling of magnetic ordering channels~\cite{yang2026reevaluatingquantumgeometriccriteria}.

To concretely illustrate our idea, we consider the quadrupole insulator model~\cite{benalcazar2017quantized}, a square-lattice tight-binding model with one atom per unit cell and four orbitals per atom, and we add strictly onsite generalized Hubbard--Kanamori interactions.
The symmetries of the many-body model allow both N\'eel AFM and AM to emerge spontaneously.
We partially fill the lowest-energy band (away from half filling) and focus on the projected narrow-band regime, in which the bandwidth is smaller than the interaction strength, while the gap to the higher bands is much larger than the interaction strength.
In this regime, the band can realize two topologically trivial atomic phases: an obstructed atomic phase and a trivial atomic phase. The two phases have identical band dispersions, but the former (latter) exhibits strong (weak) quantum geometry, providing an ideal platform for uncovering quantum-geometric effects.
In the case with strong quantum geometry, the projected Hubbard--Kanamori interactions get considerably longer-range components. 
By varying the parameters in the unprojected onsite interaction, we uncover a variety of phases, including AM, AFM, FM, and nematic states.
In contrast, in the case of weak quantum geometry, the projected interactions remain predominantly onsite, and only result in conventional phases such as the hole FM arising from hole doping the N\'eel AFM state and the fully polarized FM state.
We further provide an analytical argument for the emergence of AM in the strong quantum geometric regime.
Our results demonstrate the essential role of quantum geometry in stabilizing spontaneous AM when crystalline symmetries do not preclude the competing conventional magnetic orders.

\section{Model and symmetries}
\begin{figure}[t]
    \centering
    \includegraphics[width=\linewidth]{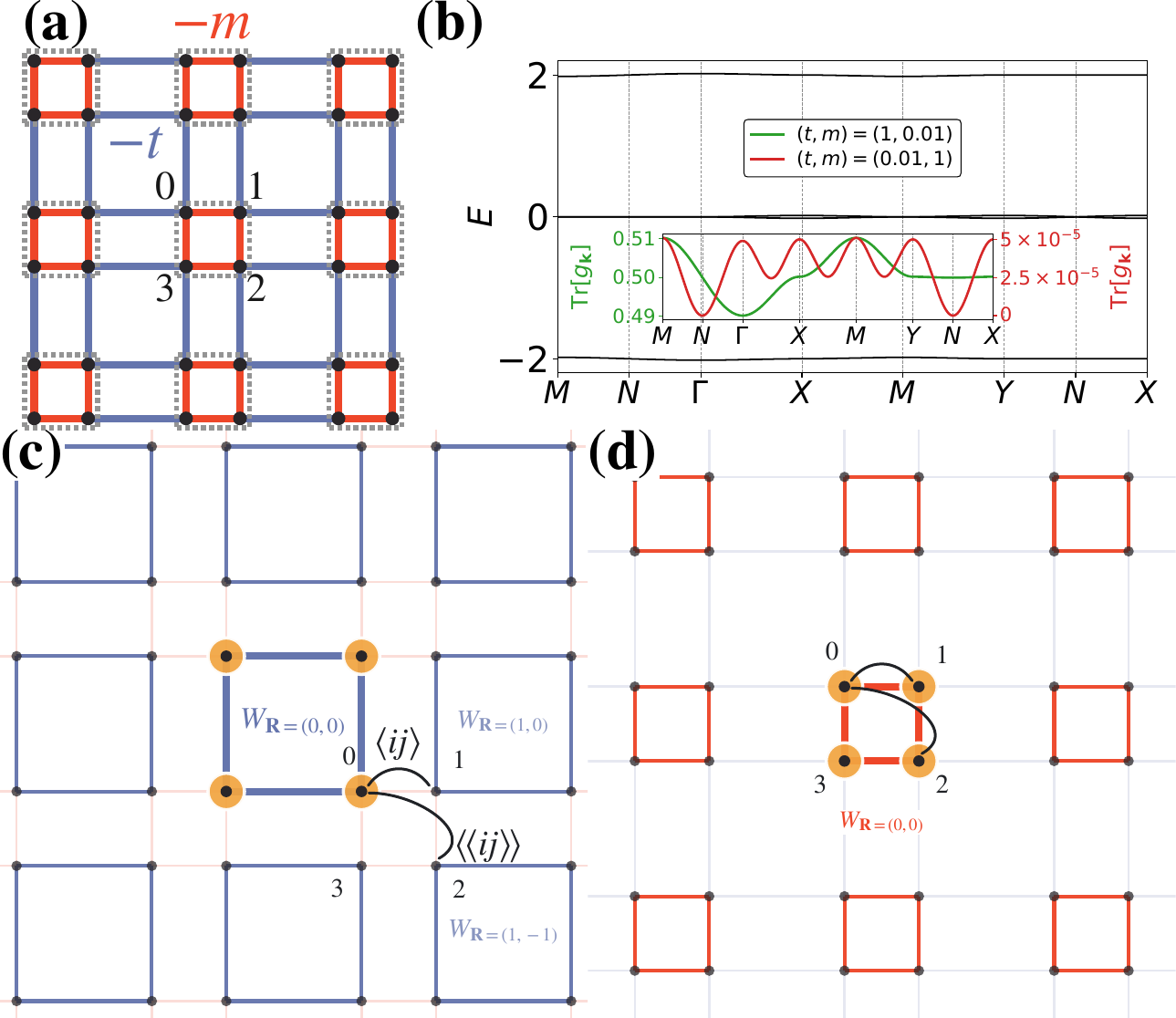}
    \caption{The zero-flux version of the quadrupole insulator model~\cite{benalcazar2017quantized} with four orbitals per unit cell.
    (a)~Illustration of the single-particle model with onsite hoppings (red) of amplitude $-m$, and nearest-neighbor hoppings (blue) of amplitude $-t$.
     The four orbitals in each unit cell (boxed by dashed lines) lie at the same point but are drawn separated for clarity.
    (b)~Identical noninteracting band structure at $(m, t) = (0.01, 1)$ and $(m, t) = (1, 0.01)$, both with a lowest narrow band. Inset shows that $\displaystyle \text{Tr}[g_{\mbf{k}}]$ of the narrow band is around $0.5$ for $m \ll t$ (green) and negligible for $m \gg t$ (red). 
    (c)~Maximally localized Wannier orbitals (blue) for $m \ll t$. Upon projection onto the narrow band, onsite interactions generate nearest-neighbor and next-nearest-neighbor terms.
    (d)~Wannier orbitals (red) for $m\gg t$. The projected interactions remain onsite to the leading order. 
    }
    \label{fig:fig1}
\end{figure}

Our model is built on a square lattice with four orbitals at the same position in each unit cell. The lattice constant and $\hbar$ are set to 1.
The Hamiltonian $H$ contains the single-particle part $H_0$ and the interaction part $H_{\mathrm{int}}$, \ie, $H=H_0 + H_{\mathrm{int}}$.
Our $H_0$ is simply a zero-flux version of the quadrupole insulator model on the square lattice~\cite{benalcazar2017quantized}:
\eqa{
\label{eq:H_0}
H_0 = {}& -\sum_{\mbf{R}\alpha s} \left(m\, c_{\mbf{R}\alpha s}^{\dagger} c_{\mbf{R}(\alpha + 1)s} + t\, c_{\mbf{R}\alpha s}^{\dagger} c_{(\mbf{R} + \boldsymbol{\delta}_{\alpha})(\alpha + 1) s}\right)\\
& + \mathrm{h.c.} \ ,
}
where $ c_{\mbf{R}\alpha s}^{\dagger}$ creates an electron with spin $ s = \uparrow,\downarrow$ in orbital $\alpha = 0, 1, 2, 3$ on the lattice site of the unit cell at $\mbf{R}$, $\mbf{R}=(l_1,l_2)$ with $l_1,l_2$ integers, $m$ is the intra-site hopping amplitude, and $t$ is the inter-site hopping amplitude. [See \cref{fig:fig1}(a).]
The orbital index $ (\alpha + 1)$ is understood modulo $4$, and nearest-neighbor hopping vectors are $\{\bsl{\delta}_{0},  \bsl{\delta}_{1}, \bsl{\delta}_{2}, \bsl{\delta}_{3}\} = \{(-1, 0), \,(0, 1), \,(1, 0), \, (0, -1)\}$.
Without affecting the physics that we aim to demonstrate, we focus on $m,t>0$.

The interaction part $H_{\mathrm{int}}$ is a generalized yet \emph{strictly onsite} variant of the Hubbard--Kanamori interaction:~\cite{kanamori1963electron,hubbard1963electron,hubbard1964electron,georges2013strong}
\begin{equation}\label{eq:H_int}
\begin{aligned}
H_{\mathrm{int}} &=  \!\sum_{\mbf{R}\alpha} U n_{\mbf{R}\alpha \uparrow} n_{\mbf{R}\alpha \downarrow}\\
&  + \!\sum_{\mbf{R}\alpha<\beta}\!\left(\frac{2U_{\alpha\beta}' \!- \!J_{H,\alpha\beta}}{2}n_{\mbf{R}\alpha}n_{\mbf{R}\beta}-2J_{H,\alpha\beta}\mbf{S}_{\mbf{R}\alpha}\!\cdot \mbf{S}_{\mbf{R}\beta} \!\right)\\
&+\!\sum_{\mbf{R}\alpha<\beta}J_{p,\alpha\beta} \left(c_{\mbf{R} \alpha\uparrow }^{\dagger}c_{\mbf{R}\alpha \downarrow}^{\dagger} c_{\mbf{R}\beta \downarrow } c_{\mbf{R}\beta \uparrow} + \mathrm{h.c.}\right) \ ,
\end{aligned}
\end{equation}   
where $ n_{\mbf{R} \alpha s} = c_{\mbf{R}\alpha s}^{\dagger} c_{\mbf{R}\alpha s}$ is the electron number operator, $ n_{\mbf{R}\alpha} = n_{\mbf{R}\alpha \uparrow} +n_{\mbf{R}\alpha \downarrow}$, and $\mbf{S}_{\mbf{R}\alpha} = (S_{\mbf{R}\alpha}^{x}, S_{\mbf{R}\alpha}^{y}, S_{\mbf{R}\alpha}^{z})$ in which $ S_{\mbf{R}\alpha}^{\mu} = \sum_{s s'}c_{\mbf{R}\alpha s}^{\dagger} (\sigma^{\mu}/2)_{ss'} c_{\mbf{R}\alpha s'}$ and $\sigma^{\mu}$ are the Pauli matrices.
Here $U$ is the intra-orbital Hubbard repulsion, $U_{\alpha\beta}'=U_{\beta\alpha}'$ is the inter-orbital density-density repulsion, $J_{H,\alpha\beta}=J_{H,\beta\alpha}$ is the Hund's coupling, and $J_{p,\alpha\beta}=J_{p,\beta\alpha}$ is the pair-hopping amplitude.
We always choose values of the interaction parameters so as to preserve the spin-$\mathrm{SU}(2)$, the spinless plane group $p4mm$, and the time reversal $\mathcal{T}$ symmetries~\cite{SM}.
We refer to the interaction in \cref{eq:H_int} as the generalized Hubbard--Kanamori interaction for the following reason.
The conventional Hubbard--Kanamori interaction requires
$
J_{H,\alpha\beta}=J_{p,\alpha\beta}=J
$, $U_{\alpha\beta}'=U'$,
and
$
U=U'+2J
$,
together with full orbital $\mathrm{SU}(2)$ rotational symmetry~\cite{wu2016topological,gastiasoro2016unconventional,brydon2011magnetic,georges2013strong,yan2026stoichiometric}. The orbital $\mathrm{SU}(2)$ requires the orbital basis to form an irreducible representation of $\mathrm{SU}(2)$ and hence to contain an odd number of orbitals (besides spin) per unit cell.
In our $H_\mathrm{int}$, there are four orbitals per unit cell, and $U$, $U_{\alpha\beta}'$, $J_{H,\alpha\beta}$, and $J_{p,\alpha\beta}$ can take values away from the conventional choice, while the rest of the interaction retains the standard Hubbard--Kanamori form.

Among the symmetries of the many-body Hamiltonian, of particular interest is the spinless fourfold rotation $C_4$, defined by $ C_{4} c_{\mbf{R}\alpha s}^{\dagger} C_{4}^{-1} = c_{R_4\mbf{R} (\alpha + 1)s}^{\dagger}$ with $ R_4(x, y) = (y, -x)$, i.e.\ a spatial rotation accompanied by the cyclic orbital permutation $\alpha \rightarrow \alpha+1 \,(\text{mod}\, 4)$.
The combined symmetry $C_4\mathcal{T}$ can protect an AM order, while $C_4^2\mathcal{T}$ (\ie\ $C_2\mathcal{T}$), $T_{(1,0)}\mathcal{T}$ and/or $T_{(0,1)}\mathcal{T}$ are characteristic of the conventional N\'eel AFM.
Here, $T_{(1,0)}$ and $T_{(0,1)}$ denote primitive lattice translations along the $x$ and $y$ directions, respectively.

\begin{figure*}[t]
    \centering
    \includegraphics[width=\linewidth]{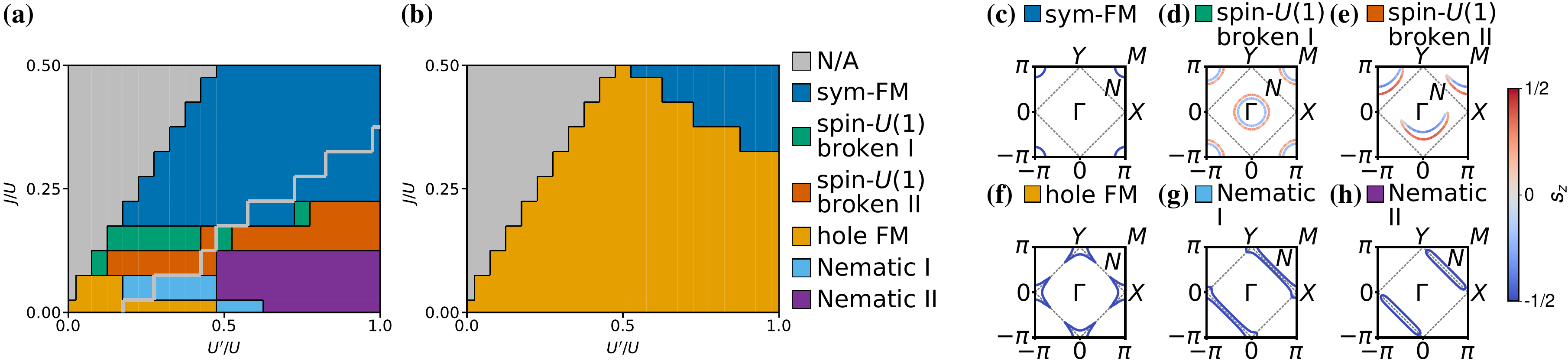}
    \caption{%
    Magnetic phases at $\displaystyle \nu = 1 - 1/24$ with interaction parameters taking the form $U'_{\alpha \beta} = U'$, and $J_{H, \alpha\beta} = J_{p, \alpha \beta} = J$.
    (a) Phase diagram with strong quantum geometry in the obstructed atomic case $m/t=0.01$ and $U/t=0.1$. 
    (b) Phase diagram with weak quantum geometry in the trivial atomic case $t/m=0.01$ and $U/m=0.1$.
    (c)--(h) Representative Fermi surfaces of all phases in (a) and (b). Dashed lines mark the $\sqrt{2}\times \sqrt{2}$ folded Brillouin zone, and colors represent the $\mbf{k}$-resolved spin polarization $s_z$ on these Fermi surfaces. The self-consistent iterations do not converge for $J>U'$, which is shaded in gray and labeled N/A.
    Below the silver line in (a), AM states (see AM I in \cref{fig:fig3}) exist as self-consistent solutions, but with a slightly higher energy than the groundstate.
    }
    \label{fig:fig2}
\end{figure*}

\section{Quantum Geometry and Longer-Range Projected Interaction}
We focus on the lowest band of the single-particle part in \cref{eq:H_0}, whose dispersion is $ E(\mbf{k}) = -\sqrt{m^2 + t^2 + 2mt \cos k_x} - \sqrt{m^2 + t^2 + 2mt \cos k_y}$, where $\mbf{k}=(k_x,k_y)$ is the Bloch momentum. 
We will consider two limiting cases: $m\ll t$ and $m \gg t$ [\cref{fig:fig1}(b)].
In both cases, the bandwidth of $E(\mbf{k})$ is $4\min(m,t)$, much smaller than the gap above it which is about $2\max(m,t)$, as $\max(m,t)\gg \min(m,t)$.
Owing to the time-reversal and inversion symmetries, the Berry curvature of the band vanishes, and thus the quantum geometric tensor~\cite{yu2025quantum} is entirely characterized by the quantum metric $\left[g_{\mbf{k}} \right]_{ij}= \Tr\left[\partial_{k_i}P_{\mbf{k}}\partial_{k_j}P_{\mbf{k}}\right]/2$ with $P_{\mbf{k}}$ the $4\times 4$ projection matrix of the lowest band. 
The integral of $\Tr\left[g_{\mbf{k}} \right]$, \ie, the Dirichlet functional~\cite{Yu_2025_WL_ideal}
\eqa{
\Tr\G = \int_{\BZ} d^2 k \Tr\left[g_{\mbf{k}} \right]\ ,
}
differs strongly in the two limits:  $\Tr\G= 2\pi^2 + O(m^2/t^2)$ for $m\ll t$, and  $\Tr\G= O(t^2/m^2)$ for $m\gg t$~\cite{SM}.
[See \cref{fig:fig1}(b) inset for numerical values.]
The difference in the quantum metric can be understood from the real-space Wannier state of the lowest-energy band.
In the limit $m\ll t$, the band is obstructed atomic: the Wannier state is centered at $\mbf{R} +(-1/2,1/2)$ instead of any atomic site and carries almost all of its weight on the four surrounding orbitals [\cref{fig:fig1}(c)].
In other words, we have $d_{\mbf{R} s}^{\dagger} = \big(c_{\mbf{R} 0 s}^{\dagger} + c_{\mbf{R}+(-1,0),1 ,s}^{\dagger} + c_{\mbf{R}+(-1, 1), 2 ,s}^{\dagger} + c_{\mbf{R}+(0, 1), 3,s}^{\dagger}\big)/2 + O(m/t)$, where  $d_{\mbf{R}s}^{\dagger}$ denoting the creation operator for the Wannier state at $\mbf{R}$ with spin $s$. 
With this gauge choice, the nonzero spread of this lowest-band Wannier state exactly comes from the quantum metric~\cite{marzari1997maximally,Wang2026SymmetryQG}, which cannot be arbitrarily small as the obstructed atomic phase has a nonzero lower bound on its quantum metric~\cite{herzog2022superfluid}. 
For $m\gg t$, the band is trivial atomic: the Wannier state is instead centered at the atomic sites, \ie, $d_{\mbf{R} s}^{\dagger} = \frac{1}{2}\sum_\alpha c_{\mbf{R} \alpha s}^{\dagger} + O(t/m)$ as shown in \cref{fig:fig1}(d), and there is no lower bound on the quantum metric, allowing $\Tr\G\rightarrow 0$ as $m/t\rightarrow \infty$.

The real-space properties manifested by the quantum metric imply very different interaction forms in the two cases when projecting the onsite generalized Hubbard--Kanamori terms in \cref{eq:H_int}.
In the obstructed atomic ($m\ll t$) case, the strong quantum geometry of the Bloch state reflected in the large spread of the Wannier state would lead to large nearest-neighbor and next-nearest-neighbor channels in the projected interaction, besides the onsite channels, as illustrated in \cref{fig:fig1}(c).
The longer-range channels include the density-density interaction ones and other non-density-density ones (such as density-assisted hopping channels).
On the other hand, the weak quantum geometry of the Bloch state in the trivial atomic ($m\gg t$) case makes the projected interaction onsite Hubbard with the subleading order no larger than $O(t^2/m^2)$, where the absence of  $O(t/m)$ is due to the inversion $C_2 = C_{4}^2$ symmetry~\cite{SM}.
Therefore, despite identical dispersion in the two cases, the difference in quantum geometry leads to dramatically different forms of the projected interaction.
In the following, we will show that such a difference would lead to very different magnetic phase diagrams, with the strong-quantum-geometry case containing various exotic phases that are absent in that of weak quantum geometry.

\begin{figure*}[t]
    \centering
    \includegraphics[width=\linewidth]{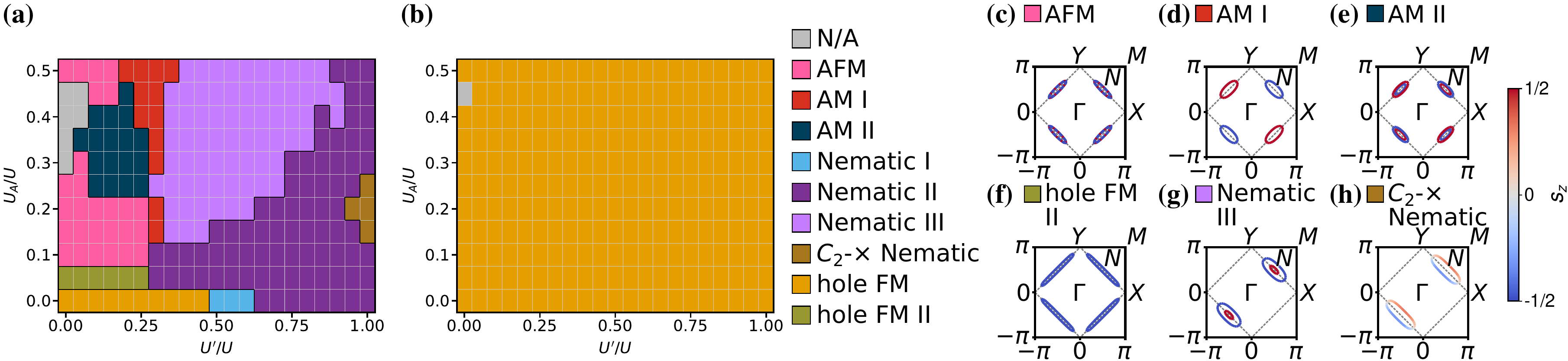}
    \caption{Magnetic phases at $\nu = 1 - 1/24$ with Hubbard--Kanamori interactions taking the form in \cref{eq:parameter_combine_orbital_aniso}.
    (a) Phase diagram in the obstructed atomic case $m/t=0.01$ and $U/t=0.1$. (b) Phase diagram in the trivial atomic case $t/m=0.01$ and $U/m=0.1$. (c)--(h) Representative Fermi surfaces of the new phases in (a) that are absent in \cref{fig:fig2}, plotted in the same format as in \cref{fig:fig2}. 
    }
    \label{fig:fig3}
\end{figure*}

\section{Competing magnetic orders and Altermagnetism}

To obtain the magnetic phase diagram, we perform the self-consistent Hartree--Fock (HF) calculations on the \emph{unprojected} interacting model~\cite{SM,KangPRL2026,KwanMoire2025,KwanAdvPhys2025}.
Since we are particularly interested in the competition with the N\'eel AFM state that has $\sqrt{2}\times \sqrt{2}$ lattice translation breaking, we allow such translation breaking in our calculations, in addition to translationally invariant states.
In addition, we neglect the direct Hartree terms that depend on the total site density $\langle\sum_{\alpha }n_{\mbf{R}\alpha}\rangle$, following a prescription used to avoid double counting in density functional theory (DFT)-derived multi-orbital models~\cite{zhou2005electron,leiria2010extended, belozerov2012monoclinic,Himmetoglu_2013_HubbardDFT,park2014computing,he2016photoinduced,jiang2016interatomic,chang2024downfolding,haas2024intersite,Dong2026ExtendedHubbardAM}.
We work at the electron filling $\nu = 1-1/24$ per lattice site so that the lowest-energy band is partially filled and away from half filling ($\nu=1$).
We consider $4\min(m, t) < U\ll 2\max(m, t)$, so that the intra-orbital $U$ is larger than the narrow band bandwidth but much smaller than the gap to the remaining bands~\cite{hu2026ferromagnetism}.  

We first consider the case with orbital-isotropic interactions where $U'_{\alpha \beta} = U', J_{H, \alpha\beta} = J_{p, \alpha \beta} = J$, and present the phase diagrams for the two limits of quantum geometry in Figs.~\ref{fig:fig2}(a) and \ref{fig:fig2}(b) when varying $U'$ and $J$ independently.
All the spinless $p4mm$, the spin-$\mathrm{SU}(2)$ and $\mathcal{T}$ symmetries are preserved.
Every solution we find in \cref{fig:fig2} is metallic, with a gapless Fermi surface, and breaks spin-$\mathrm{SU}(2)$ and $\mathcal{T}$ spontaneously.
We classify the phases by their broken symmetries and further distinguish phases within each symmetry class by Lifshitz transitions.

At large $J/U$ and $U'/U$,  both limits host a fully spin-polarized FM [\cref{fig:fig2}(c)], which we call sym-FM as it preserves the entire plane group $p4mm$ in addition to spin-$\mathrm{U}(1)$.
At small $U'/U$ and $J/U$, both limits host the hole FM [\cref{fig:fig2}(f)] with a magnetization $|S| =(1-\nu)/2$ per site, which comes from doping the N\'eel AFM since they break the primitive translations $T_{(1, 0)}$ and $T_{(0, 1)}$.
For the weak quantum geometric case of $m\gg t$, the sym-FM and hole FM are the only two phases that HF converged to in the entire phase diagram shown in \cref{fig:fig2}(b).
In sharp contrast, in the strong-quantum-geometry regime
($m\ll t$), shown in \cref{fig:fig2}(a),
two additional spin-U(1)-broken phases belonging to different symmetry classes appear~\cite{SM},
together with two additional nematic phases separated by a Lifshitz transition.
    The two spin-U(1)-broken phases have a net spin magnetization. Their in-plane total spin average value $s_{x,y}(\mbf{k})$ vanishes at every $\mbf{k}$~\cite{Dong2026ExtendedHubbardAM} because of the combined symmetry $\exp(-i\pi\sum_{\mbf{R}\alpha}S^z_{\mbf{R}\alpha})T_{(1, 0)}$, while the orbital-resolved spin average values remain nonzero at many $\bsl{k}$, similar to hidden spin polarization~\cite{yuan2023uncovering,hu2026symmetry}.
The two nematic phases are also FM, but they additionally break all the fourfold rotational symmetries in $p4mm$ unlike the hole FM [Figs.~\ref{fig:fig2}(g) and \ref{fig:fig2}(h)].
Nevertheless, \cref{fig:fig2} clearly shows that the strong quantum geometry makes the magnetic phase diagram much richer than that of the weak quantum geometry.

To explore the possibilities of other exotic magnetic phases, we now consider a different combination of the interaction parameters
\eq{
\label{eq:parameter_combine_orbital_aniso}
 U'_{\alpha \beta} &= (U' - U_A)\delta_{\langle \alpha\beta\rangle} + (U' + U_A) \delta_{\llangle \alpha\beta\rrangle}\\
 J_{p,\alpha\beta} &= 3J_{H, \alpha\beta} =  3U_A\delta_{\llangle \alpha \beta\rrangle}\ ,
}
where $\delta_{\langle \alpha \beta\rangle} = \delta_{\alpha, \beta+1}+\delta_{\alpha +1, \beta}$, $\delta_{\llangle \alpha \beta\rrangle } = \delta_{\alpha +2, \beta}$ and recall that the orbital indices are understood modulo 4. For $U_A = 0$ this reduces to the previous case in \cref{fig:fig2} with $J = 0$, and thus $U_A$ characterizes the orbital-anisotropy of the interaction.
Crucially, as mentioned above, this combination still preserves the spinless $p4mm$, the spin-$\mathrm{SU}(2)$ and $\mathcal{T}$  symmetries. 
Therefore, we do not break any symmetry in the many-body Hamiltonian to forbid a N\'eel antiferromagnet.
The converged solutions are presented in \cref{fig:fig3}, where all of them are metallic and break all of $T_{(1, 0)}, T_{(0, 1)}$ and $\mathcal{T}$ symmetries.
The spin-$\mathrm{SU}(2)$ symmetries are also reduced into smaller groups or broken entirely.
As shown in \cref{fig:fig3}(b), the hole FM now  covers the entire phase diagram for the weak quantum geometric case of $m\gg t$.
On the other hand, a much richer phase diagram again emerges for the strong quantum geometric case of $m\ll t$ in \cref{fig:fig3}(a). 
In addition to the phases present in \cref{fig:fig2}(a), we now find a spontaneous N\'eel AFM metal [\cref{fig:fig3}(c)],
the nematic III which is separated from nematic I and II by a Lifshitz transition, the $C_2$-broken nematic that further breaks spin-U(1) and $C_2$ [Figs.~\ref{fig:fig3}(g) and (h)], and a new hole FM phase [\cref{fig:fig3}(f)].
In particular, we identify two AM phases [Figs.~\ref{fig:fig3}(d) and (e)] with distinct Fermi-surface topologies, and they are both protected by $C_{4}\mathcal{T}$ and spin-$\mathrm{U}(1)$ symmetries~\cite{Jiang_2024_Spin_Group,Song_2024_Spin_Group}. 
In real space, their magnetization patterns appear identical to that of the N\'eel AFM. However, they are distinguished from the N\'eel AFM by additional nontrivial order parameters, as discussed in the following.

We now explain why the orbital-anisotropic terms characterized by $U_A$ can stabilize the AM I and AM II phases in the phase diagram in \cref{fig:fig3}(a).
The key is that within the mean-field decoupling of the $U_A$ interaction, there is a pair of uniform and staggered spin-orbital order parameters $g_0$ and $g_\pi$ which give the mean-field term
\begin{equation}\label{eq:Hzterm}
\hat{H}_z = -\sum_{\mbf{R}} (g_0 + g_\pi e^{i \mbf{q}\cdot \mbf{R}}) \hat{Q}_z(\mbf{R}),
\end{equation}
where $\hat{Q}_z(\mbf{R})=\sum_{\alpha ss'}(-1)^{\alpha}c^{\dagger}_{\mbf{R}\alpha s}(\sigma^z)_{ss'}c_{\mbf{R}(\alpha+2)s'}$ with $\alpha +2$ taking values modulo 4, and $\mbf{q} = (\pi, \pi)$.
The order parameters are given by
$g_0 = U_A\langle \sum_{\mbf{R}}\hat{Q}_z(\mbf{R})\rangle /(2 N_{\text{site}})$
and 
$g_\pi = U_A\langle \sum_{\mbf{R}} \hat{Q}_z(\mbf{R}) e^{i \mbf{q}\cdot \mbf{R}}\rangle /(2 N_{\text{site}})$,
where $N_{\text{site}}$ is the number of lattice sites.
The simultaneous appearance of both $g_0$ and $g_{\pi}$ would spontaneously break symmetries in the same way as AM I and II, whereas $g_{\pi}$ alone can be nonzero in the AFM state.
Indeed, for the AFM state at $(U'/U, U_A /U) = (0.25, 0.15)$ in \cref{fig:fig3}(a), $g_0 = 0$ and $g_\pi \approx 6.3\times 10^{-4} t$, while for the AM I state at $(U'/U, U_A/U) = (0.25, 0.45)$, $g_0 \approx 8.6 \times 10^{-4}t$ and $g_\pi \approx 1.3 \times 10^{-3}t$.
This extra $g_0$ is what distinguishes AM and AFM, and favors the AM, as it appears in the mean-field energy as $-N_{\text{site}} g_0^2/U_A$~\cite{SM}.
This energy gain requires strong quantum geometry. For $m\gg t$, the four orbitals on a site enter one Wannier state with equal weights and signs [\cref{fig:fig1}(d)], so the factor $(-1)^{\alpha}$ cancels out in $\langle\hat{Q}_z(\mbf{R})\rangle$, and hence $g_0,g_\pi=O(t/m)$ are negligible. For $m\ll t$, these orbitals belong to different Wannier states [\cref{fig:fig1}(c)], so no such cancellation occurs.
This is a concrete demonstration how the quantum geometry allows the appearance of AM.

\section{Conclusion and Discussion}
We have illustrated how the strong quantum geometry of a narrow band can allow the appearance of exotic magnetic phases, including AM, even when the interaction at the atomic level is strictly onsite and crystalline symmetries do not exclude conventional competing orders. 
We note that a nearest-neighboring density-density interaction channel was considered in \refcite{Dong2026ExtendedHubbardAM}, but the projected interaction form is very different from what we have in the strong quantum geometric case~\cite{SM}, the quantum-geometric origin was not identified in their work, and no collinear AM was found in their work.
Although the strong quantum geometry cases studied here arise in an obstructed atomic phase, we expect that the trivial atomic phases with sufficiently strong quantum geometry and thus large Wannier spread may also host unconventional magnetic orders.
Recent surveys on topological materials~\cite{Xu_2021_Real_Space_Invariants,gao2022unconventional,xu2024feOAI,robredo2025magneticTopoMat,Petralanda_2026_2D_Quantum_Chemistry} have identified numerous crystals with obstructed atomic bands using topological quantum chemistry (TQC)~\cite{bradlyn2017TQC}.
A particular example is FeSb$_2$, a narrow-gap semiconductor~\cite{tomczak2010thermopower,mazin2021prediction} whose bulk bands exhibit an obstructed atomic limit in TQC analyses of DFT calculations within the generalized gradient approximation (GGA)~\cite{gao2022unconventional,Xu_2021_Real_Space_Invariants}.
Metallic states on the (101) surface associated with the obstructed orbitals were predicted by \refcite{iwagaki2026correlated} and further supported by their surface transport measurements.
Since the surface states consist of obstructed orbitals, they can potentially host strong quantum geometry, and thus our framework suggests the possibility of AM on the surface.
Indeed, \refcite{iwagaki2026correlated} showed that electrostatic gating induces a magnetic surface phase, whose order might be AM.
Another example that our framework might be applicable to is ScCoSi, which was identified by TQC analyses of GGA calculations as a metal with obstructed atomic bands~\cite{Xu_2021_Real_Space_Invariants,gao2022unconventional}.
It has also recently been proposed as an AM based on DFT+U results~\cite{zhou2026database}.

\begin{acknowledgments}
\section*{Acknowledgments}
We thank B. Andrei Bernevig, Sankar Das Sarma, Rafael Fernandes, Miguel Gon{\c{c}}alves, Shuo Liu, Jay Deep Sau, Zhi-Da Song, Oskar Vafek, Zhenyu Xiao, Ya-Hui Zhang and Zihan Zhou for helpful discussions.
M.W.\ is pleased to acknowledge that the work reported in this paper was substantially performed using Princeton University's Research Computing resources. The work of Y.G.\ and J.Y. is supported by J.Y.'s startup funds at the University of Florida. The authors received assistance from OpenAI Codex (GPT-5.6 and GPT-6 Astra) for code debugging and literature search, and Claude Code (Opus 5.5) for polishing the Supplemental Material.
The authors independently verified all outputs and retain full responsibility for this work.
\end{acknowledgments}

\bibliography{re}

\clearpage
\onecolumngrid
\appendix

\begin{center}
    \textbf{\large Supplemental Material}
\end{center}

\vspace{1em}


\appcontentssection
    {app:symmetryRho}
    {A. Symmetries of the four-orbital model}

\appcontentssection
    {app:qgWannier}
    {B. Quantum geometry and Wannier states in the lowest band}

\appcontentssection
    {app:projHK}
    {C. Projected Hubbard--Kanamori interactions}

\appcontentssection
    {app:HFhAndChannels}
    {D. Hartree--Fock Hamiltonians and AM/AFM channels}

\appcontentssection
    {app:HFdetails}
    {E. Magnetic orders in the Hartree--Fock numerics}

\appcontentssection
    {app:TDHF}
    {F. Hartree--Fock stability and TDHF spectra for the AM phase}

\clearpage

\begin{bibunit}[apsrev4-2]

%
\makeatletter

\let\supp@cite\cite
\let\supp@citex\@citex

\let\cite\std@cite
\let\@citex\std@@citex

\DeclareRobustCommand{\suppcite}[1]{%
    \begingroup
    \let\@citex\supp@citex
    \renewcommand{\citenumfont}[1]{S##1}%
    \supp@cite{#1}%
    \endgroup
}

\makeatother

\renewcommand{\bibnumfmt}[1]{[S#1]}

\makeatletter
\def\@extra@b@citeb{@supp}
\def\@extra@binfo{@supp}
\makeatother

\section{Symmetries of the four-orbital model}
\label{app:symmetryRho}

In this section, we list the symmetries of the four-orbital model and the corresponding constraints on the one-body density matrix. For convenience, we repeat the Hamiltonian,
\begin{equation}\label{eq:app:Hall}
\begin{aligned}
H ={}& -\sum_{\mbf{R}\alpha s} \left(m\, c_{\mbf{R}\alpha s}^{\dagger} c_{\mbf{R}(\alpha + 1)s} + t\, c_{\mbf{R}\alpha s}^{\dagger} c_{(\mbf{R} + \boldsymbol{\delta}_{\alpha})(\alpha + 1) s}\right) + \mathrm{h.c.}\\
+&\!\sum_{\mbf{R}\alpha} U n_{\mbf{R}\alpha \uparrow} n_{\mbf{R}\alpha \downarrow}  + \!\sum_{\mbf{R}\alpha<\beta}\!\left(\frac{2U_{\alpha\beta}' \!- \!J_{H,\alpha\beta}}{2}n_{\mbf{R}\alpha}n_{\mbf{R}\beta}-2J_{H,\alpha\beta}\mbf{S}_{\mbf{R}\alpha}\!\cdot \mbf{S}_{\mbf{R}\beta} \!\right)+\!\sum_{\mbf{R}\alpha<\beta}J_{p,\alpha\beta} \left(c_{\mbf{R} \alpha\uparrow }^{\dagger}c_{\mbf{R}\alpha \downarrow}^{\dagger} c_{\mbf{R}\beta \downarrow } c_{\mbf{R}\beta \uparrow} + \mathrm{h.c.}\right),
\end{aligned}
\end{equation}
in which $c_{\mbf{R}\alpha s}^{\dagger}$ creates an electron with spin $s = \uparrow,\downarrow$ in orbital $\alpha = 0, 1, 2, 3$ at site $\mbf{R}$.
Following the convention of the main text, the orbital index ($\alpha + 1$) is understood modulo 4, the lattice constant is set to 1, and nearest-neighbor hopping vectors are $\{\bsl{\delta}_{0},  \bsl{\delta}_{1}, \bsl{\delta}_{2}, \bsl{\delta}_{3}\} = \{(-1, 0), \,(0, 1), \,(1, 0), \, (0, -1)\}$.

\begin{figure}[t]
    \centering
    \includegraphics[width=\linewidth]{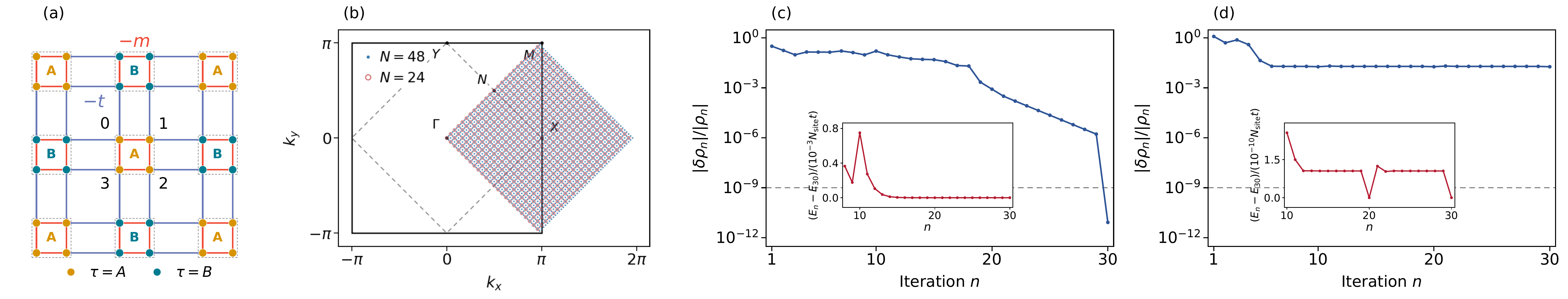}
    \caption{The real-space lattice and reciprocal-space mesh for the HF numerics, as well as the convergence of the HF numerics.
    (a)~Checkerboard sublattices $\tau=A,B$ that allow $\sqrt{2}\times\sqrt{2}$ translation breaking.
    (b)~Momentum meshes for $N_c=N \times N=24\times24$ and $48\times48$ in the reduced BZ.
    (c)--(d) DIIS residual $\|\delta\rho_n\|/\|\rho_n\|$ and energy (inset) versus iteration $n$ for (c) the hole FM in \cref{fig:fsAMdiff} at $(U'/U,J/U)=(0.6,0)$, and for (d) the nonconvergent point $(U'/U,J/U)=(0.1,0.4)$.
    Dashed lines in (c) and (d) show the threshold used for the convergence criterion, while the insets show the difference in energy density of each iteration from the 30th iteration: $(E_n-E_{30})/(10^{-3}N_\text{site}t)$ for the inset in (c), and $(E_n-E_{30})/(10^{-10}N_\text{site}t)$ for the inset in (d).
    }
    \label{fig:latticeMesh}
\end{figure}

In the Hartree--Fock (HF) calculations, we allow $\sqrt{2}\times\sqrt{2}$ translation breaking, i.e., the density matrix $\rho_{\alpha s,\beta s'}(\mbf{R},\mbf{R}')=\langle c^{\dagger}_{\mbf{R}'\beta s'}c_{\mbf{R}\alpha s}\rangle$ is only required to satisfy $\rho_{\alpha s,\beta s'}(\mbf{R}+\mbf{L},\mbf{R}'+\mbf{L})=\rho_{\alpha s,\beta s'}(\mbf{R},\mbf{R}')$ for $e^{i(\pi,\pi)\cdot\mbf{L}}=1$.
Therefore, we introduce the $\sqrt{2}\times\sqrt{2}$ superlattice spanned by $(1,-1)$ and $(1,1)$ with two sublattice sites $\tau=A,B$, as depicted in \cref{fig:latticeMesh}(a).
Let $\mbf{r}$ run over the unit cells of the superlattice. The electron creation operators in the superlattice are defined by
$c^{\dagger}_{\mbf{r}\tau\alpha s}=c^{\dagger}_{\mbf{R}=\mbf{r}+\mbf{\delta}_{\tau},\alpha s}$, where $\mbf{\delta}_{A}=(0,0)$, and $\mbf{\delta}_{B}=(1,0)$.
Define $\mbf{k}$ to be the crystal momentum of the superlattice in the reduced Brillouin zone defined by its primitive reciprocal lattice vectors $(\pi,-\pi)$ and $(\pi,\pi)$, and $c^{\dagger}_{\mbf{k}\tau\alpha s}=N_c^{-1/2}\sum_{\mbf{r}}e^{i\mbf{k}\cdot(\mbf{r}+\mbf{\delta}_{\tau})}c^{\dagger}_{\mbf{r}\tau\alpha s}$, where $N_c=N_{\text{site}}/2 = (\sum_{\mbf{R}}1)/2$ is the number of superlattice cells, which is half of the number of lattice sites $N_{\text{site}}$.
The resulting density matrix is diagonal in $\mbf{k}$, \ie, $\rho_{\tau\alpha s,\tau'\beta s'}(\mbf{k})=\langle c^{\dagger}_{\mbf{k}\tau'\beta s'}c_{\mbf{k}\tau\alpha s}\rangle$.

For both parameter sets of $U'_{\alpha \beta}, J_{H, \alpha \beta}, J_{p, \alpha \beta}$ in the main text, the Hamiltonian in \cref{eq:app:Hall} is invariant under the primitive translations $T_{(1,0)}$ and $T_{(0,1)}$, the fourfold rotation $C_4$, and the mirror $M_x:(x,y)\to(x,-y)$, which generate the plane group $p4mm$, as well as under spin-$\mathrm{SU}(2)$ rotations and time reversal $\mathcal{T}$. Their actions are listed below, with ``$\rho\to\rho'$'' denoting that a state preserving the symmetry under discussion satisfies $\rho=\rho'$.
\begin{itemize}
    \item Primitive translation $T_{(1, 0)}$
    \begin{equation}\label{eq:app:T1}
        T_{(1, 0)}c_{\mbf{R}\alpha s}^{\dagger}T_{(1, 0)}^{-1} = c_{\mbf{R}+(1, 0) \alpha s}^{\dagger}, \quad T_{(1, 0)} c_{\mbf{r}A\alpha s}^{\dagger} T_{(1, 0)}^{-1} = c_{\mbf{r}B\alpha s}^{\dagger}, \quad T_{(1, 0)} c_{\mbf{r}B\alpha s}^{\dagger} T_{(1, 0)}^{-1} = c_{\mbf{r}+(2, 0)A\alpha s}^{\dagger}.
    \end{equation}
    Therefore, $T_{(1, 0)} c_{\mbf{k}\tau\alpha s}^{\dagger} T_{(1, 0)}^{-1} = c_{\mbf{k}\bar{\tau} \alpha s}^{\dagger} e^{ -i k_x}$, where $\bar{\tau} = B, A$ for $\tau = A, B$, respectively.
    Hence, $\rho_{\tau \alpha s, \tau' \beta s'}(\mbf{k})\to \rho_{\bar{\tau}\alpha s, \bar{\tau}'\beta s'}(\mbf{k})$. 

    \item Primitive translation $T_{(0, 1)}$
    \begin{equation}\label{eq:app:T01}
        T_{(0, 1)}c_{\mbf{R}\alpha s}^{\dagger}T_{(0, 1)}^{-1} = c_{\mbf{R}+(0, 1) \alpha s}^{\dagger}, \quad T_{(0, 1)} c_{\mbf{r}A\alpha s}^{\dagger} T_{(0, 1)}^{-1} = c_{\mbf{r}+(-1, 1)B\alpha s}^{\dagger}, \quad T_{(0, 1)} c_{\mbf{r}B\alpha s}^{\dagger} T_{(0, 1)}^{-1} = c_{\mbf{r}+(1, 1)A\alpha s}^{\dagger}.
    \end{equation}
    Therefore, $T_{(0, 1)} c_{\mbf{k}\tau\alpha s}^{\dagger} T_{(0, 1)}^{-1} = c_{\mbf{k}\bar{\tau} \alpha s}^{\dagger} e^{ -i k_y}$, and $\rho_{\tau \alpha s, \tau' \beta s'}(\mbf{k})\to \rho_{\bar{\tau}\alpha s, \bar{\tau}'\beta s'}(\mbf{k})$.

    \item fourfold rotation $C_{4}$
    \begin{equation}\label{eq:app:C4}
    C_{4} c_{\mbf{R}\alpha s}^{\dagger} C_{4}^{-1} = c_{R_4\mbf{R} (\alpha + 1)s}^{\dagger}, \quad C_{4}c_{\mbf{r}A\alpha s}^{\dagger}C_{4}^{-1} = c_{R_4\mbf{r}A(\alpha+1)s}^{\dagger}, \quad C_{4} c_{\mbf{r}B\alpha s}^{\dagger}C_{4}^{-1} = c_{R_4 \mbf{r} +(-1, -1)B(\alpha +1)s }^{\dagger},
    \end{equation}
    where $R_4(x, y) = (y, -x)$. Therefore, $C_{4} c_{\mbf{k}\tau \alpha s}^{\dagger} C_{4}^{-1} = c_{R_4 \mbf{k} \tau(\alpha + 1)s}^{\dagger}$ and $\rho_{\tau \alpha s, \tau' \beta s'}(\mbf{k}) \to \rho_{\tau(\alpha+1)s, \tau'(\beta +1)s'}(R_4 \mbf{k})$.

    \item Mirror $M_x$
    \begin{equation}\label{eq:app:Mx}
    M_x c_{\mbf{R}\alpha s}^{\dagger} M_x^{-1} = c_{m_x\mbf{R} (3-\alpha)s}^{\dagger},\quad M_x c_{\mbf{r}\tau\alpha s}^{\dagger} M_x^{-1} = c_{m_x\mbf{r} \tau (3-\alpha )s}^{\dagger},
    \end{equation}
    in which $m_x(x, y) = (x, -y)$. Therefore, $M_x c_{\mbf{k}\tau \alpha s}^{\dagger} M_x^{-1} = c_{m_x\mbf{k} \tau(3-\alpha)s}^{\dagger}$ and $\rho_{\tau\alpha s, \tau' \beta s'}(\mbf{k}) \to \rho_{\tau(3-\alpha)s, \tau'(3-\beta)s'}(m_x\mbf{k})$.

    \item Time reversal $\mathcal{T}$
    \begin{equation}\label{eq:app:TR}
    \mathcal{T} c_{\mbf{R}\alpha s}^{\dagger} \mathcal{T}^{-1} = \eta_{s} c_{\mbf{R}\alpha \bar{s}}^{\dagger}, \quad
    \mathcal{T} c_{\mbf{r}\tau \alpha s}^{\dagger} \mathcal{T}^{-1} = \eta_s c_{\mbf{r}\tau \alpha \bar{s}}^{\dagger},
    \end{equation}
    in which $\eta_s = 1, -1$ for $s = \uparrow, \downarrow$ respectively, and $\bar{s} = \downarrow, \uparrow$ for $s = \uparrow, \downarrow$ respectively.
    Therefore, $\mathcal{T}c_{\mbf{k}\tau \alpha s}^{\dagger} \mathcal{T}^{-1} = \eta_s c_{-\mbf{k}\tau \alpha \bar{s}}^{\dagger}$ and $\rho_{\tau\alpha s, \tau'\beta s'}(\mbf{k})\to \eta_s \eta_{s'}\rho^{*}_{\tau \alpha \bar{s}, \tau' \beta \bar{s}'}(-\mbf{k})$.
    
\end{itemize}

We also test the combined symmetries $C_{4}T_{(1,0)}$, $C_{4}^2$, $T_{(1,0)}\mathcal{T}$, $C_{4}\mathcal{T}$, $C_{4}T_{(1,0)}\mathcal{T}$, and $C_{4}^2\mathcal{T}$ in later calculations, under which $\rho_{\tau\alpha s,\tau'\beta s'}(\mbf{k})\to\rho_{\bar{\tau}(\alpha + 1)s, \bar{\tau}'(\beta+1)s'}(R_4\mbf{k})$, $\rho_{\tau(\alpha+2)s, \tau'(\beta+2)s'}(-\mbf{k})$, $\eta_s \eta_{s'}\rho_{\bar\tau\alpha\bar{s}, \bar{\tau}'\beta \bar{s}'}^{*}(-\mbf{k})$, $\eta_s \eta_{s'}\rho_{\tau(\alpha+1)\bar{s}, \tau'(\beta+1)\bar{s}'}^{*}(-R_4\mbf{k})$, $\eta_s \eta_{s'}\rho_{\bar{\tau}(\alpha+1)\bar{s}, \bar{\tau}'(\beta+1)\bar{s}'}^{*}(-R_4\mbf{k})$, and $\eta_s \eta_{s'} \rho_{\tau (\alpha+2)\bar{s}, \tau' (\beta+2)\bar{s}'}^{*}(\mbf{k})$, respectively.

\section{Quantum geometry and Wannier states in the lowest band}
\label{app:qgWannier}
In this section, we derive the quantum geometry and the Wannier states of the lowest band of $H_0$. With $c^{\dagger}_{\mbf{k}\alpha s}=N_{\text{site}}^{-1/2}\sum_{\mbf{R}}e^{i\mbf{k}\cdot\mbf{R}}c^{\dagger}_{\mbf{R}\alpha s}$, the hopping Hamiltonian reads
\begin{equation}\label{eq:app:H0momentum}
\begin{aligned}
H_0 ={}&   -\sum_{\mbf{R}\alpha s} \left(m\, c_{\mbf{R}\alpha s}^{\dagger} c_{\mbf{R}(\alpha + 1)s} + t\, c_{\mbf{R}\alpha s}^{\dagger} c_{(\mbf{R} + \boldsymbol{\delta}_{\alpha})(\alpha + 1) s}\right) + \mathrm{h.c.}\\
={}& -\sum_{\mbf{k}\alpha\beta  s}c_{\mbf{k}\alpha s}^{\dagger} \begin{pmatrix}
0 &m + te^{-i k_x } & 0 & m + te^{i k_y }\\
m + t e^{i k_x } & 0 & m + te^{i k_y } & 0\\
0 & m+ te^{-i k_y } & 0  & m + t e^{i k_x }\\
m + t e^{-i k_y } & 0 & m + t e^{-i k_x } & 0
\end{pmatrix}_{\alpha \beta} c_{\mbf{k}\beta s}.
\end{aligned}
\end{equation}
The lowest band has the dispersion $E(\mbf{k})=-\sqrt{m^2+t^2+2mt\cos k_x}-\sqrt{m^2+t^2+2mt\cos k_y}$ and the bandwidth $4\min(m,t)$. Its minimal direct gap to the other three bands is $2|m-t|$, so it is isolated for all $m,t>0$ except for $m=t$. Its Bloch eigenvector can be chosen as $u(\mbf{k}) = e^{i \theta_{\mbf{k}}} (e^{i(\alpha_{\mbf{k}} + \beta_{\mbf{k}})/2}, e^{i(-\alpha_{\mbf{k}} + \beta_{\mbf{k}})/2}, e^{-i(\alpha_{\mbf{k}} + \beta_{\mbf{k}})/2}, e^{i(\alpha_{\mbf{k}} - \beta_{\mbf{k}})/2})^{T}/2$ with $\alpha_{\mbf{k}} = \arg (m + te^{-i k_x})$ and $\beta_{\mbf{k}} = \arg(m + t e^{i k_y})$, and in which $\theta_\mbf{k}$ depends on specific gauge we choose. The band projector is
\begin{equation}\label{eq:app:pk}
\begin{aligned}
&P(\mbf{k})
= u(\mbf{k}) u^{\dagger}(\mbf{k})=\frac{1}{4}\begin{pmatrix}
1 & f(k_x) &f(k_x)f(-k_y) & f(-k_y)\\ f(-k_x) & 1 & f(-k_y) & f(-k_x) f(-k_y) \\ f(-k_x)f(k_y) & f(k_y) & 1 & f(-k_x) \\ f(k_y) & f(k_x ) f(k_y) & f(k_x) & 1
\end{pmatrix},
\end{aligned}
\end{equation}
in which $f(x) = (m + te^{-ix})/\sqrt{m^2 + t^2 + 2mt \cos x}$. 

For $m\neq t$, the quantum geometric tensor $Q_{ij}(\mbf{k}) = \partial_{k_i} u^{\dagger}(\mbf{k}) \partial_{k_j} u(\mbf{k}) - [\partial_{k_i} u^{\dagger}(\mbf{k}) u(\mbf{k})] [u^{\dagger}(\mbf{k})\partial_{k_j} u(\mbf{k})]$ for $i,j = x, y$ is~\suppcite{yu2025quantum},
\begin{equation}\label{eq:app:qgViaP}
\begin{aligned}
&Q_{ij}(\mbf{k}) = \text{Tr}(P(\mbf{k}) \partial_{k_i}P(\mbf{k}) \partial_{k_j}P(\mbf{k}))  = \frac{-1}{4}\left(\begin{smallmatrix}
(f'(k_x)/f(k_x))^2 & 0 \\ 0& (f'(k_y)/f(k_y))^2
\end{smallmatrix}\right) = \delta_{ij} \frac{t^2(m\cos k_i+t)^2}{4(m^2 + t^2 + 2mt \cos k_i)^2}.
\end{aligned}
\end{equation}
Since $Q_{ij}(\mbf{k})$ is real and diagonal, the quantum metric $g_{ij}(\mbf{k})=[Q_{ij}(\mbf{k})+Q_{ji}(\mbf{k})]/2=\text{Tr}[\partial_{k_i}P(\mbf{k})\partial_{k_j}P(\mbf{k})]/2$ equals $Q_{ij}(\mbf{k})$, and the Berry curvature $F_{ij}(\mbf{k})=i[Q_{ij}(\mbf{k})-Q_{ji}(\mbf{k})]$ vanishes. The Dirichlet functional~\suppcite{Yu_2025_WL_ideal} is,
\begin{equation}\label{eq:app:Trgk}
\Tr \G = \int_{\text{BZ}} d^2\mbf{k} \text{Tr}[g(\mbf{k})] = \pi t^2 \int_{0}^{2\pi}dx \frac{(m\cos x + t)^2}{(m^2 + t^2 + 2mt \cos x)^2}=\begin{cases}
\displaystyle \pi^2\frac{2t^2-m^2}{t^2-m^2}, & m<t,\\[6pt]
\displaystyle \pi^2\frac{t^2}{m^2-t^2}, & m>t.
\end{cases}
\end{equation}
which diverges as $m\to t$, where the gap closes. In the limits $m\ll t$ and $m\gg t$, the bands are equally narrow, yet $\Tr\G=2\pi^2+O(m^2/t^2)$ and $O(t^2/m^2)$, respectively, bearing vastly different quantum geometry. 

\begin{figure}[t]
    \centering
    \includegraphics[width=0.6\linewidth]{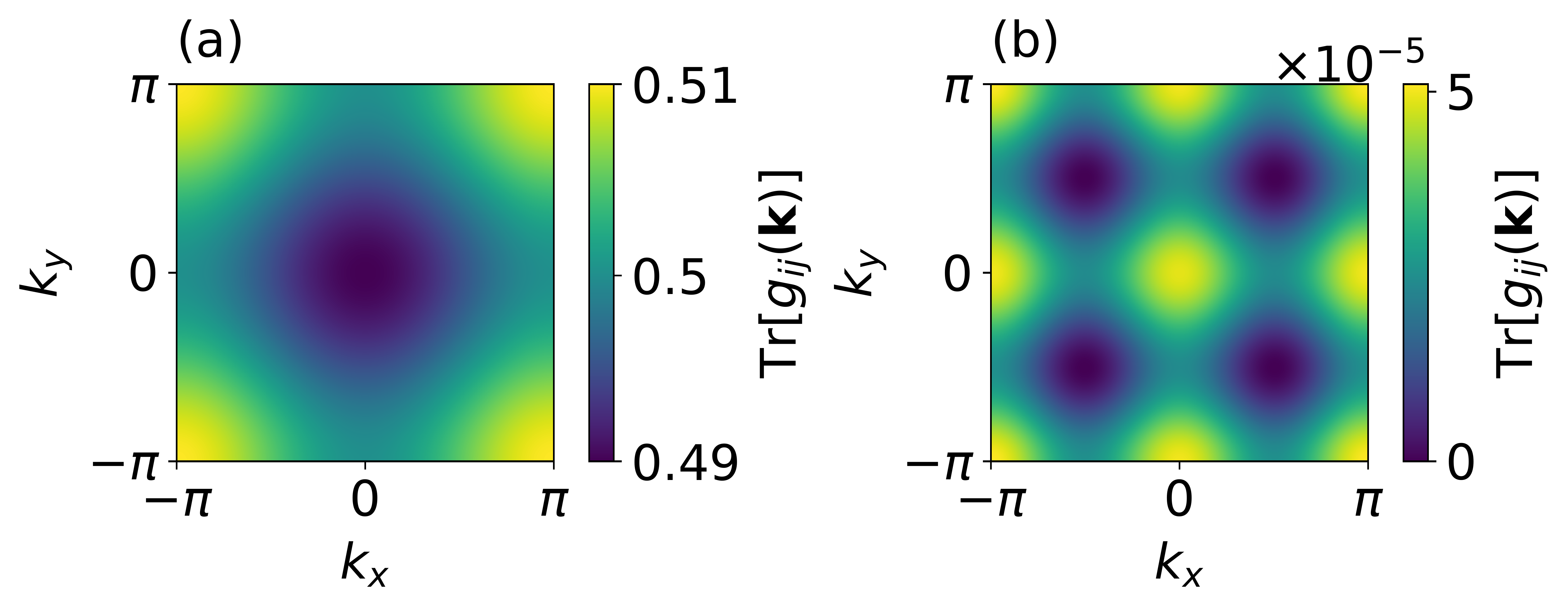}
    \caption{The quantum metric $\text{Tr}[g_{ij}(\mbf{k})]$ across the Brillouin zone for (a) the strong quantum geometry case, $(m, t) = (0.01, 1)$, and (b) the weak quantum geometry case, $(m, t) = (1, 0.01)$.}
    \label{fig:trGkinBZ}
\end{figure}

\begin{figure}[t]
    \centering
    \includegraphics[width=\linewidth]{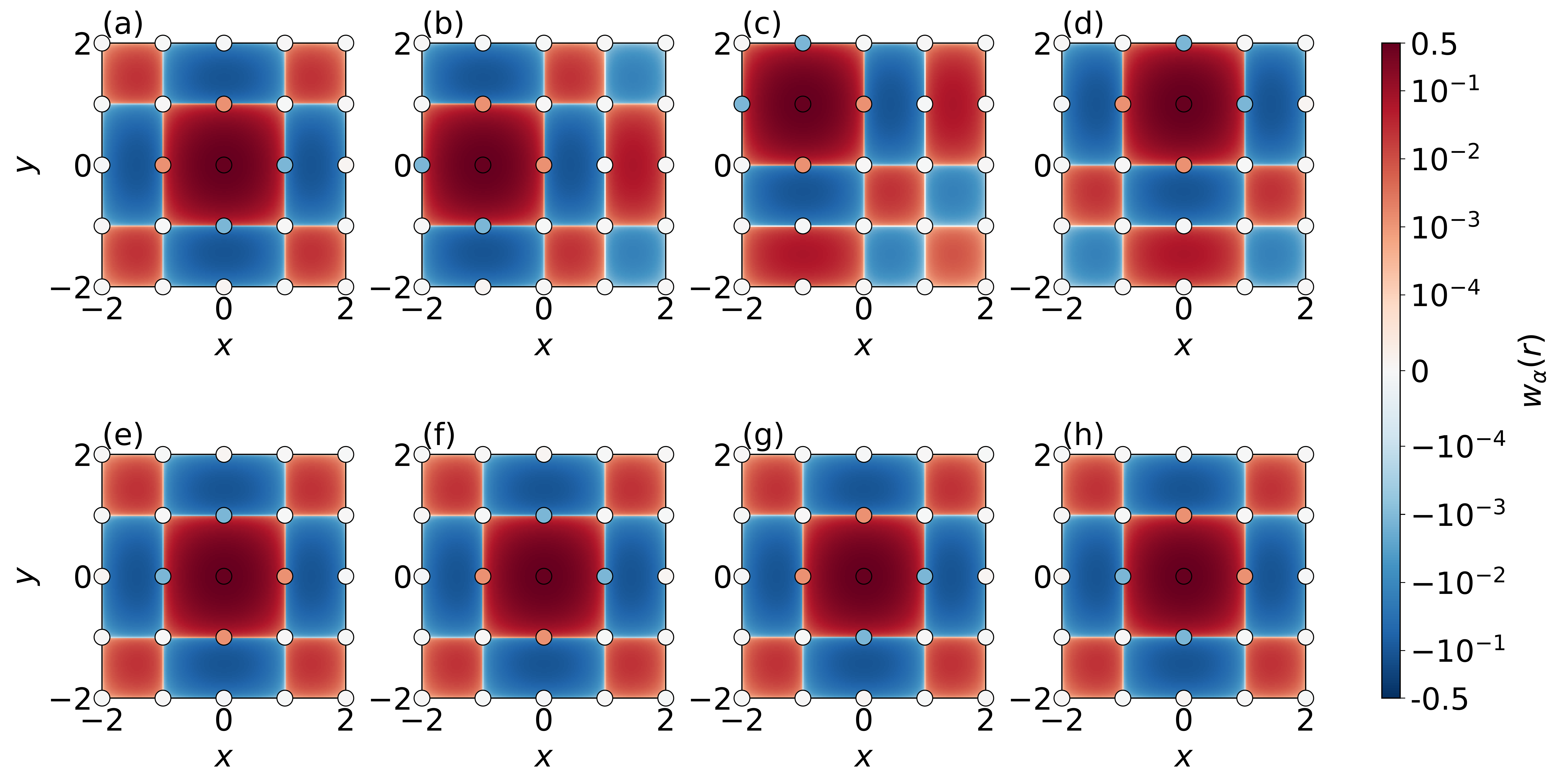}
    \caption{Wannier wavefunctions $w_{\alpha}(\mbf{r})$ with the minimal localization functional, \ie\ minimal spread, $\Omega =\Tr\G/(2\pi)^2$. Circles mark the lattice sites, each colored by its $w_{\alpha}(\mbf{R})$ that enters $d^{\dagger}_{\mbf{R}s}$. The background shows $w_{\alpha}(\mbf{r})$ at continuous $\mbf{r}$. (a)--(d) $\alpha=0,1,2,3$ at $(m,t)=(0.01,1)$. (e)--(h) $\alpha=0,1,2,3$ at $(m,t)=(1,0.01)$. }
    \label{fig:wannierRealSpace}
\end{figure}

We now construct the Wannier states for $m\neq t$. In a smooth gauge of $u(\mbf{k})$, the Wannier function is $w(\mbf{r})=\int_{\text{BZ}}\frac{d^2k}{(2\pi)^2}e^{i\mbf{k}\cdot\mbf{r}}u(\mbf{k})$, and $d^{\dagger}_{\mbf{R}s}=\sum_{\mbf{R}'\alpha}w_{\alpha}(\mbf{R}'-\mbf{R})c^{\dagger}_{\mbf{R}'\alpha s}$ creates the Wannier state associated with $\mbf{R}$. Its spread $\Omega=\langle r^2\rangle-\langle\mbf{r}\rangle^2$~\suppcite{marzari1997maximally,yu2025quantum} splits into the gauge-invariant part $\Tr\G/(2\pi)^2$ and the gauge-dependent part $\int_{\text{BZ}}\frac{d^2k}{(2\pi)^2}|\mbf{A}(\mbf{k})-\overline{\mbf{A}}|^2$~\suppcite{marzari1997maximally}, where $\mbf{A}(\mbf{k})=i u^{\dagger}(\mbf{k})\nabla_{\mbf{k}}u(\mbf{k})$ is the Berry connection, $\overline{\mbf{A}}$ is its average over the BZ, and $\overline{\mbf{A}}=\langle\mbf{r}\rangle$ is the Wannier center. Because the Berry curvature vanishes, we can choose a gauge with a constant $\mbf{A}(\mbf{k})$, which removes the gauge-dependent part and gives maximally localized Wannier states with $\Omega=\Tr\G/(2\pi)^2$~\suppcite{Wang2026SymmetryQG}. For $m<t$, we take $e^{i\theta_{\mbf{k}}}=e^{i(k_x-k_y)/2}$, with $u(\mbf{k})$ continued smoothly across the BZ, which gives $\mbf{A}=(-1/2,1/2)$, i.e., a Wannier center at the plaquette center $\mbf{R}+(-1/2,1/2)$. For $m>t$, we take $\theta_{\mbf{k}}=0$, which gives $\mbf{A}=0$ and a Wannier center on the atomic site. In both gauges $w_\alpha(\mbf{r})$ is real.
To first order in $m/t$ or $t/m$, with $\hat{\mbf{x}}=(1,0)$ and $\hat{\mbf{y}}=(0,1)$,
\begin{align}
\label{eq:app:limitdR}
d_{\mbf{R}s}^{\dagger}
=
\left\{
\begin{aligned}
&\frac{1}{2}\Big(
c_{\mbf{R},0s}^{\dagger}
+c_{\mbf{R}-\hat{\mbf{x}},1s}^{\dagger}
+c_{\mbf{R}-\hat{\mbf{x}}+\hat{\mbf{y}},2s}^{\dagger}
+c_{\mbf{R}+\hat{\mbf{y}},3s}^{\dagger}
\Big)
+\frac{m}{8t}\Big[
c_{\mbf{R}-\hat{\mbf{x}},0s}^{\dagger}
-c_{\mbf{R}+\hat{\mbf{x}},0s}^{\dagger}
+c_{\mbf{R}+\hat{\mbf{y}},0s}^{\dagger}
-c_{\mbf{R}-\hat{\mbf{y}},0s}^{\dagger}\\
&\quad
+c_{\mbf{R},1s}^{\dagger}
-c_{\mbf{R}-2\hat{\mbf{x}},1s}^{\dagger}
+c_{\mbf{R}-\hat{\mbf{x}}+\hat{\mbf{y}},1s}^{\dagger}
-c_{\mbf{R}-\hat{\mbf{x}}-\hat{\mbf{y}},1s}^{\dagger}
+c_{\mbf{R}+\hat{\mbf{y}},2s}^{\dagger}
-c_{\mbf{R}-2\hat{\mbf{x}}+\hat{\mbf{y}},2s}^{\dagger}
+c_{\mbf{R}-\hat{\mbf{x}},2s}^{\dagger}
-c_{\mbf{R}-\hat{\mbf{x}}+2\hat{\mbf{y}},2s}^{\dagger}\\
&\quad
+c_{\mbf{R}-\hat{\mbf{x}}+\hat{\mbf{y}},3s}^{\dagger}
-c_{\mbf{R}+\hat{\mbf{x}}+\hat{\mbf{y}},3s}^{\dagger}
+c_{\mbf{R},3s}^{\dagger}
-c_{\mbf{R}+2\hat{\mbf{y}},3s}^{\dagger}
\Big]
+O\left(\frac{m^2}{t^2}\right),
\quad m\ll t,\\[6pt]
&\frac{1}{2}\sum_{\alpha=0}^{3}
c_{\mbf{R}\alpha s}^{\dagger}
+\frac{t}{8m}\Big[
c_{\mbf{R}+\hat{\mbf{x}},0s}^{\dagger}
-c_{\mbf{R}-\hat{\mbf{x}},0s}^{\dagger}
-c_{\mbf{R}+\hat{\mbf{y}},0s}^{\dagger}
+c_{\mbf{R}-\hat{\mbf{y}},0s}^{\dagger}\\
&\quad
-c_{\mbf{R}+\hat{\mbf{x}},1s}^{\dagger}
+c_{\mbf{R}-\hat{\mbf{x}},1s}^{\dagger}
-c_{\mbf{R}+\hat{\mbf{y}},1s}^{\dagger}
+c_{\mbf{R}-\hat{\mbf{y}},1s}^{\dagger}
-c_{\mbf{R}+\hat{\mbf{x}},2s}^{\dagger}
+c_{\mbf{R}-\hat{\mbf{x}},2s}^{\dagger}
+c_{\mbf{R}+\hat{\mbf{y}},2s}^{\dagger}
-c_{\mbf{R}-\hat{\mbf{y}},2s}^{\dagger}\\
&\quad
+c_{\mbf{R}+\hat{\mbf{x}},3s}^{\dagger}
-c_{\mbf{R}-\hat{\mbf{x}},3s}^{\dagger}
+c_{\mbf{R}+\hat{\mbf{y}},3s}^{\dagger}
-c_{\mbf{R}-\hat{\mbf{y}},3s}^{\dagger}
\Big]
+O\left(\frac{t^2}{m^2}\right),
\quad m\gg t.
\end{aligned}
\right.
\end{align}
\Cref{fig:trGkinBZ,fig:wannierRealSpace} show $\text{Tr}[g(\mbf{k})]$ and $w_{\alpha}(\mbf{r})$ in the two limits, respectively.

\section{Projected Hubbard--Kanamori interactions}
\label{app:projHK}
We now discuss the projection of the generalized Hubbard--Kanamori interaction $H_{\text{int}}$ in \cref{eq:H_int} onto the lowest band, to illustrate how the projected interaction channels are related to quantum geometry. 
Using the real Wannier functions in \cref{app:qgWannier}, the projected electron operators are given by $Pc^{\dagger}_{\mbf{R}\alpha s}P=\sum_{\mbf{R}'}w_{\alpha}(\mbf{R}-\mbf{R}')d^{\dagger}_{\mbf{R}'s}$, which to first order reads
\begin{align}
\label{eq:app:PcRP}
&\left.
\begin{aligned}
Pc_{\mbf{R}0s}^{\dagger}P
&=\frac{1}{2}d_{\mbf{R}s}^{\dagger}
+\frac{m}{8t}\Big(
d_{\mbf{R}+\hat{\mbf{x}},s}^{\dagger}
-d_{\mbf{R}-\hat{\mbf{x}},s}^{\dagger}
+d_{\mbf{R}-\hat{\mbf{y}},s}^{\dagger}
-d_{\mbf{R}+\hat{\mbf{y}},s}^{\dagger}
\Big)\\
Pc_{\mbf{R}1s}^{\dagger}P
&=\frac{1}{2}d_{\mbf{R}+\hat{\mbf{x}},s}^{\dagger}
+\frac{m}{8t}\Big(
d_{\mbf{R}s}^{\dagger}
-d_{\mbf{R}+2\hat{\mbf{x}},s}^{\dagger}
+d_{\mbf{R}+\hat{\mbf{x}}-\hat{\mbf{y}},s}^{\dagger}
-d_{\mbf{R}+\hat{\mbf{x}}+\hat{\mbf{y}},s}^{\dagger}
\Big)\\
Pc_{\mbf{R}2s}^{\dagger}P
&=\frac{1}{2}d_{\mbf{R}+\hat{\mbf{x}}-\hat{\mbf{y}},s}^{\dagger}
+\frac{m}{8t}\Big(
d_{\mbf{R}-\hat{\mbf{y}},s}^{\dagger}
-d_{\mbf{R}+2\hat{\mbf{x}}-\hat{\mbf{y}},s}^{\dagger}
+d_{\mbf{R}+\hat{\mbf{x}},s}^{\dagger}
-d_{\mbf{R}+\hat{\mbf{x}}-2\hat{\mbf{y}},s}^{\dagger}
\Big)\\
Pc_{\mbf{R}3s}^{\dagger}P
&=\frac{1}{2}d_{\mbf{R}-\hat{\mbf{y}},s}^{\dagger}
+\frac{m}{8t}\Big(
d_{\mbf{R}+\hat{\mbf{x}}-\hat{\mbf{y}},s}^{\dagger}
-d_{\mbf{R}-\hat{\mbf{x}}-\hat{\mbf{y}},s}^{\dagger}
+d_{\mbf{R}s}^{\dagger}
-d_{\mbf{R}-2\hat{\mbf{y}},s}^{\dagger}
\Big)
\end{aligned}
\ \right\}
+O\left(\frac{m^2}{t^2}\right),
\quad m\ll t,
\nonumber\\[6pt]
&\left.
\begin{aligned}
Pc_{\mbf{R}0s}^{\dagger}P
&=\frac{1}{2}d_{\mbf{R}s}^{\dagger}
+\frac{t}{8m}\Big(
d_{\mbf{R}-\hat{\mbf{x}},s}^{\dagger}
-d_{\mbf{R}+\hat{\mbf{x}},s}^{\dagger}
+d_{\mbf{R}+\hat{\mbf{y}},s}^{\dagger}
-d_{\mbf{R}-\hat{\mbf{y}},s}^{\dagger}
\Big)\\
Pc_{\mbf{R}1s}^{\dagger}P
&=\frac{1}{2}d_{\mbf{R}s}^{\dagger}
+\frac{t}{8m}\Big(
d_{\mbf{R}+\hat{\mbf{x}},s}^{\dagger}
-d_{\mbf{R}-\hat{\mbf{x}},s}^{\dagger}
+d_{\mbf{R}+\hat{\mbf{y}},s}^{\dagger}
-d_{\mbf{R}-\hat{\mbf{y}},s}^{\dagger}
\Big)\\
Pc_{\mbf{R}2s}^{\dagger}P
&=\frac{1}{2}d_{\mbf{R}s}^{\dagger}
+\frac{t}{8m}\Big(
d_{\mbf{R}+\hat{\mbf{x}},s}^{\dagger}
-d_{\mbf{R}-\hat{\mbf{x}},s}^{\dagger}
+d_{\mbf{R}-\hat{\mbf{y}},s}^{\dagger}
-d_{\mbf{R}+\hat{\mbf{y}},s}^{\dagger}
\Big)\\
Pc_{\mbf{R}3s}^{\dagger}P
&=\frac{1}{2}d_{\mbf{R}s}^{\dagger}
+\frac{t}{8m}\Big(
d_{\mbf{R}-\hat{\mbf{x}},s}^{\dagger}
-d_{\mbf{R}+\hat{\mbf{x}},s}^{\dagger}
+d_{\mbf{R}-\hat{\mbf{y}},s}^{\dagger}
-d_{\mbf{R}+\hat{\mbf{y}},s}^{\dagger}
\Big)
\end{aligned}
\ \right\}
+O\left(\frac{t^2}{m^2}\right),
\quad m\gg t.
\end{align}

In the normal-ordered form, 
\begin{eqnarray}
\hat{H}_{\text{int}} & = & U\sum_{\mbf{R}\alpha} c_{\mbf{R}\alpha \uparrow}^{\dagger} c_{\mbf{R}\alpha \downarrow}^{\dagger} c_{\mbf{R}\alpha \downarrow} c_{\mbf{R}\alpha \uparrow} + \sum_{\mbf{R}\alpha<\beta, ss'}U'_{\alpha \beta} c_{\mbf{R}\alpha s}^{\dagger} c_{\mbf{R}\beta s'}^{\dagger} c_{\mbf{R}\beta s'} c_{\mbf{R}\alpha s} \nonumber\\
&&-\sum_{\mbf{R}\alpha<\beta}J_{H, \alpha \beta}(c_{\mbf{R}\alpha \uparrow}^{\dagger} c_{\mbf{R}\beta \uparrow}^{\dagger} c_{\mbf{R}\beta \uparrow} c_{\mbf{R}\alpha \uparrow} + c_{\mbf{R}\alpha \downarrow}^{\dagger} c_{\mbf{R}\beta \downarrow}^{\dagger} c_{\mbf{R}\beta \downarrow} c_{\mbf{R}\alpha \downarrow} + c_{\mbf{R}\alpha \uparrow}^{\dagger} c_{\mbf{R}\beta \downarrow}^{\dagger} c_{\mbf{R}\beta \uparrow} c_{\mbf{R}\alpha \downarrow} + c_{\mbf{R}\alpha \downarrow}^{\dagger} c_{\mbf{R}\beta \uparrow}^{\dagger} c_{\mbf{R}\beta \downarrow} c_{\mbf{R}\alpha \uparrow}) \nonumber\\
&&+ \sum_{\mbf{R}\alpha <\beta}J_{p, \alpha \beta} ( c_{\mbf{R} \alpha\uparrow }^{\dagger}c_{\mbf{R} \alpha\downarrow}^{\dagger} c_{\mbf{R} \beta\downarrow }  c_{\mbf{R} \beta\uparrow } + c_{\mbf{R} \beta \uparrow}^{\dagger} c_{\mbf{R}\beta\downarrow}^{\dagger} c_{\mbf{R}\alpha\downarrow} c_{\mbf{R} \alpha\uparrow}).
\end{eqnarray}
In this form, the projection is obtained by replacing each operator by its projected counterpart, $c^{\dagger}c^{\dagger}cc\,\rightarrow(Pc^{\dagger}P)(Pc^{\dagger}P)(PcP)(PcP)$. Other orderings would generate peculiar one-body terms because $\{Pc_{\mbf{R}\alpha s}P,Pc^{\dagger}_{\mbf{R}'\beta s'}P\}\neq\delta_{\mbf{R}\mbf{R}'}\delta_{\alpha\beta}\delta_{ss'}$. Keeping $Pc^{\dagger}P$ to first order gives $\hat{H}_{\text{int}}^{\text{proj}}$ to first order in $m/t$ ($t/m$), with corrections of $O(m^2/t^2)$ [$O(t^2/m^2)$]. 

We write $\hat{H}_{\text{int}}^{\text{proj}}=\hat{H}^{(0)}_{m\ll t}+(m/t)\hat{H}^{(1)}_{m\ll t}+O(m^2/t^2)$ for $m\ll t$, and $\hat{H}_{\text{int}}^{\text{proj}}=\hat{H}^{(0)}_{t\ll m}+(t/m)\hat{H}^{(1)}_{t\ll m}+O(t^2/m^2)$ for $m\gg t$. At zeroth order, the projected interaction already contains nearest-neighbor (NN) and next-nearest-neighbor (NNN) terms for $m\ll t$, but is purely onsite for $m\gg t$,
\begin{align}\label{eq:app:Hm<<t0}
\hat{H}_{m\ll t}^{(0)}
={}&\frac{U}{4}\sum_{\mbf{R}}n_{\mbf{R}\uparrow}^{d}n_{\mbf{R}\downarrow}^{d}
+\frac{1}{16}\sum_{\mbf{R}}\Big\{
\Big[U'_{01}+U'_{23}-\frac{J_{H,01}+J_{H,23}}{2}\Big]n_{\mbf{R}}^{d}n_{\mbf{R}+\hat{\mbf{x}}}^{d}
+\Big[U'_{03}+U'_{12}-\frac{J_{H,03}+J_{H,12}}{2}\Big]n_{\mbf{R}}^{d}n_{\mbf{R}+\hat{\mbf{y}}}^{d}
\nonumber\\
&+\Big[U'_{13}-\frac{J_{H,13}}{2}\Big]n_{\mbf{R}}^{d}n_{\mbf{R}+\hat{\mbf{x}}+\hat{\mbf{y}}}^{d}
+\Big[U'_{02}-\frac{J_{H,02}}{2}\Big]n_{\mbf{R}}^{d} n_{\mbf{R}+\hat{\mbf{x}}-\hat{\mbf{y}}}^{d}
-2J_{H,13}\,\mbf{S}_{\mbf{R}}^{d}\cdot\mbf{S}_{\mbf{R}+\hat{\mbf{x}}+\hat{\mbf{y}}}^{d}
-2J_{H,02}\,\mbf{S}_{\mbf{R}}^{d}\cdot\mbf{S}_{\mbf{R}+\hat{\mbf{x}}-\hat{\mbf{y}}}^{d}
\nonumber\\
&-2\left(J_{H,01}+J_{H,23}\right)\mbf{S}_{\mbf{R}}^{d}\cdot\mbf{S}_{\mbf{R}+\hat{\mbf{x}}}^{d}
-2\left(J_{H,03}+J_{H,12}\right)\mbf{S}_{\mbf{R}}^{d}\cdot\mbf{S}_{\mbf{R}+\hat{\mbf{y}}}^{d}
+\Big(\Big[\left(J_{p,01}+J_{p,23}\right)d_{\mbf{R}+\hat{\mbf{x}},\uparrow}^{\dagger}d_{\mbf{R}+\hat{\mbf{x}},\downarrow}^{\dagger}
\nonumber\\
&+\left(J_{p,03}+J_{p,12}\right)d_{\mbf{R}+\hat{\mbf{y}},\uparrow}^{\dagger}d_{\mbf{R}+\hat{\mbf{y}},\downarrow}^{\dagger}
+J_{p,13}\,d_{\mbf{R}+\hat{\mbf{x}}+\hat{\mbf{y}},\uparrow}^{\dagger}d_{\mbf{R}+\hat{\mbf{x}}+\hat{\mbf{y}},\downarrow}^{\dagger}
+J_{p,02}\,d_{\mbf{R}+\hat{\mbf{x}}-\hat{\mbf{y}},\uparrow}^{\dagger}d_{\mbf{R}+\hat{\mbf{x}}-\hat{\mbf{y}},\downarrow}^{\dagger}\Big]
d_{\mbf{R}\downarrow}d_{\mbf{R}\uparrow}+\text{H.c.}\Big)\Big\},
\end{align}
\begin{align}\label{eq:app:Ht<<m0}
\hat{H}_{t\ll m}^{(0)} = \sum_{\mbf{R}}\left[\frac{U}{4} + \frac{\sum_{\alpha<\beta}( U'_{\alpha\beta} + J_{H, \alpha \beta} + J_{p, \alpha \beta})}{8}\right] n_{\mbf{R}\uparrow}^{d} n_{\mbf{R}\downarrow}^{d},
\end{align}
where $n^{d}_{\mbf{R}s}=d^{\dagger}_{\mbf{R}s}d_{\mbf{R}s}$, $n^{d}_{\mbf{R}}=n^{d}_{\mbf{R}\uparrow}+n^{d}_{\mbf{R}\downarrow}$, and $S^{d,\mu}_{\mbf{R}}=\sum_{ss'}d^{\dagger}_{\mbf{R}s}(\sigma^{\mu}/2)_{ss'}d_{\mbf{R}s'}$. In both limits, $U$ only gives the onsite term $(U/4)\sum_{\mbf{R}}n^{d}_{\mbf{R}\uparrow}n^{d}_{\mbf{R}\downarrow}$. For $m\ll t$, each inter-orbital coupling generates an interaction along the bond that connects the two orbitals in \cref{fig:fig1}(c), e.g., $U'_{01}$ gives $n^{d}_{\mbf{R}}n^{d}_{\mbf{R}+\hat{\mbf{x}}}$, and $U'_{02}$ gives $n^{d}_{\mbf{R}}n^{d}_{\mbf{R}+\hat{\mbf{x}}-\hat{\mbf{y}}}$. For $m\gg t$, the inter-orbital couplings only enhance the onsite Hubbard repulsion, with equal weights for $U'_{\alpha\beta}$, $J_{H,\alpha\beta}$, and $J_{p,\alpha\beta}$. The first-order term for $m\gg t$ is,
\begin{align}\label{eq:app:H1t<<m}
\frac{t}{m}\hat{H}_{t\ll m}^{(1)} ={}&\frac{t}{32m}\sum_{\mbf{R}s}\Big[
\left(U'_{12}+J_{H,12}+J_{p,12}-U'_{03}-J_{H,03}-J_{p,03}\right)
n_{\mbf{R}s}
\big(d_{\mbf{R}\bar{s}}^{\dagger}(d_{\mbf{R} + \hat{\mbf{x}} \bar{s}} - d_{\mbf{R} - \hat{\mbf{x}} \bar{s}}) +\text{H.c.}\big)
\nonumber\\
&\quad
+\left(U'_{01}+J_{H,01}+J_{p,01} - U'_{23}-J_{H,23}-J_{p,23}\right)
n_{\mbf{R}s}
\big(d_{\mbf{R}\bar{s}}^{\dagger}(d_{\mbf{R} + \hat{\mbf{y}} \bar{s}} - d_{\mbf{R} - \hat{\mbf{y}} \bar{s}}) +\text{H.c.}\big)
\Big],
\end{align}
where $\bar{s}$ is the spin opposite to $s$. Neither $U$ nor the diagonal couplings ($\alpha\beta=02,13$) contribute at this order. Inversion $C_2=C_4^2$ requires $U'_{\alpha\beta}=U'_{(\alpha+2)(\beta+2)}, J_{H, \alpha \beta} = J_{H, (\alpha+2)(\beta+2)}, J_{p, \alpha\beta} = J_{p, (\alpha+2)(\beta+2)}$, so both coefficients in \cref{eq:app:H1t<<m} vanish and $P\hat{H}_{\text{int}}P=\hat{H}^{(0)}_{t\ll m}+O(t^2/m^2)$. For $m\ll t$, the first-order terms do not vanish. For the two parameter sets of the main text, $U'_{\alpha\beta}=U'$, $J_{H,\alpha\beta}=J_{p,\alpha\beta}=J$ (denoted $U$-$U'$-$J$) and \cref{eq:parameter_combine_orbital_aniso} (denoted $U$-$U'$-$U_A$), we obtain for $m\ll t$, 
\begin{align}
\label{eq:app:PHPJ}
&\hat{H}_{\text{int}, U\text{-}U'\text{-}J}^{\text{proj}}=\frac{U}{4}\sum_{\mbf{R}}n_{\mbf{R}\uparrow}n_{\mbf{R}\downarrow}+\frac{1}{8}\sum_{\langle\mbf{R}\mbf{R}'\rangle}\Big[\Big(U'-\frac{J}{2}\Big)n_{\mbf{R}}n_{\mbf{R}'}-2J\,\mbf{S}_{\mbf{R}}\cdot\mbf{S}_{\mbf{R}'}+J\big(d_{\mbf{R}\uparrow}^{\dagger}d_{\mbf{R}\downarrow}^{\dagger}d_{\mbf{R}'\downarrow}d_{\mbf{R}'\uparrow}+\text{H.c.}\big)\Big]
\nonumber\\
&+\frac{1}{16}\sum_{\llangle\mbf{R}\mbf{R}'\rrangle}\Big[\Big(U'-\frac{J}{2}\Big)n_{\mbf{R}}n_{\mbf{R}'}-2J\,\mbf{S}_{\mbf{R}}\cdot\mbf{S}_{\mbf{R}'}+J\big(d_{\mbf{R}\uparrow}^{\dagger}d_{\mbf{R}\downarrow}^{\dagger}d_{\mbf{R}'\downarrow}d_{\mbf{R}'\uparrow}+\text{H.c.}\big)\Big]
\nonumber\\
&+\frac{m}{t}\bigg\{\frac{1}{32}\sum_{\langle\mbf{R}\mbf{R}'\rangle}\Big[\sum_{s}\Big(\left(U'+2J\right)n_{\mbf{R}'\bar{s}}d_{\mbf{R}s}^{\dagger}d_{\mbf{R}'s}-\Big(U'-\frac{J}{2}\Big)n_{\mbf{R}'}d_{\mbf{R}s}^{\dagger}d_{2\mbf{R}-\mbf{R}',s}\Big)+J\sum_{ss'}\mbf{S}_{\mbf{R}'}\cdot d_{\mbf{R}s}^{\dagger}\boldsymbol{\sigma}_{ss'}d_{2\mbf{R}-\mbf{R}',s'}
\nonumber\\
&-J\big(d_{\mbf{R}\uparrow}^{\dagger}d_{2\mbf{R}-\mbf{R}',\downarrow}^{\dagger}-d_{\mbf{R}\downarrow}^{\dagger}d_{2\mbf{R}-\mbf{R}',\uparrow}^{\dagger}\big)d_{\mbf{R}'\downarrow}d_{\mbf{R}'\uparrow}+\text{H.c.}+(\mbf{R}\leftrightarrow\mbf{R}')\Big]+\frac{1}{64}\sum_{\llangle\mbf{R}\mbf{R}'\rrangle}\sum_{\substack{\mbf{R}'':\langle\mbf{R}\mbf{R}''\rangle,\\ \langle\mbf{R}''\mbf{R}'\rangle}}\Big[
\nonumber\\
&\Big(U'-\frac{J}{2}\Big)\sum_{s}n_{\mbf{R}'}d_{\mbf{R}s}^{\dagger}(d_{\mbf{R}''s}-d_{2\mbf{R}-\mbf{R}'',s})-J\sum_{ss'}\mbf{S}_{\mbf{R}'}\cdot d_{\mbf{R}s}^{\dagger}\boldsymbol{\sigma}_{ss'}(d_{\mbf{R}''s'}-d_{2\mbf{R}-\mbf{R}'',s'})+J\big[d_{\mbf{R}\uparrow}^{\dagger}(d_{\mbf{R}''\downarrow}^{\dagger}-d_{2\mbf{R}-\mbf{R}'',\downarrow}^{\dagger})
\nonumber\\
&-d_{\mbf{R}\downarrow}^{\dagger}(d_{\mbf{R}''\uparrow}^{\dagger}-d_{2\mbf{R}-\mbf{R}'',\uparrow}^{\dagger})\big]d_{\mbf{R}'\downarrow}d_{\mbf{R}'\uparrow}+\text{H.c.}+(\mbf{R}\leftrightarrow\mbf{R}')\Big]\bigg\}+O\Big(\frac{m^2}{t^2}\Big),
\end{align}
\begin{align}
\label{eq:app:PHPUA}
&\hat{H}_{\text{int}, U\text{-}U'\text{-}U_A}^{\text{proj}}=\frac{U}{4}\sum_{\mbf{R}}n_{\mbf{R}\uparrow}n_{\mbf{R}\downarrow}+\frac{U'-U_A}{8}\sum_{\langle\mbf{R}\mbf{R}'\rangle}n_{\mbf{R}}n_{\mbf{R}'}+\frac{1}{16}\sum_{\llangle\mbf{R}\mbf{R}'\rrangle}\Big[\Big(U'+\frac{U_A}{2}\Big)n_{\mbf{R}}n_{\mbf{R}'}-2U_A\,\mbf{S}_{\mbf{R}}\cdot\mbf{S}_{\mbf{R}'}
\nonumber\\
&+3U_A\big(d_{\mbf{R}\uparrow}^{\dagger}d_{\mbf{R}\downarrow}^{\dagger}d_{\mbf{R}'\downarrow}d_{\mbf{R}'\uparrow}+\text{H.c.}\big)\Big]+\frac{m}{t}\bigg\{\frac{U'-U_A}{32}\sum_{\langle\mbf{R}\mbf{R}'\rangle}\sum_{s}\big(n_{\mbf{R}'\bar{s}}d_{\mbf{R}s}^{\dagger}d_{\mbf{R}'s}-n_{\mbf{R}'}d_{\mbf{R}s}^{\dagger}d_{2\mbf{R}-\mbf{R}',s}+\text{H.c.}+(\mbf{R}\leftrightarrow\mbf{R}')\big)
\nonumber\\
&+\frac{1}{64}\sum_{\llangle\mbf{R}\mbf{R}'\rrangle}\sum_{\substack{\mbf{R}'':\langle\mbf{R}\mbf{R}''\rangle,\\ \langle\mbf{R}''\mbf{R}'\rangle}}\Big[\Big(U'+\frac{U_A}{2}\Big)\sum_{s}n_{\mbf{R}'}d_{\mbf{R}s}^{\dagger}(d_{\mbf{R}''s}-d_{2\mbf{R}-\mbf{R}'',s})-U_A\sum_{ss'}\mbf{S}_{\mbf{R}'}\cdot d_{\mbf{R}s}^{\dagger}\boldsymbol{\sigma}_{ss'}(d_{\mbf{R}''s'}-d_{2\mbf{R}-\mbf{R}'',s'})
\nonumber\\
&+3U_A\big[d_{\mbf{R}\uparrow}^{\dagger}(d_{\mbf{R}''\downarrow}^{\dagger}-d_{2\mbf{R}-\mbf{R}'',\downarrow}^{\dagger})-d_{\mbf{R}\downarrow}^{\dagger}(d_{\mbf{R}''\uparrow}^{\dagger}-d_{2\mbf{R}-\mbf{R}'',\uparrow}^{\dagger})\big]d_{\mbf{R}'\downarrow}d_{\mbf{R}'\uparrow}+\text{H.c.}+(\mbf{R}\leftrightarrow\mbf{R}')\Big]\bigg\}+O\Big(\frac{m^2}{t^2}\Big),
\end{align}
where $\langle\mbf{R}\mbf{R}'\rangle$ ($\llangle\mbf{R}\mbf{R}'\rrangle$) runs over NN (NNN) bonds, each counted once, and $\mbf{R}''$ as specified in $\mbf{R}'':\langle\mbf{R}\mbf{R}''\rangle,  \langle\mbf{R}''\mbf{R}'\rangle$ runs over the two common NN sites of $\mbf{R}$ and $\mbf{R}'$. These projected interactions differ from the single-orbital extended Hubbard model of Ref.~\suppcite{Dong2026ExtendedHubbardAM}, which contains only an onsite repulsion and an NN density-density interaction. Besides the NN density-density term, \cref{eq:app:PHPJ,eq:app:PHPUA} contain NN spin-exchange, NN pair-hopping, and three analogous NNN terms at zeroth order. For $t\ll m$, we obtain,
\begin{align}
\label{eq:app:PHPJatomic}
&\hat{H}_{\text{int},U\text{-}U'\text{-}J}^{\text{proj}}=\frac{U+3U'+6J}{4}\sum_{\mbf{R}}n_{\mbf{R}\uparrow}n_{\mbf{R}\downarrow}+O\Big(\frac{t^2}{m^2}\Big),
\end{align}
\begin{align}
\label{eq:app:PHPUAatomic}
&\hat{H}_{\text{int},U\text{-}U'\text{-}U_A}^{\text{proj}}=\frac{U+3U'+3U_A}{4}\sum_{\mbf{R}}n_{\mbf{R}\uparrow}n_{\mbf{R}\downarrow}+O\Big(\frac{t^2}{m^2}\Big).
\end{align}

\section{Hartree--Fock Hamiltonians and AM/AFM channels}
\label{app:HFhAndChannels}
In this section, we give the Hartree--Fock Hamiltonians for the $U$-$U'$-$J$ and $U$-$U'$-$U_A$ parameter sets defined in the main text [\cref{eq:parameter_combine_orbital_aniso}] and in \cref{app:projHK}. In the HF approximation, the energy $E[\rho]=\langle\hat{H}_0+\hat{H}_{\text{int}}\rangle$ of a Slater determinant is a functional of its one-body density matrix $\rho_{\alpha s,\beta s'}(\mbf{R},\mbf{R}')=\langle c^{\dagger}_{\mbf{R}'\beta s'}c_{\mbf{R}\alpha s}\rangle$, and the HF Hamiltonian $\hat{H}^{\text{HF}}=\sum_{\mbf{R}\mbf{R}'\alpha s\beta s'}c^{\dagger}_{\mbf{R}\alpha s}h^{\text{HF}}_{\alpha s,\beta s'}(\mbf{R},\mbf{R}')c_{\mbf{R}'\beta s'}$ is defined by the functional derivative $h^{\text{HF}}_{\alpha s,\beta s'}(\mbf{R},\mbf{R}')=\delta E/\delta\rho_{\beta s',\alpha s}(\mbf{R}',\mbf{R})$~\suppcite{KangPRL2026}. 
A self-consistent solution $\rho$ is a stationary point of $E[\rho]$ which satisfies $[h^{\text{HF}},\rho]=0$ and $\rho^2=\rho$, with the $N_e$ lowest levels of $h^{\text{HF}}$ occupied so that $\Tr\rho=N_e$. The HF ground state is the lowest-energy self-consistent solution. Since $\hat{H}_{\text{int}}$ is strictly onsite, the interaction part of $h^{\text{HF}}(\mbf{R},\mbf{R}')$ is nonzero only for $\mbf{R}=\mbf{R}'$.

We first specify how the direct Hartree term of the total site density is removed~\suppcite{zhou2005electron,leiria2010extended,belozerov2012monoclinic,Himmetoglu_2013_HubbardDFT,park2014computing,he2016photoinduced,jiang2016interatomic,chang2024downfolding,haas2024intersite,Dong2026ExtendedHubbardAM}. Hartree terms arise only from the density-density interaction terms $\hat{H}_n$ of $\hat{H}_{\text{int}}$, which read,
\begin{align}\label{eq:app:directHUJ}
\hat{H}_{U\text{-}U'\text{-}J, n} = U\sum_{\mbf{R}\alpha} n_{\mbf{R} \alpha \uparrow} n_{\mbf{R} \alpha\downarrow} + U'\sum_{\mbf{R} \alpha<\beta}\left(n_{\mbf{R} \alpha \uparrow} + n_{\mbf{R} \alpha \downarrow}\right)\left(n_{\mbf{R} \beta \uparrow} + n_{\mbf{R} \beta \downarrow}\right) - J\sum_{\mbf{R} \alpha<\beta}\left(n_{\mbf{R}\alpha \uparrow} n_{\mbf{R}\beta \uparrow} + n_{\mbf{R} \alpha \downarrow} n_{\mbf{R} \beta \downarrow}\right),
\end{align}
\begin{align}\label{eq:app:directHUUA}
\hat{H}_{U\text{-}U'\text{-}U_A, n} &= U\sum_{\mbf{R}\alpha} n_{\mbf{R} \alpha \uparrow} n_{\mbf{R} \alpha\downarrow} + U'\sum_{\mbf{R} \alpha<\beta}\left(n_{\mbf{R} \alpha \uparrow} + n_{\mbf{R} \alpha \downarrow}\right)\left(n_{\mbf{R} \beta \uparrow} + n_{\mbf{R} \beta \downarrow}\right)\nonumber\\
&-U_A\sum_{\mbf{R}(\alpha\beta) = (01), (12), (23), (03)}\left(n_{\mbf{R} \alpha \uparrow} + n_{\mbf{R} \alpha \downarrow}\right)\left(n_{\mbf{R} \beta \uparrow} + n_{\mbf{R} \beta \downarrow}\right) + U_A\sum_{\mbf{R}(\alpha \beta) = (02), (13)} (n_{\mbf{R}\alpha \uparrow} n_{\mbf{R}\beta \downarrow} + n_{\mbf{R}\alpha \downarrow} n_{\mbf{R}\beta \uparrow}),
\end{align}
where $n_{\mbf{R}\alpha s}=c^{\dagger}_{\mbf{R}\alpha s}c_{\mbf{R}\alpha s}$, $n_{\mbf{R}\alpha}=n_{\mbf{R}\alpha\uparrow}+n_{\mbf{R}\alpha\downarrow}$, and $n_{\mbf{R}}=\sum_{\alpha}n_{\mbf{R}\alpha}$. In terms of the orbital polarization $A_{\mbf{R}\alpha}=n_{\mbf{R}\alpha}-n_{\mbf{R}}/4$ and the spin polarization $B_{\mbf{R}\alpha}=n_{\mbf{R}\alpha\uparrow}-n_{\mbf{R}\alpha\downarrow}$, $n_{\mbf{R}\alpha\uparrow(\downarrow)}=n_{\mbf{R}}/8+A_{\mbf{R}\alpha}/2\pm B_{\mbf{R}\alpha}/2$, and the $\hat{H}_n$ terms become,
\begin{align}\label{eq:app:directHUJab}
\hat{H}_{U\text{-}U'\text{-}J, n} = \frac{U + 6U' - 3J}{16}\sum_{\mbf{R}} n_{\mbf{R}}^2+\frac{U - 2U' + J}{4}\sum_{\mbf{R}\alpha} A_{\mbf{R}\alpha}^2 - \frac{U}{4}\sum_{\mbf{R}\alpha} B_{\mbf{R}\alpha}^2 - \frac{J}{2}\sum_{\mbf{R}\alpha<\beta}B_{\mbf{R}\alpha} B_{\mbf{R}\beta},
\end{align}
\begin{align}\label{eq:app:directHUUAab}
\hat{H}_{U\text{-}U'\text{-}U_A, n} &=\frac{U + 6U' - 3U_{A}}{16}\sum_{\mbf{R}} n_{\mbf{R}}^2+ \frac{U - 2 U' - U_A}{4}\sum_{\mbf{R}\alpha} A_{\mbf{R}\alpha}^2 + \frac{3U_A}{4}\sum_{\mbf{R}}(A_{\mbf{R}0} + A_{\mbf{R}2})^2 +  \frac{3U_A}{4}\sum_{\mbf{R}}(A_{\mbf{R}1} + A_{\mbf{R}3})^2 \nonumber\\
&- \frac{U}{4}\sum_{\mbf{R}\alpha} B_{\mbf{R}\alpha}^2 - \frac{U_A}{2}\sum_{\mbf{R}} B_{\mbf{R}0}B_{\mbf{R}2} - \frac{U_A}{2}\sum_{\mbf{R}}B_{\mbf{R}1} B_{\mbf{R}3}.
\end{align}
The Hartree terms follow by replacing, in each quadratic form, one factor of $n$, $A$, or $B$ by its expectation value.
For the resulting terms, we drop the ones proportional to $\sum_{\mbf{R}}\langle n_{\mbf{R}}\rangle n_{\mbf{R}}$ to avoid double counting the direct Hartree terms, assuming that their effect is already incorporated in $\hat{H}_0$~\suppcite{Dong2026ExtendedHubbardAM}. The remaining Hartree terms are,
\begin{align}\label{eq:app:directHUJab2}
\hat{H}_{U\text{-}U'\text{-}J, n}^{\text{Hartree}} =\frac{U - 2U' + J}{2}\sum_{\mbf{R}\alpha}\langle A_{\mbf{R}\alpha}\rangle A_{\mbf{R}\alpha} - \frac{U}{2}\sum_{\mbf{R}\alpha} \langle B_{\mbf{R}\alpha}\rangle B_{\mbf{R}\alpha} - \frac{J}{2}\sum_{\mbf{R}\alpha<\beta}(\langle B_{\mbf{R}\alpha}\rangle  B_{\mbf{R}\beta} + \langle B_{\mbf{R}\beta}\rangle B_{\mbf{R}\alpha} ),
\end{align}
\begin{align}\label{eq:app:directHUUAab2}
\hat{H}_{U\text{-}U'\text{-}U_A, n}^{\text{Hartree}} = {}& \frac{U - 2 U' - U_A}{2}\sum_{\mbf{R}\alpha} \langle A_{\mbf{R}\alpha}\rangle A_{\mbf{R}\alpha} + \frac{3U_A}{2}\sum_{\mbf{R}}\langle (A_{\mbf{R}0} + A_{\mbf{R}2})\rangle (A_{\mbf{R}0} + A_{\mbf{R}2}) \nonumber\\
&+  \frac{3U_A}{2}\sum_{\mbf{R}}\langle (A_{\mbf{R}1} + A_{\mbf{R}3})\rangle(A_{\mbf{R}1} + A_{\mbf{R}3}) - \frac{U}{2}\sum_{\mbf{R}\alpha} \langle B_{\mbf{R}\alpha}\rangle B_{\mbf{R}\alpha} \nonumber\\
& - \frac{U_A}{2}\sum_{\mbf{R}} (\langle B_{\mbf{R}0}\rangle B_{\mbf{R}2} + \langle B_{\mbf{R}2}\rangle B_{\mbf{R}0}) - \frac{U_A}{2}\sum_{\mbf{R}}(\langle B_{\mbf{R}1}\rangle  B_{\mbf{R}3} + \langle B_{\mbf{R}3}\rangle B_{\mbf{R}1} ).
\end{align}

Including the Fock terms, the HF Hamiltonians after this subtraction are,
\begin{align}\label{eq:app:HHFUU'J}
\hat{H}^{\text{HF}}_{U\text{-}U'\text{-}J} &= \hat{H}_0 + \hat{H}^{\text{HF}}_{U} + \hat{H}^{\text{HF}}_{U'} + \hat{H}^{\text{HF}}_{J},\\
\hat{H}^{\text{HF}}_{U\text{-}U'\text{-}U_A} &= \hat{H}_0 + \hat{H}^{\text{HF}}_{U} + \hat{H}^{\text{HF}}_{U'} + \hat{H}^{\text{HF}}_{U_A},
\end{align}
where $\hat{H}_0=H_0$ in \cref{eq:H_0}, 
and
\begin{align}\label{eq:app:HHFU}
\hat{H}^{\text{HF}}_{U} ={}& \frac{U}{2} \sum_{\mbf{R} \alpha} \left(n_{\mbf{R}  \alpha \uparrow} + n_{\mbf{R}  \alpha \downarrow} - \frac{n_{\mbf{R} }}{4}\right) \left(\langle n_{\mbf{R} \alpha \uparrow}\rangle + \langle n_{\mbf{R} \alpha \downarrow}\rangle - \frac{\langle n_{\mbf{R} }\rangle}{4}\right)\nonumber\\
&+\frac{U}{2} \sum_{\mbf{R} \alpha}\left(n_{\mbf{R}  \alpha \uparrow} - n_{\mbf{R} \alpha \downarrow}\right)\left(\langle n_{\mbf{R}  \alpha \downarrow}\rangle - \langle n_{\mbf{R}  \alpha \uparrow}\rangle\right)-U \sum_{\mbf{R}  \alpha} \bigg(c_{\mbf{R} \alpha \uparrow}^{\dagger} c_{\mbf{R} \alpha \downarrow}\langle c_{\mbf{R} \alpha \downarrow}^{\dagger} c_{\mbf{R} \alpha \uparrow}\rangle + c_{\mbf{R}  \alpha \downarrow}^{\dagger} c_{\mbf{R}  \alpha \uparrow} \langle c_{\mbf{R}  \alpha \uparrow}^{\dagger} c_{\mbf{R}  \alpha \downarrow}\rangle\bigg),
\end{align}
\begin{align}\label{eq:app:HHFU'}
\hat{H}^{\text{HF}}_{U'} ={}& -U' \sum_{\mbf{R} \alpha} \left(n_{\mbf{R}  \alpha \uparrow} + n_{\mbf{R}  \alpha \downarrow} - \frac{n_{\mbf{R} }}{4}\right) \left(\langle n_{\mbf{R} \alpha \uparrow}\rangle + \langle n_{\mbf{R} \alpha \downarrow}\rangle - \frac{\langle n_{\mbf{R} }\rangle}{4}\right)\nonumber\\
&-U' \sum_{\mbf{R}, \alpha < \beta} \bigg( c_{\mbf{R}  \alpha \uparrow}^{\dagger} c_{\mbf{R}  \beta \uparrow} \langle c_{\mbf{R}  \beta \uparrow}^{\dagger} c_{\mbf{R}  \alpha \uparrow}\rangle + c_{\mbf{R}  \beta \uparrow}^{\dagger} c_{\mbf{R} \alpha \uparrow} \langle c_{\mbf{R} \alpha \uparrow}^{\dagger} c_{\mbf{R} \beta \uparrow}\rangle + c_{\mbf{R} \alpha \uparrow}^{\dagger} c_{\mbf{R} \beta \downarrow} \langle c_{\mbf{R} \beta \downarrow}^{\dagger} c_{\mbf{R} \alpha \uparrow}\rangle + c_{\mbf{R} \beta \downarrow}^{\dagger} c_{\mbf{R} \alpha \uparrow} \langle c_{\mbf{R} \alpha \uparrow}^{\dagger} c_{\mbf{R} \beta \downarrow}\rangle\nonumber\\
&+c_{\mbf{R} \alpha \downarrow}^{\dagger} c_{\mbf{R} \beta \uparrow} \langle c_{\mbf{R} \beta \uparrow}^{\dagger} c_{\mbf{R} \alpha \downarrow}\rangle + c_{\mbf{R} \beta \uparrow}^{\dagger} c_{\mbf{R} \alpha \downarrow} \langle c_{\mbf{R} \alpha \downarrow}^{\dagger} c_{\mbf{R} \beta \uparrow}\rangle + c_{\mbf{R} \alpha \downarrow}^{\dagger} c_{\mbf{R} \beta \downarrow} \langle c_{\mbf{R} \beta \downarrow}^{\dagger} c_{\mbf{R} \alpha \downarrow}\rangle + c_{\mbf{R} \beta \downarrow}^{\dagger} c_{\mbf{R} \alpha \downarrow} \langle c_{\mbf{R} \alpha \downarrow}^{\dagger} c_{\mbf{R} \beta \downarrow}\rangle\bigg),
\end{align}
\begin{align}\label{eq:app:HHFJ}
\hat{H}^{\text{HF}}_{J}={}&\frac{J}{2} \sum_{\mbf{R} \alpha} \left(n_{\mbf{R}  \alpha \uparrow} + n_{\mbf{R}  \alpha \downarrow} - \frac{n_{\mbf{R} }}{4}\right) \left(\langle n_{\mbf{R} \alpha \uparrow}\rangle + \langle n_{\mbf{R} \alpha \downarrow}\rangle - \frac{\langle n_{\mbf{R} }\rangle}{4}\right)\nonumber\\
&- \frac{J}{2}\sum_{\mbf{R},  \alpha < \beta} \left[\left(n_{\mbf{R} \alpha \uparrow} - n_{\mbf{R} \alpha \downarrow}\right)\left(\langle n_{\mbf{R} \beta \uparrow}\rangle - \langle n_{\mbf{R} \beta \downarrow}\rangle\right) + \left(n_{\mbf{R} \beta \uparrow} - n_{\mbf{R} \beta \downarrow}\right)\left(\langle n_{\mbf{R} \alpha \uparrow}\rangle - \langle n_{\mbf{R} \alpha \downarrow}\rangle\right)\right]\nonumber\\
&-J \sum_{\mbf{R}, \alpha <\beta} \bigg(c_{\mbf{R} \alpha \uparrow}^{\dagger} c_{\mbf{R} \alpha \downarrow}\langle c_{\mbf{R} \beta \downarrow}^{\dagger} c_{\mbf{R} \beta \uparrow}\rangle + c_{\mbf{R} \beta \downarrow}^{\dagger} c_{\mbf{R} \beta \uparrow}\langle c_{\mbf{R} \alpha \uparrow}^{\dagger} c_{\mbf{R} \alpha \downarrow}\rangle + c_{\mbf{R} \alpha \downarrow}^{\dagger} c_{\mbf{R}\alpha \uparrow}\langle c_{\mbf{R} \beta \uparrow}^{\dagger} c_{\mbf{R} \beta \downarrow}\rangle + c_{\mbf{R} \beta \uparrow}^{\dagger} c_{\mbf{R} \beta \downarrow}\langle c_{\mbf{R} \alpha \downarrow}^{\dagger} c_{\mbf{R} \alpha \uparrow}\rangle\nonumber\\
&\qquad +c_{\mbf{R} \alpha \uparrow}^{\dagger} c_{\mbf{R} \beta \downarrow}\langle c_{\mbf{R} \alpha \downarrow}^{\dagger} c_{\mbf{R} \beta \uparrow}\rangle + c_{\mbf{R} \alpha \downarrow}^{\dagger} c_{\mbf{R} \beta \uparrow}\langle c_{\mbf{R} \alpha \uparrow}^{\dagger} c_{\mbf{R} \beta \downarrow}\rangle + c_{\mbf{R} \beta \uparrow}^{\dagger} c_{\mbf{R} \alpha \downarrow}\langle c_{\mbf{R} \beta \downarrow}^{\dagger} c_{\mbf{R} \alpha \uparrow}\rangle + c_{\mbf{R} \beta \downarrow}^{\dagger} c_{\mbf{R} \alpha \uparrow}\langle c_{\mbf{R} \beta \uparrow}^{\dagger} c_{\mbf{R} \alpha \downarrow}\rangle\bigg)\nonumber\\
&+ J \sum_{\mbf{R}, \alpha < \beta}\bigg( c_{\mbf{R}\alpha \uparrow}^{\dagger} c_{\mbf{R} \beta \uparrow}\langle c_{\mbf{R} \beta \uparrow}^{\dagger} c_{\mbf{R} \alpha \uparrow}\rangle + c_{\mbf{R} \beta \uparrow}^{\dagger} c_{\mbf{R} \alpha \uparrow}\langle c_{\mbf{R} \alpha \uparrow}^{\dagger} c_{\mbf{R} \beta \uparrow}\rangle + c_{\mbf{R}\alpha \downarrow}^{\dagger} c_{\mbf{R} \beta \downarrow}\langle c_{\mbf{R} \beta \downarrow}^{\dagger} c_{\mbf{R} \alpha \downarrow}\rangle + c_{\mbf{R} \beta \downarrow}^{\dagger} c_{\mbf{R}\alpha \downarrow}\langle c_{\mbf{R}\alpha \downarrow}^{\dagger} c_{\mbf{R} \beta \downarrow}\rangle\nonumber\\
&\qquad +c_{\mbf{R} \alpha \uparrow}^{\dagger} c_{\mbf{R} \beta \uparrow}\langle c_{\mbf{R}\beta \downarrow}^{\dagger} c_{\mbf{R} \alpha \downarrow}\rangle + c_{\mbf{R} \beta \downarrow}^{\dagger} c_{\mbf{R} \alpha \downarrow}\langle c_{\mbf{R} \alpha \uparrow}^{\dagger} c_{\mbf{R} \beta \uparrow}\rangle + c_{\mbf{R} \alpha \downarrow}^{\dagger} c_{\mbf{R} \beta \downarrow}\langle c_{\mbf{R} \beta \uparrow}^{\dagger} c_{\mbf{R} \alpha \uparrow}\rangle + c_{\mbf{R} \beta \uparrow}^{\dagger} c_{\mbf{R} \alpha \uparrow}\langle c_{\mbf{R} \alpha \downarrow}^{\dagger} c_{\mbf{R} \beta \downarrow}\rangle\nonumber\\
&\qquad +c_{\mbf{R} \alpha \uparrow}^{\dagger} c_{\mbf{R} \beta \uparrow}\langle c_{\mbf{R} \alpha \downarrow}^{\dagger} c_{\mbf{R} \beta \downarrow}\rangle + c_{\mbf{R} \alpha \downarrow}^{\dagger} c_{\mbf{R} \beta \downarrow}\langle c_{\mbf{R} \alpha \uparrow}^{\dagger} c_{\mbf{R} \beta \uparrow}\rangle + c_{\mbf{R}\beta \uparrow}^{\dagger} c_{\mbf{R} \alpha \uparrow}\langle c_{\mbf{R} \beta \downarrow}^{\dagger} c_{\mbf{R} \alpha \downarrow}\rangle + c_{\mbf{R} \beta \downarrow}^{\dagger} c_{\mbf{R} \alpha \downarrow}\langle c_{\mbf{R} \beta \uparrow}^{\dagger} c_{\mbf{R} \alpha \uparrow}\rangle\bigg),
\end{align}
\begin{align}
\label{eq:app:HHFUA}
&\hat H_{U_A}^{\mathrm{HF}}=-\frac{U_A}{2}\sum_{\mbf{R}\alpha} \left(n_{\mbf{R}\alpha\uparrow}+n_{\mbf{R}\alpha\downarrow}-\frac{n_{\mbf{R}}}{4}\right)\left(\langle n_{\mbf{R}\alpha\uparrow}\rangle+\langle n_{\mbf{R}\alpha\downarrow}\rangle-\frac{\langle n_{\mbf{R}}\rangle}{4}\right)\nonumber\\
&+\frac{3U_A}{2}\sum_{\mbf{R},\,(\alpha\beta)=(02),(13)} \left(n_{\mbf{R}\alpha}+n_{\mbf{R}\beta}-\frac{n_{\mbf{R}}}{2}\right)\left(\langle n_{\mbf{R}\alpha}\rangle+\langle n_{\mbf{R}\beta}\rangle-\frac{\langle n_{\mbf{R}}\rangle}{2}\right)\nonumber\\
&-\frac{U_A}{2}\sum_{\mbf{R},\,(\alpha\beta)=(02),(13)}\bigg[ \left(n_{\mbf{R}\alpha\uparrow}-n_{\mbf{R}\alpha\downarrow}\right)\left(\langle n_{\mbf{R}\beta\uparrow}\rangle-\langle n_{\mbf{R}\beta\downarrow}\rangle\right)+\left(n_{\mbf{R}\beta\uparrow}-n_{\mbf{R}\beta\downarrow}\right)\left(\langle n_{\mbf{R}\alpha\uparrow}\rangle-\langle n_{\mbf{R}\alpha\downarrow}\rangle\right)\bigg]\nonumber\\
&-U_A\sum_{\mbf{R},\,(\alpha\beta)=(02),(13)}\bigg( c_{\mbf{R} \alpha \uparrow}^{\dagger} c_{\mbf{R} \alpha \downarrow}\langle c_{\mbf{R} \beta \downarrow}^{\dagger} c_{\mbf{R} \beta \uparrow}\rangle + c_{\mbf{R} \beta \downarrow}^{\dagger} c_{\mbf{R} \beta \uparrow}\langle c_{\mbf{R} \alpha \uparrow}^{\dagger} c_{\mbf{R} \alpha \downarrow}\rangle + c_{\mbf{R} \alpha \downarrow}^{\dagger} c_{\mbf{R} \alpha \uparrow}\langle c_{\mbf{R} \beta \uparrow}^{\dagger} c_{\mbf{R} \beta \downarrow}\rangle + c_{\mbf{R} \beta \uparrow}^{\dagger} c_{\mbf{R} \beta \downarrow}\langle c_{\mbf{R} \alpha \downarrow}^{\dagger} c_{\mbf{R} \alpha \uparrow}\rangle\nonumber\\
&\qquad + 3c_{\mbf{R} \alpha \uparrow}^{\dagger} c_{\mbf{R} \beta \downarrow}\langle c_{\mbf{R} \alpha \downarrow}^{\dagger} c_{\mbf{R} \beta \uparrow}\rangle + 3c_{\mbf{R} \alpha \downarrow}^{\dagger} c_{\mbf{R} \beta \uparrow}\langle c_{\mbf{R} \alpha \uparrow}^{\dagger} c_{\mbf{R} \beta \downarrow}\rangle + 3c_{\mbf{R} \beta \uparrow}^{\dagger} c_{\mbf{R} \alpha \downarrow}\langle c_{\mbf{R} \beta \downarrow}^{\dagger} c_{\mbf{R} \alpha \uparrow}\rangle + 3c_{\mbf{R} \beta \downarrow}^{\dagger} c_{\mbf{R} \alpha \uparrow}\langle c_{\mbf{R} \beta \uparrow}^{\dagger} c_{\mbf{R} \alpha \downarrow}\rangle\bigg)\nonumber\\
&+U_A\sum_{\mbf{R},\,(\alpha\beta)=(01),(12),(23),(03)}\bigg( c_{\mbf{R} \alpha \uparrow}^{\dagger} c_{\mbf{R} \beta \uparrow}\langle c_{\mbf{R} \beta \uparrow}^{\dagger} c_{\mbf{R} \alpha \uparrow}\rangle + c_{\mbf{R} \beta \uparrow}^{\dagger} c_{\mbf{R} \alpha \uparrow}\langle c_{\mbf{R} \alpha \uparrow}^{\dagger} c_{\mbf{R} \beta \uparrow}\rangle + c_{\mbf{R} \alpha \uparrow}^{\dagger} c_{\mbf{R} \beta \downarrow}\langle c_{\mbf{R} \beta \downarrow}^{\dagger} c_{\mbf{R} \alpha \uparrow}\rangle \nonumber\\
&\qquad\qquad\qquad\qquad\qquad\qquad + c_{\mbf{R} \beta \downarrow}^{\dagger} c_{\mbf{R} \alpha \uparrow}\langle c_{\mbf{R} \alpha \uparrow}^{\dagger} c_{\mbf{R} \beta \downarrow}\rangle + c_{\mbf{R} \alpha \downarrow}^{\dagger} c_{\mbf{R} \beta \uparrow}\langle c_{\mbf{R} \beta \uparrow}^{\dagger} c_{\mbf{R} \alpha \downarrow}\rangle + c_{\mbf{R} \beta \uparrow}^{\dagger} c_{\mbf{R} \alpha \downarrow}\langle c_{\mbf{R} \alpha \downarrow}^{\dagger} c_{\mbf{R} \beta \uparrow}\rangle \nonumber\\
&\qquad\qquad\qquad\qquad\qquad\qquad + c_{\mbf{R} \alpha \downarrow}^{\dagger} c_{\mbf{R} \beta \downarrow}\langle c_{\mbf{R} \beta \downarrow}^{\dagger} c_{\mbf{R} \alpha \downarrow}\rangle + c_{\mbf{R} \beta \downarrow}^{\dagger} c_{\mbf{R} \alpha \downarrow}\langle c_{\mbf{R} \alpha \downarrow}^{\dagger} c_{\mbf{R} \beta \downarrow}\rangle\bigg)\nonumber\\
&-U_A\sum_{\mbf{R},\,(\alpha\beta)=(02),(13)}\bigg(\colorbox{black!10}{$\displaystyle c_{\mbf{R} \alpha \uparrow}^{\dagger} c_{\mbf{R} \beta \uparrow}\langle c_{\mbf{R} \beta \uparrow}^{\dagger} c_{\mbf{R} \alpha \uparrow}\rangle + c_{\mbf{R} \beta \uparrow}^{\dagger} c_{\mbf{R} \alpha \uparrow}\langle c_{\mbf{R} \alpha \uparrow}^{\dagger} c_{\mbf{R} \beta \uparrow}\rangle $} + c_{\mbf{R} \alpha \uparrow}^{\dagger} c_{\mbf{R} \beta \downarrow}\langle c_{\mbf{R} \beta \downarrow}^{\dagger} c_{\mbf{R} \alpha \uparrow}\rangle + c_{\mbf{R} \beta \downarrow}^{\dagger} c_{\mbf{R} \alpha \uparrow}\langle c_{\mbf{R} \alpha \uparrow}^{\dagger} c_{\mbf{R} \beta \downarrow}\rangle\nonumber\\
&\qquad + c_{\mbf{R} \alpha \downarrow}^{\dagger} c_{\mbf{R} \beta \uparrow}\langle c_{\mbf{R} \beta \uparrow}^{\dagger} c_{\mbf{R} \alpha \downarrow}\rangle + c_{\mbf{R} \beta \uparrow}^{\dagger} c_{\mbf{R} \alpha \downarrow}\langle c_{\mbf{R} \alpha \downarrow}^{\dagger} c_{\mbf{R} \beta \uparrow}\rangle + \colorbox{black!10}{$\displaystyle c_{\mbf{R} \alpha \downarrow}^{\dagger} c_{\mbf{R} \beta \downarrow}\langle c_{\mbf{R} \beta \downarrow}^{\dagger} c_{\mbf{R} \alpha \downarrow}\rangle + c_{\mbf{R} \beta \downarrow}^{\dagger} c_{\mbf{R} \alpha \downarrow}\langle c_{\mbf{R} \alpha \downarrow}^{\dagger} c_{\mbf{R} \beta \downarrow}\rangle$}\bigg)\nonumber\\
&+U_A\sum_{\mbf{R},\,(\alpha\beta)=(02),(13)}\bigg( \colorbox{black!10}{$\displaystyle c_{\mbf{R} \alpha \uparrow}^{\dagger} c_{\mbf{R} \beta \uparrow}\langle c_{\mbf{R} \beta \uparrow}^{\dagger} c_{\mbf{R} \alpha \uparrow}\rangle + c_{\mbf{R} \beta \uparrow}^{\dagger} c_{\mbf{R} \alpha \uparrow}\langle c_{\mbf{R} \alpha \uparrow}^{\dagger} c_{\mbf{R} \beta \uparrow}\rangle + c_{\mbf{R} \alpha \downarrow}^{\dagger} c_{\mbf{R} \beta \downarrow}\langle c_{\mbf{R} \beta \downarrow}^{\dagger} c_{\mbf{R} \alpha \downarrow}\rangle + c_{\mbf{R} \beta \downarrow}^{\dagger} c_{\mbf{R} \alpha \downarrow}\langle c_{\mbf{R} \alpha \downarrow}^{\dagger} c_{\mbf{R} \beta \downarrow}\rangle$}\nonumber\\
&\qquad \colorbox{black!10}{$\displaystyle + c_{\mbf{R} \alpha \uparrow}^{\dagger} c_{\mbf{R} \beta \uparrow}\langle c_{\mbf{R} \beta \downarrow}^{\dagger} c_{\mbf{R} \alpha \downarrow}\rangle + c_{\mbf{R} \beta \downarrow}^{\dagger} c_{\mbf{R} \alpha \downarrow}\langle c_{\mbf{R} \alpha \uparrow}^{\dagger} c_{\mbf{R} \beta \uparrow}\rangle + c_{\mbf{R} \alpha \downarrow}^{\dagger} c_{\mbf{R} \beta \downarrow}\langle c_{\mbf{R} \beta \uparrow}^{\dagger} c_{\mbf{R} \alpha \uparrow}\rangle + c_{\mbf{R} \beta \uparrow}^{\dagger} c_{\mbf{R} \alpha \uparrow}\langle c_{\mbf{R} \alpha \downarrow}^{\dagger} c_{\mbf{R} \beta \downarrow}\rangle$}\nonumber\\
&\qquad\colorbox{black!10}{$\displaystyle  + 3c_{\mbf{R} \alpha \uparrow}^{\dagger} c_{\mbf{R} \beta \uparrow}\langle c_{\mbf{R} \alpha \downarrow}^{\dagger} c_{\mbf{R} \beta \downarrow}\rangle + 3c_{\mbf{R} \alpha \downarrow}^{\dagger} c_{\mbf{R} \beta \downarrow}\langle c_{\mbf{R} \alpha \uparrow}^{\dagger} c_{\mbf{R} \beta \uparrow}\rangle + 3c_{\mbf{R} \beta \uparrow}^{\dagger} c_{\mbf{R} \alpha \uparrow}\langle c_{\mbf{R} \beta \downarrow}^{\dagger} c_{\mbf{R} \alpha \downarrow}\rangle + 3c_{\mbf{R} \beta \downarrow}^{\dagger} c_{\mbf{R} \alpha \downarrow}\langle c_{\mbf{R} \beta \uparrow}^{\dagger} c_{\mbf{R} \alpha \uparrow}\rangle$}\bigg).
\end{align}

Within these HF Hamiltonians, we identify the terms contributing to the $\hat{H}_z$ channel in \cref{eq:Hzterm}, which explains the emergence of, and the competition between the AFM and AM phases. We define the local operators
\begin{equation}\label{eq:app:Q0Qz}
o_{\mbf{R}\alpha}=\sum_{ss'}c^{\dagger}_{\mbf{R}\alpha s}(\sigma^z)_{ss'}c_{\mbf{R}(\alpha+2)s'},\qquad
\hat{Q}_0(\mbf{R})=\sum_{\alpha}o_{\mbf{R}\alpha},\qquad\hat{Q}_z(\mbf{R})=\sum_{\alpha}(-1)^{\alpha}o_{\mbf{R}\alpha},
\end{equation}
so that $\hat{H}_z=-\sum_{\mbf{R}}(g_0+g_{\pi}e^{i\mbf{q}\cdot\mbf{R}})\hat{Q}_z(\mbf{R})$.
These two operators $\hat{Q}_0(\mbf{R})$ and $\hat{Q}_z(\mbf{R})$ are orthogonal as measured by the Hilbert--Schmidt inner product of their orbital matrices.
Collecting all terms in $\hat{H}^{\text{HF}}$ that contain only $o_{\mbf{R}\alpha}$ and $\langle o_{\mbf{R}\alpha} \rangle$, we obtain contribution to the $\hat{H}_z$ channel from the Hamiltonian below for each of the two interacting parameter sets used, denoted by $\left.\hat{H}^{\text{HF}}\right|_{\mathcal{O}_z}$.
This includes, for example, the $U_A$ terms shaded in \cref{eq:app:HHFUA}.
\begin{align}\label{eq:app:Qzchannel}
\left.\hat{H}^{\text{HF}}_{U\text{-}U'\text{-}J}\right|_{\mathcal{O}_z} \!={}&\!  -\frac{U' \!+\! J}{8}\sum_{\mbf{R}} \langle \hat{Q}_z(\mbf{R})\rangle \hat{Q}_z(\mbf{R}) -\frac{U' \!+\! J}{8}\sum_{\mbf{R}} \langle \hat{Q}_0(\mbf{R})\rangle \hat{Q}_0(\mbf{R})
+ \frac{U'\!-\!J}{4}\sum_{\mbf{R}\alpha}\left(\langle o_{\mbf{R}\alpha} \rangle \!-\! \langle o_{\mbf{R}\alpha} \rangle^* \right) o_{\mbf{R}\alpha}, \\
\left.\hat{H}^{\text{HF}}_{U\text{-}U'\text{-}U_A}\right|_{\mathcal{O}_z} \!={}&\!  -\frac{U' \!+\! 4U_A}{8}\sum_{\mbf{R}} \langle \hat{Q}_z(\mbf{R})\rangle \hat{Q}_z(\mbf{R}) -\frac{U' \!+\! 4U_A}{8}\sum_{\mbf{R}} \langle \hat{Q}_0(\mbf{R})\rangle \hat{Q}_0(\mbf{R}) + \frac{U'\!-\!2U_A}{4} \sum_{\mbf{R}\alpha}\left(\langle o_{\mbf{R}\alpha} \rangle \!-\! \langle o_{\mbf{R}\alpha} \rangle^* \right) o_{\mbf{R}\alpha}.
\end{align}
We further define $\langle Y_{\mbf{R}\alpha}\rangle = \langle o_{\mbf{R}\alpha}\rangle-\langle o_{\mbf{R}\alpha}\rangle^{*}$, which is purely imaginary.

So far, no symmetry constraint has been imposed on $\rho$. If a state preserves the $\sqrt{2}\times\sqrt{2}$ translations $T_{m(1,-1)+n(1,1)}$ for $m,n\in\mathbb{Z}$, then it also preserves $T_{(2m,0)}$ and $T_{(0,2m)}$.
Hence, $\langle Y_{-\mbf{R},\alpha}\rangle=\langle Y_{\mbf{R}\alpha}\rangle$.
If the state also preserves $C_2$, then $\langle Y_{\mbf{R}\alpha}\rangle=-\langle Y_{-\mbf{R},\alpha}\rangle$.
Therefore, when both symmetries are present, $\langle Y_{\mbf{R}\alpha}\rangle=0$.
In contrast, $\hat{Q}_z$ and $\hat{Q}_0$ are even under $C_2$ and need not vanish.
Define
\begin{equation}\label{Q0Qz_0pi}
Q_a^{(0)}=\frac{1}{N_{\text{site}}}\sum_{\mbf{R}}
  \langle\hat{Q}_a(\mbf{R})\rangle,
  \qquad
  Q_a^{(\pi)}=\frac{1}{N_{\text{site}}}\sum_{\mbf{R}}
  e^{i\mbf{q}\cdot\mbf{R}}\langle\hat{Q}_a(\mbf{R})\rangle,
  \qquad a\in\{0,z\}.
\end{equation}
Under $T_{m(1,-1)+n(1,1)}$ one can make the decomposition $\langle\hat{Q}_a(\mbf{R})\rangle=Q_a^{(0)}+Q_a^{(\pi)}e^{i\mbf{q}\cdot\mbf{R}}$ with $a\in\{0,z\}$ and $\mbf{q}=(\pi,\pi)$, so that
\begin{equation}\label{Q0Qz_0pi_Tsimp}
Q_a^{(0)}=[\langle\hat{Q}_a(\mbf{0})\rangle+\langle\hat{Q}_a(\hat{\mbf{x}})\rangle]/2,\qquad Q_a^{(\pi)}=[\langle\hat{Q}_a(\mbf{0})\rangle-\langle\hat{Q}_a(\hat{\mbf{x}})\rangle]/2,
\qquad a\in\{0,z\}.
\end{equation}
Both $T_{m(1,-1)+n(1,1)}$ and $C_2$ are present for all phases in \cref{fig:fig2,fig:fig3} except the spin-$\mathrm{U}(1)$ broken II and $C_2$-broken nematic phases (\cref{tab:symmetryStrongQG}). When both symmetries are present, the channel contributions become
\begin{align}
\label{eq:app:QzchannelSym_J}
\left.\hat{H}^{\text{HF}}_{U\text{-}U'\text{-}J}\right|_{\mathcal{O}_z, \text{Sym}} ={}&-\sum_{\mbf{R}} \frac{U' + J}{8}\left[\frac{\langle \sum_{\mbf{R}'}\hat{Q}_z(\mbf{R}')(1 + e^{i \mbf{q}\cdot (\mbf{R} + \mbf{R}')})\rangle}{N_{\text{site}}}\hat{Q}_z(\mbf{R})+\frac{\langle \sum_{\mbf{R}'}\hat{Q}_0(\mbf{R}')(1 + e^{i \mbf{q}\cdot (\mbf{R} + \mbf{R}')})\rangle}{N_{\text{site}}}\hat{Q}_0(\mbf{R})\right]\nonumber\\
={}&-\sum_{\mbf{R}} \frac{U' + J}{8}\left[\frac{\langle \hat{Q}_z(\mbf{0})\rangle + \langle \hat{Q}_z(\hat{\mbf{x}})\rangle}{2} + \frac{\langle \hat{Q}_z(\mbf{0})\rangle - \langle \hat{Q}_z(\hat{\mbf{x}})\rangle}{2} e^{i \mbf{q}\cdot \mbf{R}}\right] \hat{Q}_z(\mbf{R})\nonumber\\
&- \sum_{\mbf{R}} \frac{U' + J}{8}\left[\frac{\langle \hat{Q}_0(\mbf{0})\rangle + \langle \hat{Q}_0(\hat{\mbf{x}})\rangle}{2} + \frac{\langle \hat{Q}_0(\mbf{0})\rangle - \langle \hat{Q}_0(\hat{\mbf{x}})\rangle}{2} e^{i \mbf{q}\cdot \mbf{R}}\right] \hat{Q}_0(\mbf{R}),\\
\left.\hat{H}^{\text{HF}}_{U\text{-}U'\text{-}U_A}\right|_{\mathcal{O}_z, \text{Sym}} ={}&-  \sum_{\mbf{R}} \frac{U' + 4U_A}{8}\left[\frac{\langle \sum_{\mbf{R}'}\hat{Q}_z(\mbf{R}')(1 + e^{i \mbf{q}\cdot (\mbf{R} + \mbf{R}')})\rangle}{N_{\text{site}}}\hat{Q}_z(\mbf{R}) +\frac{\langle \sum_{\mbf{R}'}\hat{Q}_0(\mbf{R}')(1 + e^{i \mbf{q}\cdot (\mbf{R} + \mbf{R}')})\rangle}{N_{\text{site}}}\hat{Q}_0(\mbf{R})\right]\nonumber\\
={}& -\sum_{\mbf{R}} \frac{U' + 4U_A}{8}\left[\frac{\langle \hat{Q}_z(\mbf{0})\rangle + \langle \hat{Q}_z(\hat{\mbf{x}})\rangle}{2} + \frac{\langle \hat{Q}_z(\mbf{0})\rangle - \langle \hat{Q}_z(\hat{\mbf{x}})\rangle}{2} e^{i \mbf{q}\cdot \mbf{R}}\right] \hat{Q}_z(\mbf{R})\nonumber\\
&- \sum_{\mbf{R}} \frac{U' + 4U_A}{8}\left[\frac{\langle \hat{Q}_0(\mbf{0})\rangle + \langle \hat{Q}_0(\hat{\mbf{x}})\rangle}{2} + \frac{\langle \hat{Q}_0(\mbf{0})\rangle - \langle \hat{Q}_0(\hat{\mbf{x}})\rangle}{2} e^{i \mbf{q}\cdot \mbf{R}}\right] \hat{Q}_0(\mbf{R}),\label{eq:app:QzchannelSym_UA}
\end{align}
The corresponding contribution to the HF energy is,
\begin{align}\label{eq:app:QzchannelEsym_J}
\left. \langle\hat{H}_0 + \hat{H}_{\text{int}}\rangle_{U\text{-}U'\text{-}J}\right|_{\mathcal{O}_z, \text{Sym}} ={}& -\frac{U' + J}{16 N_{\text{site}}}\left[\langle \sum_{\mbf{R}} \hat{Q}_z(\mbf{R})\rangle^2 + \langle \sum_{\mbf{R}} \hat{Q}_z(\mbf{R}) e^{i \mbf{q}\cdot \mbf{R}}\rangle^2 + \langle \sum_{\mbf{R}} \hat{Q}_0(\mbf{R})\rangle^2 + \langle \sum_{\mbf{R}} \hat{Q}_0(\mbf{R}) e^{i \mbf{q}\cdot \mbf{R}}\rangle^2 \right]\nonumber\\
={}&-\frac{U' + J}{32}N_{\text{site}}\left[\langle \hat{Q}_z(\mbf{0})\rangle^2 + \langle \hat{Q}_z(\hat{\mbf{x}})\rangle^2 + \langle \hat{Q}_0(\mbf{0})\rangle^2 + \langle \hat{Q}_0(\hat{\mbf{x}})\rangle^2\right],\\
\left. \langle\hat{H}_0 + \hat{H}_{\text{int}}\rangle_{U\text{-}U'\text{-}U_A}\right|_{\mathcal{O}_z, \text{Sym}} ={}& -\frac{U' + 4U_A}{16 N_{\text{site}}}\left[\langle \sum_{\mbf{R}} \hat{Q}_z(\mbf{R})\rangle^2 + \langle \sum_{\mbf{R}} \hat{Q}_z(\mbf{R}) e^{i \mbf{q}\cdot \mbf{R}}\rangle^2 + \langle \sum_{\mbf{R}} \hat{Q}_0(\mbf{R})\rangle^2 + \langle \sum_{\mbf{R}} \hat{Q}_0(\mbf{R}) e^{i \mbf{q}\cdot \mbf{R}}\rangle^2 \right]\nonumber\\
={}&-\frac{U' + 4U_A}{32}N_{\text{site}}\left[\langle \hat{Q}_z(\mbf{0})\rangle^2 + \langle \hat{Q}_z(\hat{\mbf{x}})\rangle^2 + \langle \hat{Q}_0(\mbf{0})\rangle^2 + \langle \hat{Q}_0(\hat{\mbf{x}})\rangle^2\right].
\label{eq:app:QzchannelEsym_UA}
\end{align}

Thus, $U_A$ can enhance this channel four times as strongly as $U'$ or $J$. For the AFM, AM I, and AM II, $\langle\hat{Q}_0(\mbf{R})\rangle=0$ by symmetry, because its uniform part is odd under $C_4\mathcal{T}$ and $T_{(1,0)}\mathcal{T}$, and its staggered part is odd under $C_4T_{(1,0)}$ and $C_4\mathcal{T}$.
Hence, only $\hat{Q}_z$ contributes for these phases. Then, keeping only the $U_A$ part of \cref{eq:app:QzchannelSym_UA} reproduces $\hat{H}_z$ in \cref{eq:Hzterm} with $g_0=(U_A/2)Q_z^{(0)}$ and $g_{\pi}=(U_A/2)Q_z^{(\pi)}$, and the $U_A$ part of \cref{eq:app:QzchannelEsym_UA} equals $-N_{\text{site}}(g_0^2+g_{\pi}^2)/U_A$ when $\langle\hat{Q}_0\rangle=0$. 

\begin{table*}[t]
\caption{Parities of the uniform and staggered sums of $\hat{Q}_z(\mbf{R})=\sum_{\alpha ss'}(-1)^{\alpha}c^{\dagger}_{\mbf{R}\alpha s}(\sigma^z)_{ss'}c_{\mbf{R}(\alpha+2)s'}$ and of their combination with $ab\neq0$, which enter $\hat{H}_z$ in \cref{eq:Hzterm}. The staggered wave vector is $\mbf{q}=(\pi,\pi)$. The parity symbols denote $+$: invariant; $-$: sign reversal; and $\times$: the uniform and staggered components transform differently. Spin-$\mathrm{U}(1)$ refers to rotations about the $z$ axis. The staggered sum has the same broken symmetries as the AFM, and the combination has those of AM I and AM II (\cref{tab:symmetryStrongQG}).}
\label{tab:QzSymmetrySigns}
\centering
\setlength{\extrarowheight}{2.5pt}
\begin{tabular*}{\textwidth}{@{\extracolsep{\fill}}c|ccccccccccc}
\hline\hline

 & spin-$\U(1)$
 & $T_{(1,0)}$
 & $C_4$
 & $M_x$
 & $\mathcal{T}$
 & $C_4 T_{(1,0)}$
 & $C_4^2$
 & $T_{(1,0)}\mathcal{T}$
 & $C_4\mathcal{T}$
 & $C_4 T_{(1,0)}\mathcal{T}$
 & $C_4^2\mathcal{T}$ \\
\hline\hline
$\sum_{\mbf{R}}\hat{Q}_z(\mbf{R})$
 & $+$ & $+$ & $-$ & $-$ & $-$ & $-$ & $+$ & $-$ & $+$ & $+$ & $-$ \\
\hline
$\sum_{\mbf{R}}\hat{Q}_z(\mbf{R}) e^{i \mbf{q}\cdot \mbf{R}}$
 & $+$ & $-$ & $-$ & $-$ & $-$ & $+$ & $+$ & $+$ & $+$ & $-$ & $-$ \\
\hline
$-\sum_{\mbf{R}}\hat{Q}_z(\mbf{R}) (a + b e^{i \mbf{q}\cdot \mbf{R}})$
& $+$ & $\times$ & $-$ & $-$ & $-$ & $\times$ & $+$ & $\times$ & $+$ & $\times$ & $-$ \\
\hline\hline
\end{tabular*}
\end{table*}

\begin{table*}[t]
\caption{Order parameters $Q_z^{(0)}$ and $Q_z^{(\pi)}$, defined in \cref{Q0Qz_0pi}, of the five self-consistent solutions under the $U$-$U'$-$J$ interaction parameter set at $(U'/U,J/U)=(0.6,0)$ and $m/t=0.01$, shown in Figs.~\ref{fig:fsAMdiff}(a) and \ref{fig:bandStrucJ}. They are ordered by increasing energy.}
\label{tab:QzFiveJ}
\centering
\renewcommand{\arraystretch}{1.25}
\setlength{\extrarowheight}{1pt}
\resizebox{\textwidth}{!}{%
\begin{tabular}{l|ccccc}
\hline\hline
\rule{0pt}{2.3em}
 & Nematic I & Nematic II & \shortstack{spin-$\U(1)$ broken\\``AM''} & hole FM & AM I \\
\hline\hline
$Q_z^{(0)} = [\langle\hat Q_z(\mbf{0})\rangle+\langle\hat Q_z(\hat{\mbf{x}})\rangle]/2 = \langle \sum_{\mbf{R}} \hat{Q}_z(\mbf{R})\rangle/N_{\text{site}}$ & $0.015767$ & $0.025440$ & $0$ & $0$ & $0.033267$ \\
\hline
$Q_z^{(\pi)}=[\langle\hat Q_z(\mbf{0})\rangle-\langle\hat Q_z(\hat{\mbf{x}})\rangle]/2 = \langle \sum_{\mbf{R}} \hat{Q}_z(\mbf{R}) e^{i \mbf{q}\cdot \mbf{R}}\rangle/N_{\text{site}}$ & $-0.088500$ & $0.098537$ & $0$ & $-0.078166$ & $0.106354$ \\
\hline\hline
\end{tabular}%
}
\end{table*}

\begin{table*}[t]
\caption{Same as \cref{tab:QzFiveJ} for the five solutions in Figs.~\ref{fig:fsAMdiffUA}(a) and \ref{fig:bandStrucUA} at $(U'/U,U_A/U)=(0.25,0.45)$.}
\label{tab:QzFiveUA}
\centering
\renewcommand{\arraystretch}{1.25}
\setlength{\extrarowheight}{1pt}
\resizebox{\textwidth}{!}{%
\begin{tabular}{l|ccccc}
\hline\hline
\rule{0pt}{2.3em}
 & AM I & \shortstack{spin-$\U(1)$ broken\\``Nematic III''} & Nematic IV & AFM & partial sym-FM \\
\hline\hline
$Q_{z}^{(0)} = [\langle\hat Q_z(\mbf{0})\rangle+\langle\hat Q_z(\hat{\mbf{x}})\rangle]/2 = \langle \sum_{\mbf{R}} \hat{Q}_z(\mbf{R})\rangle/N_{\text{site}}$ & $0.038202$ & $-0.000106$ & $-0.007848$ & $0$ & $0$ \\
\hline
$Q_{z}^{(\pi)} = [\langle\hat Q_z(\mbf{0})\rangle-\langle\hat Q_z(\hat{\mbf{x}})\rangle]/2 = \langle \sum_{\mbf{R}} \hat{Q}_z(\mbf{R}) e^{i \mbf{q}\cdot \mbf{R}}\rangle/N_{\text{site}}$ & $0.057292$ & $-0.000792$ & $0.058257$ & $-0.058606$ & $0$ \\
\hline\hline
\end{tabular}%
}
\end{table*}

\Cref{tab:QzSymmetrySigns} lists the parities of $\sum_{\mbf{R}}\hat{Q}_z(\mbf{R})$ and $\sum_{\mbf{R}}\hat{Q}_z(\mbf{R})e^{i\mbf{q}\cdot\mbf{R}}$.
Their expectation values, equivalently $Q_z^{(0)}$ and $Q_z^{(\pi)}$, can be nonzero only if the state breaks all symmetries under which it is odd.
\Cref{tab:QzFiveJ,tab:QzFiveUA} list $Q_z^{(0)}$ and $Q_z^{(\pi)}$ for the competing solutions at representative interaction parameters for strong quantum geometry, further discussed later in \cref{fig:fsAMdiff,fig:fsAMdiffUA}, respectively. Together with \cref{tab:symmetryStrongQG}, these data confirm the symmetry analysis of the order parameters. AFM preserves $C_4T_{(1,0)}$, under which the uniform sum is odd, so it forces $g_0=0$.
On the other hand, AM I and AM II allow both $g_0$ and $g_{\pi}$ to be nonzero; so do all the nematic phases discussed in this work,
which break all the symmetries broken by the AMs and more.
No single operator with definite parities has the symmetry of AM I and AM II, since it cannot be odd under $C_4T_{(1,0)}$, $\mathcal{T}$, and their product $C_{4}T_{(1, 0)}\mathcal{T}$ simultaneously, as demanded by the AM phases. 
Therefore, if the channel $\hat{H}_z$ is favored, \eg at large $U_A$, the lowest-energy state must break at least the symmetries broken by the staggered component (AFM) if $g_0=0$ or those by AM if both $g_0$ and $g_{\pi}$ are nonzero. Within this channel, the energy gain $-N_{\text{site}}(g_0^2+g_{\pi}^2)/U_A$ favors the AM.
The competition for the ground state is, however, decided by the total HF energy. For example, in \cref{tab:QzFiveJ} at $(U'/U,J/U)=(0.6,0)$, AM I attains the largest $g_0^2+g_{\pi}^2$ among competing solutions but has the highest energy. 

\section{Magnetic orders in the Hartree--Fock numerics}
\label{app:HFdetails}

\begin{table*}[t]
\caption{Magnetization, occupations at high-symmetry momenta, and symmetries of the magnetic phases for strong quantum geometry ($m\ll t$). occuN and occuX are the numbers of occupied states at $\mbf{k}_N=(\pi/2,\pi/2)$ and $\mbf{k}_X=(\pi,0)$, respectively. The symbol $\checkmark$ ($\times$) denotes a preserved (broken) symmetry \emph{as resolved by the finite-size HF calculations}. A pair $(a,b)$ denotes the range of $|\mathbf{S}|/N_{\text{site}}$ within a phase, and $\{1,2\}$ denotes an occupation that varies within a phase.}
\label{tab:symmetryStrongQG}
\centering
\renewcommand{\arraystretch}{1.25}
\resizebox{\textwidth}{!}{%
\begin{tabular}{l|ccc|ccccccccccc}
\hline\hline
 & $|\mathbf{S}|/N_{\mathrm{site}}$
 & occuN
 & occuX
 & spin-$\U(1)$
 & $T_{(1,0)}$
 & $C_4$
 & $M_x$
 & $\mathcal{T}$
 & $C_4 T_{(1,0)}$
 & $C_4^2$
 & $T_{(1,0)}\mathcal{T}$
 & $C_4\mathcal{T}$
 & $C_4 T_{(1,0)}\mathcal{T}$
 & $C_4^2\mathcal{T}$ \\
\hline\hline
sym-FM                       & $\nu/2$& 2& 2& $\checkmark$ & $\checkmark$& $\checkmark$& $\checkmark$& $\times$& $\checkmark$& $\checkmark$& $\times$& $\times$& $\times$& $\times$\\
\hline
spin-\U(1) broken I       & $\left(\tfrac{1-\nu}{2},\,\tfrac{\nu}{2}\right)$& 2& 2& $\times$& $\times$& $\times$& $\times$& $\times$& $\checkmark$& $\checkmark$& $\times$& $\times$& $\times$& $\times$\\
\hline
spin-\U(1) broken II      & $\left(\tfrac{1-\nu}{2},\,\tfrac{\nu}{2}\right)$& 2& 2& $\times$& $\times$& $\times$& $\times$& $\times$& $\times$& $\times$& $\times$& $\times$& $\times$& $\times$\\
\hline
hole FM                  & $(1-\nu)/2$& $\{1, 2\}$& 1& $\checkmark$& $\times$& $\times$& $\times$& $\times$& $\checkmark$& $\checkmark$& $\times$& $\times$& $\times$& $\times$\\
\hline
hole FM II               &$(1-\nu)/2$& 1& 2& $\checkmark$& $\times$& $\times$& $\times$& $\times$& $\checkmark$& $\checkmark$& $\times$& $\times$& $\times$& $\times$\\
\hline
Nematic I                & $(1-\nu)/2$& $\{1, 2\}$& 1& $\checkmark$& $\times$& $\times$& $\times$& $\times$& $\times$& $\checkmark$& $\times$& $\times$& $\times$& $\times$\\
\hline
Nematic II               & $(1-\nu)/2$& $\{1, 2\}$& 2& $\checkmark$& $\times$& $\times$& $\times$& $\times$& $\times$& $\checkmark$& $\times$& $\times$& $\times$& $\times$\\
\hline
Nematic III              & $(0, \tfrac{1-\nu}{2})$& $\{1, 2\}$& 2& $\checkmark$& $\times$& $\times$& $\times$& $\times$& $\times$& $\checkmark$& $\times$& $\times$& $\times$& $\times$\\
\hline
$C_2$-broken Nematic               & $(0, \tfrac{1-\nu}{2})$& $\{1, 2\}$& 2& $\times$& $\times$& $\times$& $\times$& $\times$& $\times$& $\times$& $\times$& $\times$& $\times$& $\times$\\
\hline
AFM                      & 0& 0& 2& $\checkmark$& $\times$& $\times$& $\times$& $\times$& $\checkmark$& $\checkmark$& $\checkmark$& $\checkmark$& $\times$& $\times$\\
\hline
AM I                     & 0& 1& 2& $\checkmark$& $\times$& $\times$& $\times$& $\times$& $\times$& $\checkmark$& $\times$& $\checkmark$& $\times$& $\times$\\
\hline
AM II                    & 0& 0& 2& $\checkmark$& $\times$& $\times$& $\times$& $\times$& $\times$& $\checkmark$& $\times$& $\checkmark$& $\times$& $\times$\\
\hline\hline
\end{tabular}%
}
\end{table*}

\begin{figure}[t]
    \centering
    \includegraphics[width=\linewidth]{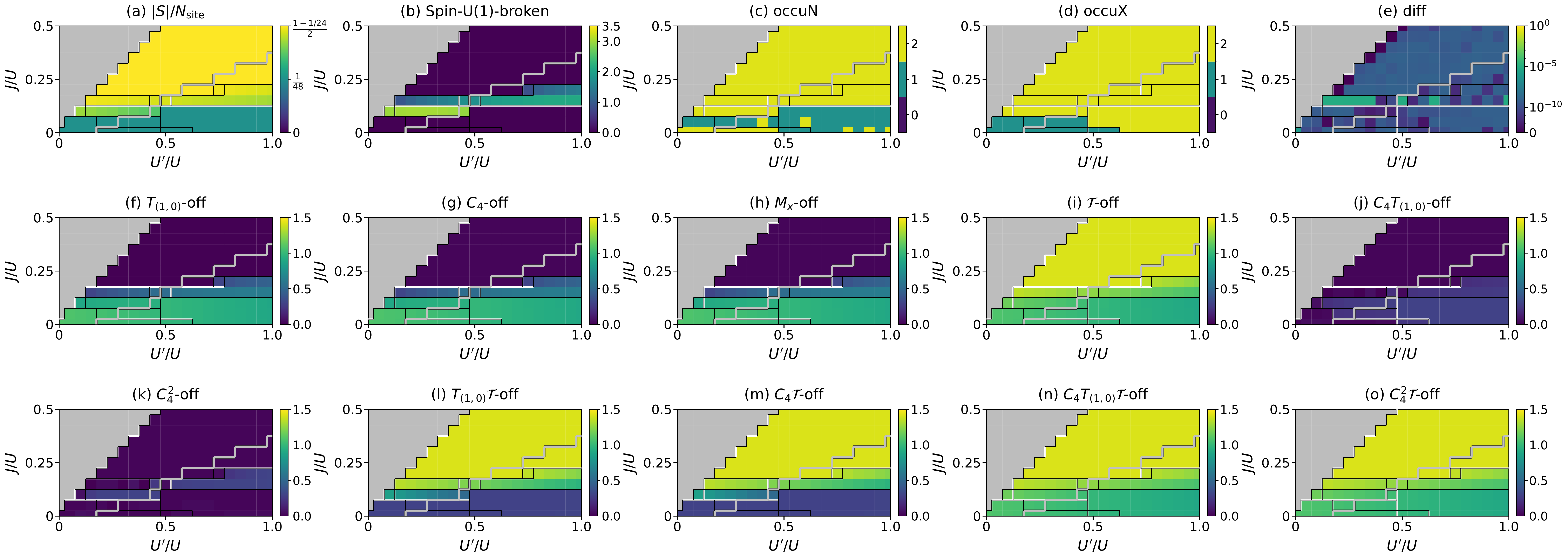}
    \caption{Characterization of the HF solutions in \cref{fig:fig2}(a) ($m/t=0.01$, $U$-$U'$-$J$): (a) magnetization, (b) spin-$\mathrm{U}(1)$ breaking $\sum|\rho_{\uparrow\downarrow}|$, (c) occupation at $N$, (d) occupation at $X$, (e) final $||\delta\rho||/||\rho||$, and (f)--(o) symmetry deviations $\|\rho-\rho'\|/\|\rho\|$ under the respective symmetry operations: (f) $T_{(1,0)}$, (g) $C_4$, (h) $M_x$, (i) $\mathcal{T}$, (j) $C_4 T_{(1,0)}$, (k) $C^2_4$, (l) $T_{(1,0)}\mathcal{T}$, (m) $C_4\mathcal{T}$, (n) $C_4 T_{(1,0)} \mathcal{T}$, and (o) $C_4^2 \mathcal{T}$.}
    \label{fig:strongQGdiagUJ}
\end{figure}

We solve the HF equations self-consistently for $\rho_{\tau\alpha s,\tau'\beta s'}(\mbf{k})$ in the two-sublattice basis introduced in \cref{app:symmetryRho} [\cref{fig:latticeMesh}(a)], which allows checkerboard ($\sqrt{2}\times\sqrt{2}$) translation breaking while keeping $\rho$ diagonal in $\mbf{k}$. The momenta $\mbf{k}=(n_1/N)(\pi,-\pi)+(n_2/N)(\pi,\pi)$ with $n_{1,2}=0,\dots,N-1$ cover the reduced BZ [\cref{fig:latticeMesh}(b)]. We use $N=48$ (2304 momenta, including the high-symmetry points) in the main text and throughout the Supplementary Materials, except in \cref{app:TDHF} where we used $N=24$ for stability analysis and time-dependent Hartree-Fock calculations. Although the parameters lie in the regime where band projection is justified, we solve the full, unprojected four-band model; projected HF calculations yield similar phases.

We use both the optimal damping algorithm (ODA)~\suppcite{CancesIJQC2000,Wang2023MagneticOrders} and direct inversion in the iterative subspace (DIIS)~\suppcite{PulayCPL1980,PulayJCC1982} as complementary iterative methods. ODA tends to find the lowest-energy state but converges slowly; DIIS converges quickly but does not always reach the lowest-energy state and is often unstable for gapless states~\suppcite{Dong2026MetallicWigner}. Each iteration builds the HF Hamiltonian from the current one-body density matrix, using \cref{app:HFhAndChannels}. Since $\hat{H}_{\text{int}}$ is strictly onsite, no momentum convolution is needed for the Fock terms~\suppcite{Wolf2024Magnetism,KwanAdvPhys2025}. At each parameter point, we run ODA and DIIS starting from at least three random initial density matrices, both without and with spin-$\mathrm{U}(1)$ symmetry enforced during the iterations. Additionally, we run DIIS starting from the five solutions in \cref{fig:fsAMdiff} and from the AM I solution in \cref{fig:fsAMdiffUA}. An iteration is considered converged when $\|\delta\rho\|/\|\rho\|<10^{-5}$ under ODA, or $<10^{-9}$ under DIIS, where $\delta\rho$ is the change of $\rho$ in one iteration and $\|\cdot\|$ is the Frobenius norm. Each converged $\rho$ is spin-rotated globally so that its net magnetization points along $z$. The phases in \cref{fig:fig2,fig:fig3} are the lowest-energy solutions among all distinct converged solutions at each parameter point.

\begin{figure}[t]
    \centering
    \includegraphics[width=\linewidth]{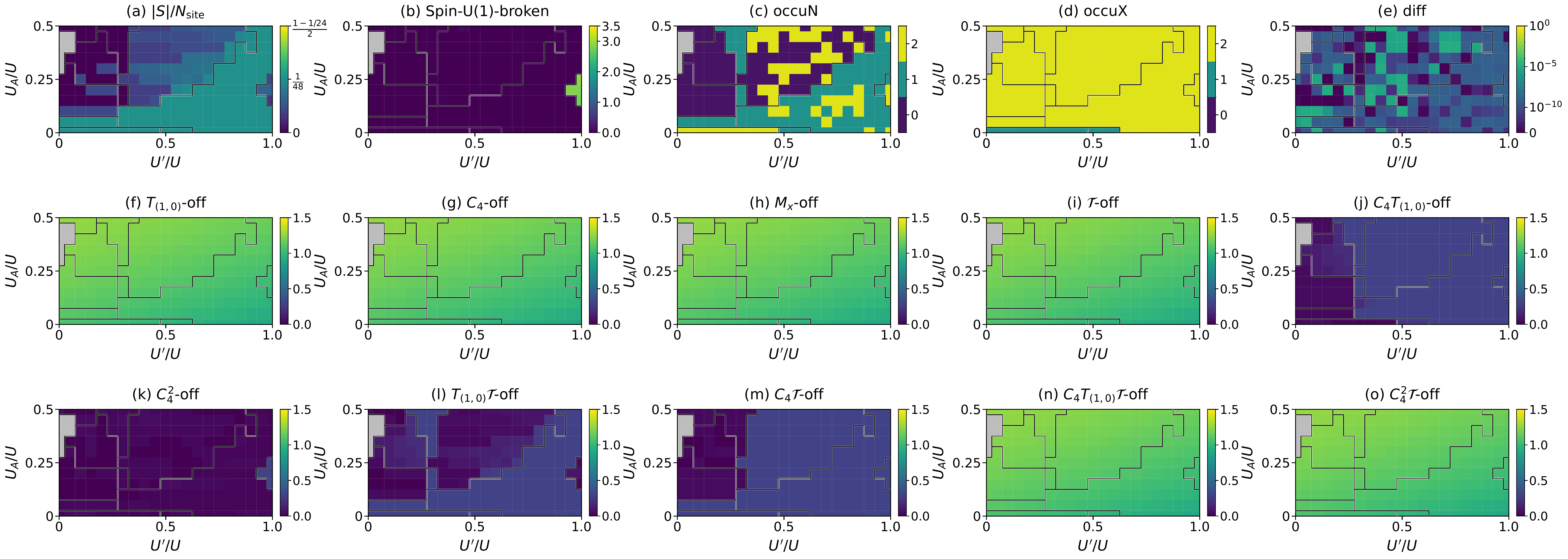}
    \caption{Same as \cref{fig:strongQGdiagUJ} for \cref{fig:fig3}(a) ($m/t=0.01$, $U$-$U'$-$U_A$). }
    \label{fig:strongQGdiagUUA}
\end{figure}

\begin{figure}[t]
    \centering
    \includegraphics[width=\linewidth]{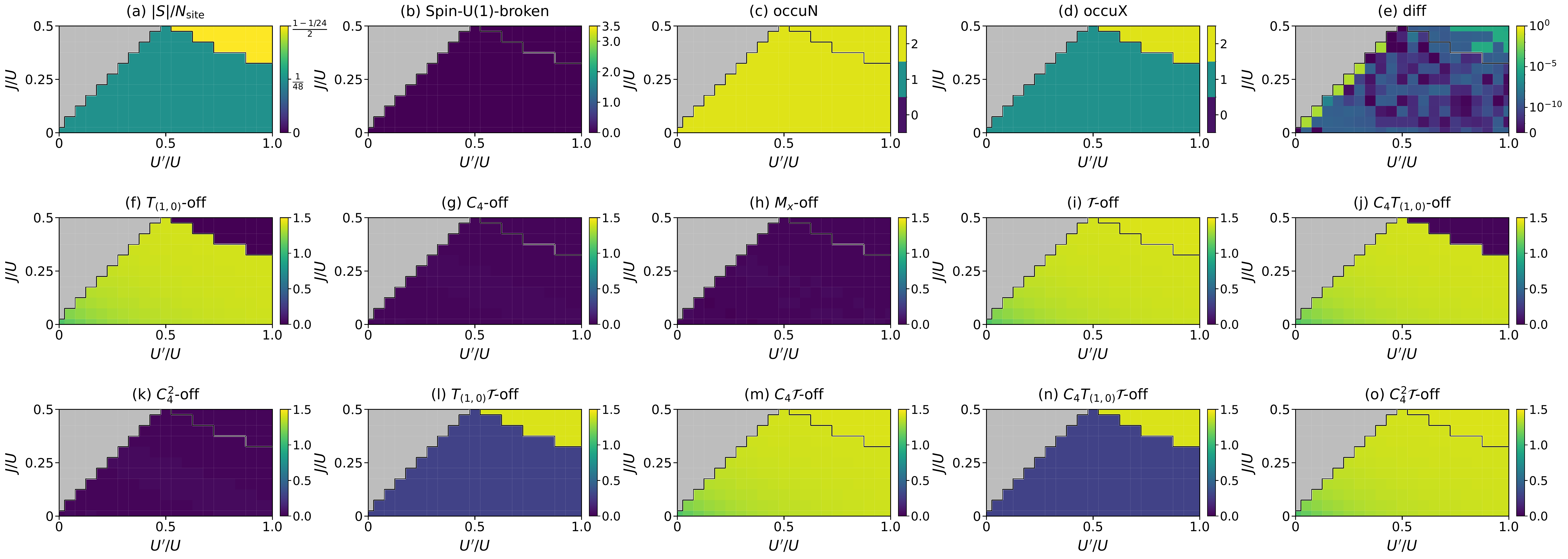}
    \caption{Same as \cref{fig:strongQGdiagUJ} for \cref{fig:fig2}(b) ($t/m=0.01$, $U$-$U'$-$J$). }
    \label{fig:weakQGdiagUJ}
\end{figure}

\begin{figure}[t]
    \centering
    \includegraphics[width=\linewidth]{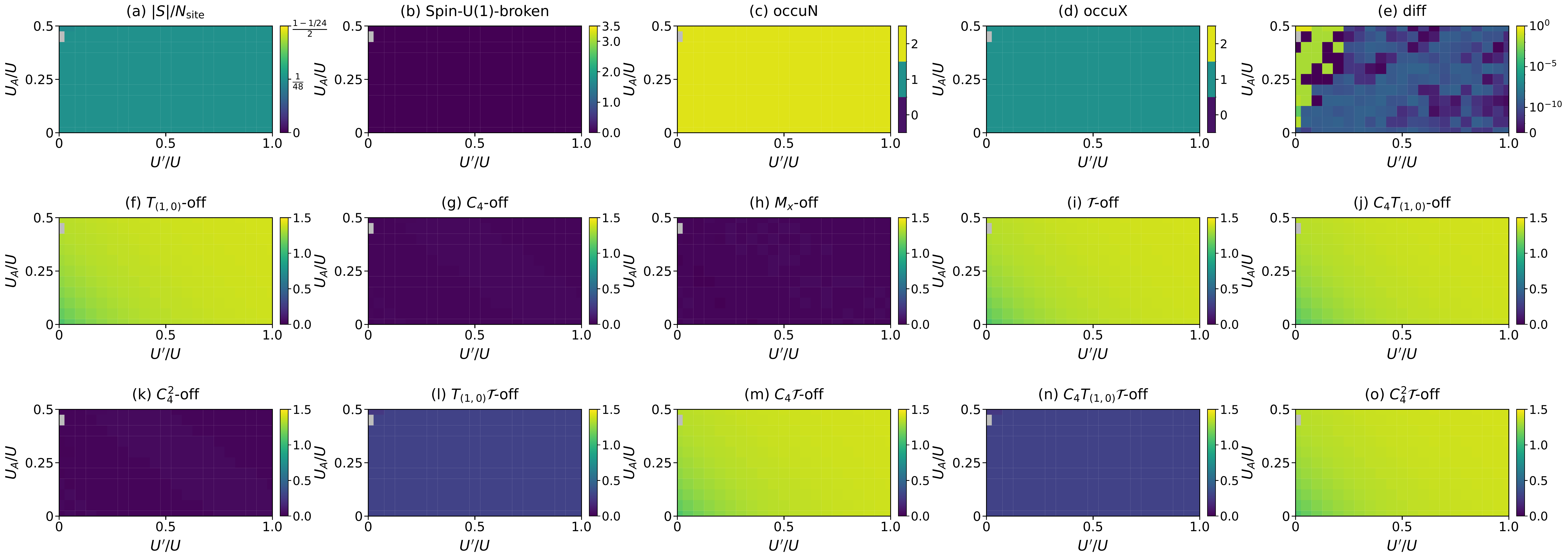}
    \caption{Same as \cref{fig:strongQGdiagUJ} for \cref{fig:fig3}(b) ($t/m=0.01$, $U$-$U'$-$U_A$). }
    \label{fig:weakQGdiagUUA}
\end{figure}

\begin{figure}[t]
    \centering
    \includegraphics[width=\linewidth]{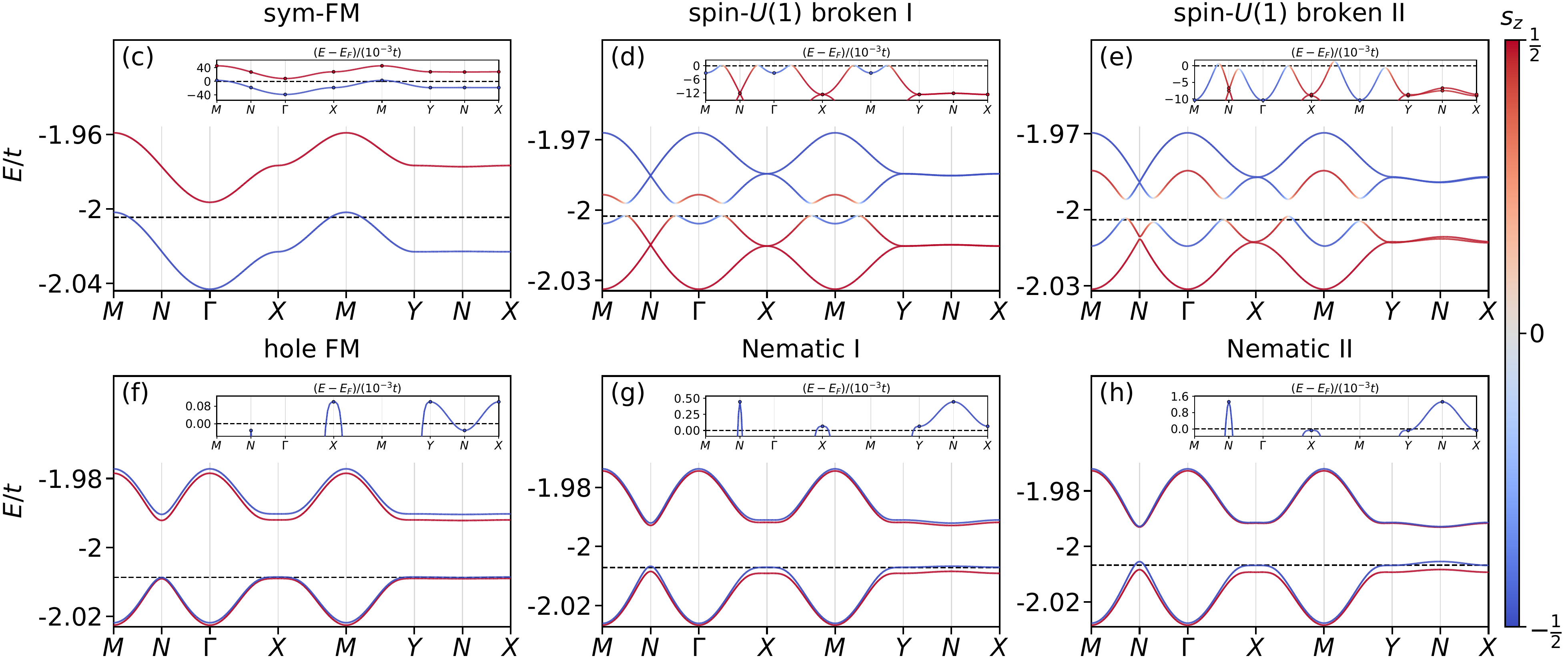}
    \caption{HF dispersion of the lowest band for the Fermi surfaces in \cref{fig:fig2}(c)--(h). The dashed line marks the Fermi level, and the insets show magnified views near the Fermi level.}
    \label{fig:bandStrucJ_main}
\end{figure}

\begin{figure}[t]
    \centering
    \includegraphics[width=\linewidth]{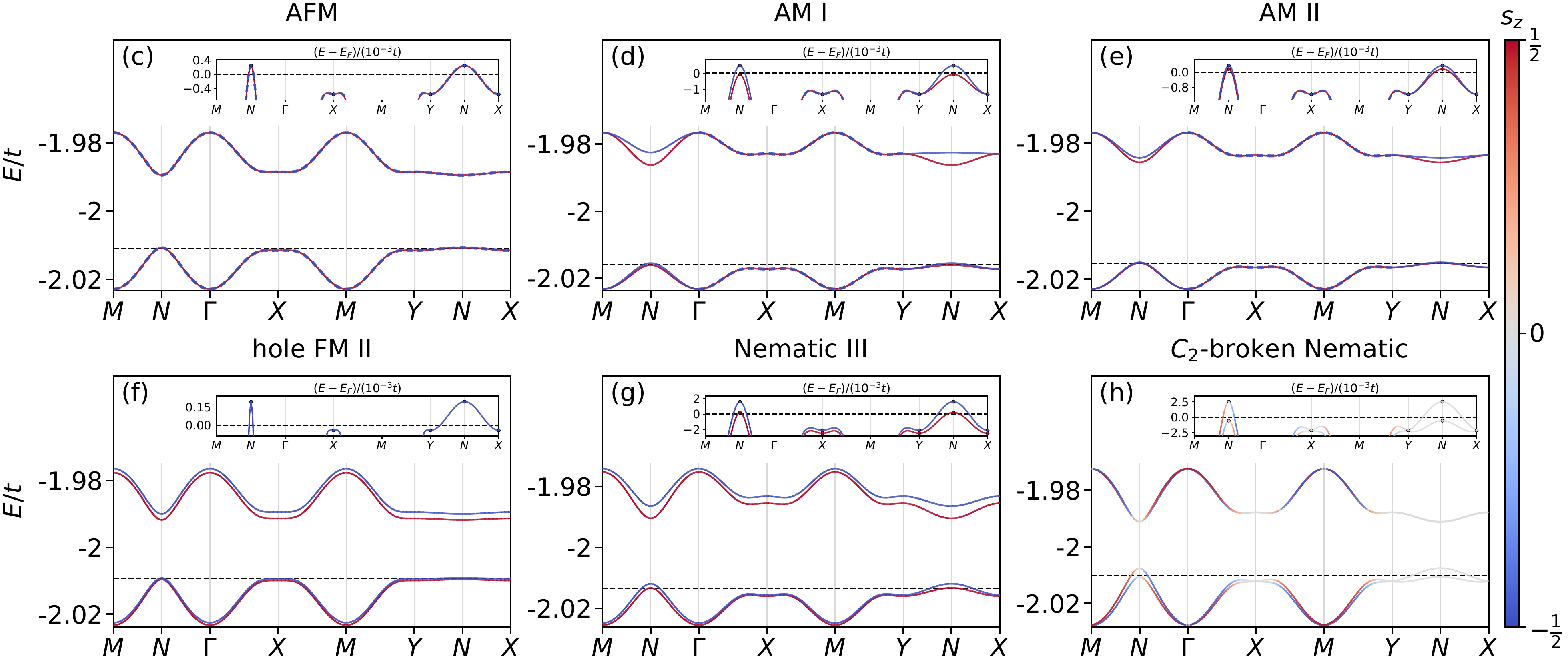}
    \caption{HF dispersion of the lowest band for the Fermi surfaces in \cref{fig:fig3}(c)--(h). The dashed line marks the Fermi level, and the insets show magnified views near the Fermi level.}
    \label{fig:bandStrucUA_main}
\end{figure}

\begin{figure}[t]
    \centering
    \includegraphics[width=\linewidth]{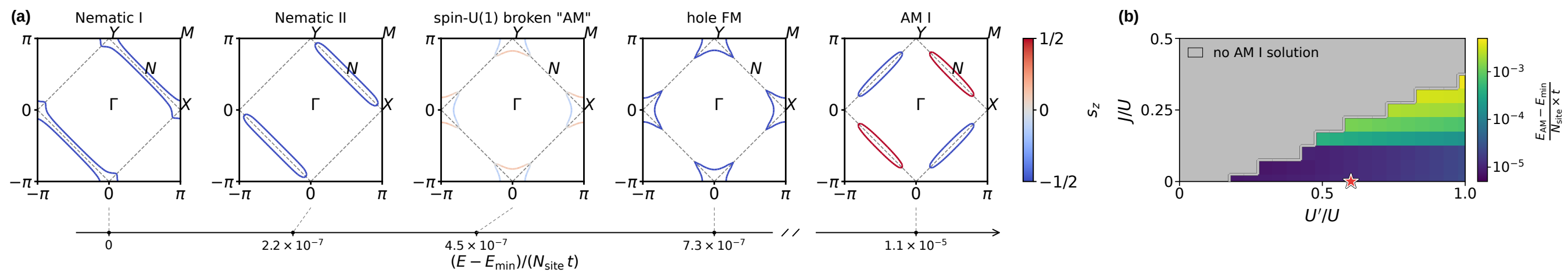}
    \caption{Competing magnetic orders and existence of AM I self-consistent solutions in the strong quantum geometry case, with interaction parameters $U'_{\alpha \beta} = U'$, and $J_{H, \alpha\beta} = J_{p, \alpha \beta} = J$. (a)~Fermi surfaces of five self-consistent solutions and their relative energies at $U'/U = 0.6, J/U = 0$, plotted in the same format as \cref{fig:fig2}(b) arranged along the HF energy axis. (b)~Energy difference between AM I type self-consistent solutions and lowest-energy states in the phase diagram. Star shows the parameter point in (a).
    The spin-$\mathrm{U}(1)$-broken ``AM'' is a metastable state with an $s_z$-splitting Fermi surface resembling that of AM I, but it breaks spin-$\mathrm{U}(1)$ and has $Q_z^{(0)}=Q_z^{(\pi)}=0$ in \cref{tab:QzFiveJ}, so it is not AM by symmetry characterization.
    }
    \label{fig:fsAMdiff}
\end{figure}

\begin{figure}
    \centering
    \includegraphics[width=\linewidth]{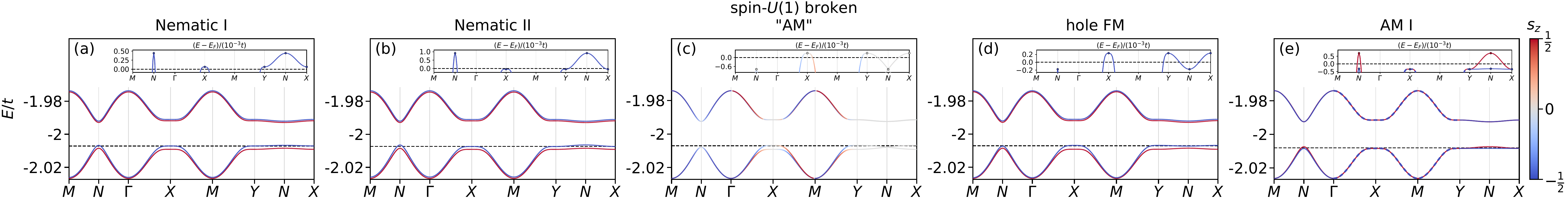}
    \caption{HF dispersions of the lowest band for the five solutions in \cref{fig:fsAMdiff}(a).  Color shows spin polarization $s_z$. Dashed line marks the Fermi level, and the insets show magnified views near the Fermi level.}
    \label{fig:bandStrucJ}
\end{figure}

\begin{figure}
    \centering
    \includegraphics[width=\linewidth]{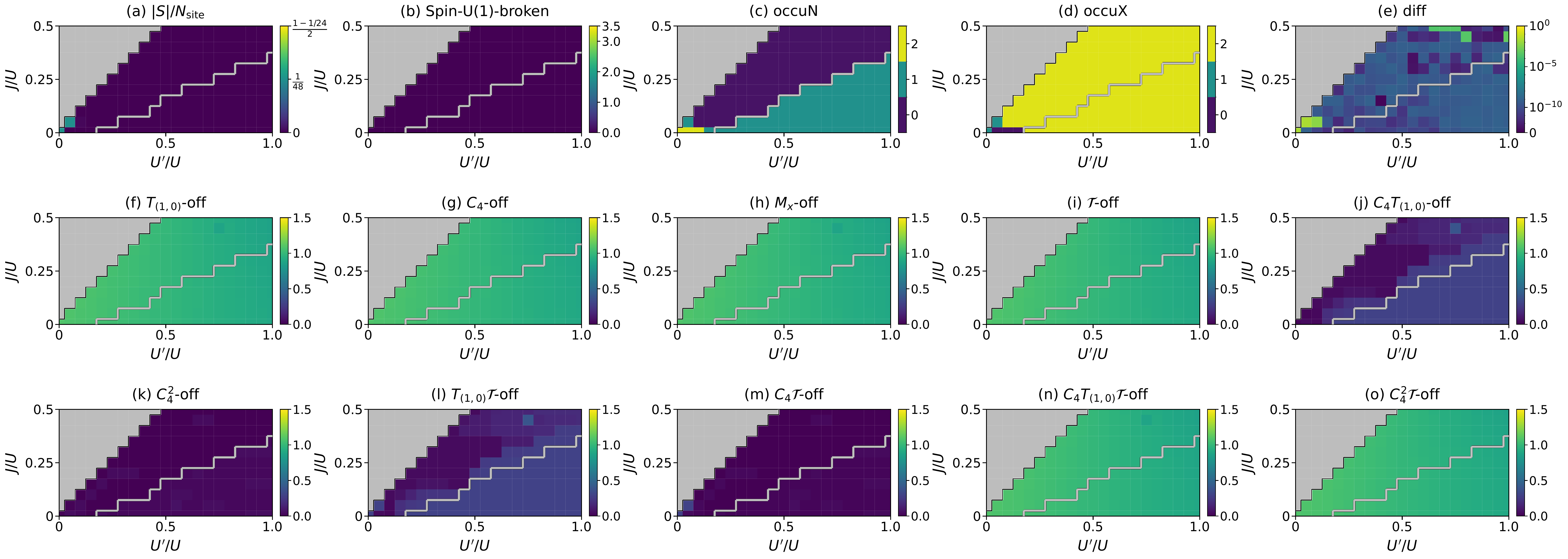}
    \caption{Magnetizations, broken symmetries, high-symmetry momentum occupation, and convergence of constrained Hartree--Fock calculations in strong quantum geometry at $\displaystyle \nu = 1 - 1/24$ with interaction parameters $U'_{\alpha \beta} = U'$, and $J_{H, \alpha\beta} = J_{p, \alpha \beta} = J$. DIIS iterations start from the AM I solution at $(U'/U,J/U)=(0.6,0)$ (\cref{fig:fsAMdiff}) with spin-$\mathrm{U}(1)$ symmetry enforced. Solutions below the silver line agree with the characterization of AM I in \cref{tab:symmetryStrongQG}. }
    \label{fig:AMsolution}
\end{figure}

\begin{figure}
    \centering
    \includegraphics[width=\linewidth]{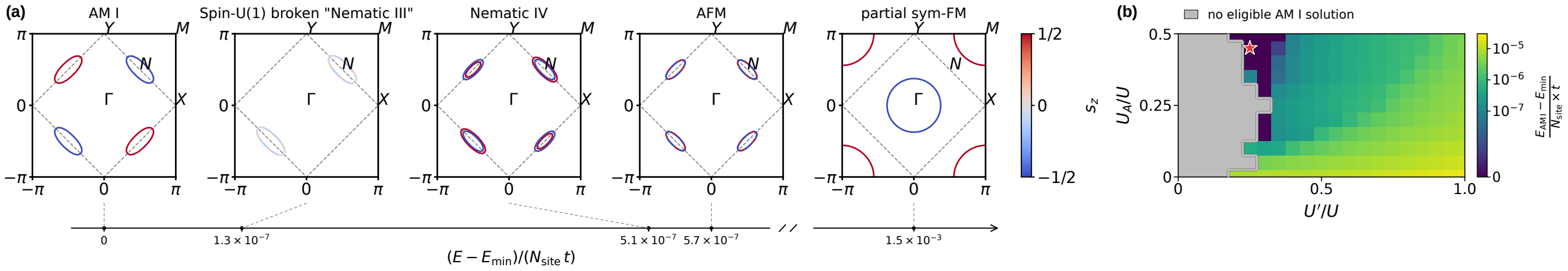}
    \caption{
    Competing magnetic orders and AM I as the lowest-energy state in the strong quantum geometry case, with interaction parameters $U'_{\alpha \beta} = (U' - U_A)\delta_{\langle \alpha\beta\rangle} + (U' + U_A) \delta_{\llangle \alpha\beta\rrangle}, J_{p,\alpha\beta} = 3J_{H, \alpha\beta} =  3U_A\delta_{\llangle \alpha \beta\rrangle}$, displayed in the same format as in \cref{fig:fsAMdiff}. (a)~Five self-consistent solutions and their relative energies at $U'/U = 0.25, U_A/U = 0.45$. (b)~Energy difference between AM I type self-consistent solutions and lowest-energy states. Star shows the parameter point in (a). Spin-U(1) broken ``Nematic III'', Nematic IV, and partial sym-FM are never ground states in the phase diagram studied. The spin-$\mathrm{U}(1)$-broken ``Nematic III'' has a similar Fermi surface to that of Nematic III, but breaks spin-$\mathrm{U}(1)$. Nematic IV belongs to the same symmetry class as Nematic I--III in \cref{tab:QzFiveUA}, but differs by Lifshitz transitions. The partial sym-FM has a smaller magnetization than the sym-FM. 
    }
    \label{fig:fsAMdiffUA}
\end{figure}

\begin{figure}[t]
    \centering
    \includegraphics[width=\linewidth]{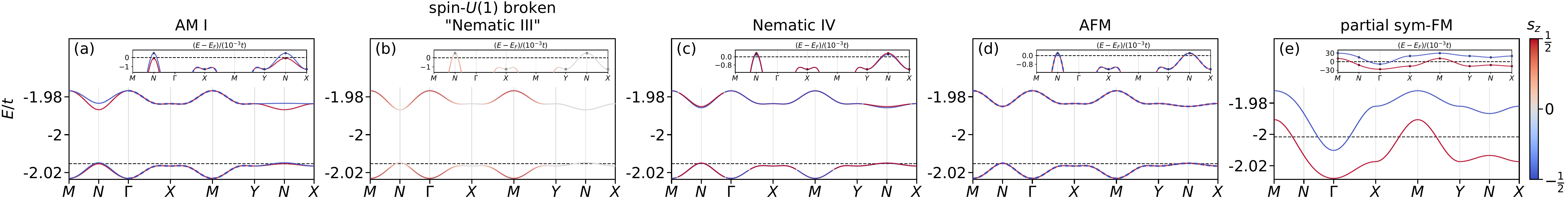}
    \caption{
    HF dispersions of the lowest band for the five solutions in \cref{fig:fsAMdiffUA}(a).  Color shows spin polarization $s_z$. Dashed line marks the Fermi level, and insets show the zoom-in at Fermi level.}
    \label{fig:bandStrucUA}
\end{figure}

\begin{figure}[t]
    \centering
    \includegraphics[width=\linewidth]{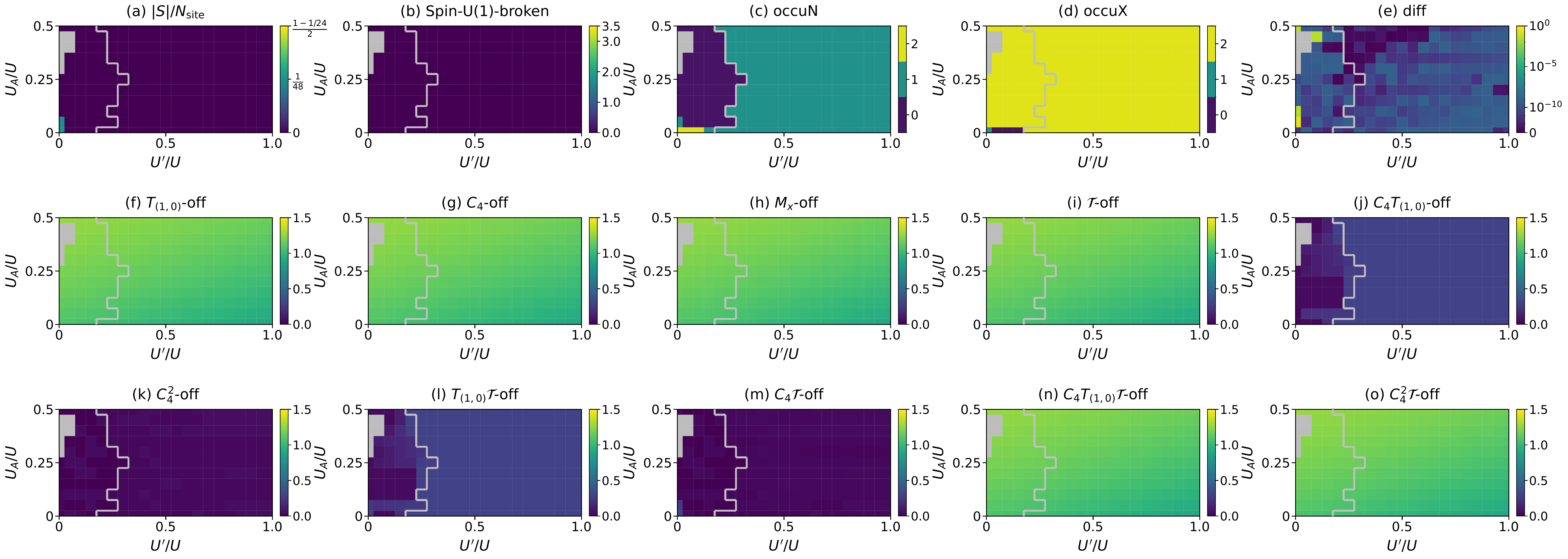}
    \caption{HF solutions using DIIS iterations starting from the AM I solution at $(U'/U,U_A/U)=(0.25,0.45)$ (\cref{fig:fsAMdiffUA}) with spin-$\mathrm{U}(1)$ symmetry enforced. Solutions to the right of the silver line agree with the characterization of AM I in \cref{tab:symmetryStrongQG}. }
    \label{fig:AMsolutionUA}
\end{figure}

We identify the phases by their magnetization, their occupations at $N$ and $X$, and their symmetries (\cref{tab:symmetryStrongQG,tab:symmetryWeakQG}); the hole FM has different symmetries for $m\ll t$ and $m\gg t$. Spin-$\mathrm{U}(1)$ breaking is measured by $\sum_{\mbf{k}\tau\alpha\tau'\beta}|\rho_{\tau\alpha\uparrow,\tau'\beta\downarrow}(\mbf{k})|$ after the spin rotation above. For all discrete symmetries, we compute the deviation $\|\rho-\rho'\|/\|\rho\|$, where $\rho'$ is the transformed density matrix in \cref{app:symmetryRho}. \Cref{fig:strongQGdiagUJ,fig:strongQGdiagUUA,fig:weakQGdiagUJ,fig:weakQGdiagUUA} show these quantities across the phase diagrams of the main text. Small nonzero deviations of preserved symmetries are attributed to the finite momentum mesh; e.g., 
AFM at $(U'/U,U_A/U)=(0.25,0.15)$ has a small $C_4T_{(1,0)}$ deviation $||\rho - \rho'||/||\rho|| = 0.043$, negligible compared with that $||\rho - \rho'||/||\rho|| = 0.295$ of Nematic III at $(0.5,0.45)$, as seen in \cref{fig:strongQGdiagUUA}(j). \Cref{fig:bandStrucJ_main,fig:bandStrucUA_main} show the dispersions of the lowest HF band for the Fermi surfaces in \cref{fig:fig2}(c)--(h) and \cref{fig:fig3}(c)--(h), computed at $(U'/U,J/U)=(0.60,0.30)$, $(0.30,0.15)$, $(0.80,0.15)$, $(0.05,0)$, $(0.60,0)$, $(0.80,0)$ and $(U'/U,U_A/U)=(0.25,0.15)$, $(0.25,0.50)$, $(0.20,0.45)$, $(0.20,0.05)$, $(0.80,0.40)$, $(1.00,0.20)$, respectively. \Cref{fig:fsAMdiff,fig:fsAMdiffUA} compare the competing solutions at one parameter point of each phase diagram, with their dispersions in \cref{fig:bandStrucJ,fig:bandStrucUA}.
In particular, we test the metastability of AM I solutions.
Starting from the AM I solutions and enforcing spin-$\mathrm{U}(1)$ symmetry, DIIS yields AM I solutions over the regions bounded by the silver line in \cref{fig:AMsolution,fig:AMsolutionUA}; in the $U$-$U'$-$J$ case they have slightly higher energy than the ground state in the region below the silver line in \cref{fig:fig2}(a).

The iterations do not converge in the region $J>U'$, labeled N/A in \cref{fig:fig2}(a,b). As shown in \cref{fig:latticeMesh}(d), the residual $\|\delta\rho\|/\|\rho\|$ here stays around $10^{-2}$, whereas a typical run for $J<U'$ reaches $10^{-9}$ within about 30 iterations [\cref{fig:latticeMesh}(c)].
The mechanism underlying this nonconvergence can be illustrated by considering a single electron in two degenerate orbitals of the same spin.
In this case the effective interaction channel is $(U'-J)\sum_{\mbf{R},\alpha<\beta,s} n_{\mbf{R}\alpha s}n_{\mbf{R}\beta s}$, which is attractive for $J>U'$. Its HF Hamiltonian in this attractive channel is $h^{\text{HF}}[\rho]=(U'-J)(I_2-\rho)$, so for $J>U'$ the unoccupied HF orbital has the lower energy, and in the next iteration it will be filled leaving the previously occupied orbital empty.
This leads to a two-cycle oscillation, $\rho\to I_2-\rho\to\rho$. Similar occupation switching is seen between nearly degenerate same-spin Bloch states near the Fermi level at $(U'/U,J/U)=(0.1,0.4)$. The few N/A points in \cref{fig:fig3} are also points where the iterations do not converge due to similar two-cycle oscillations.

\begin{table*}[t]
\caption{Same as \cref{tab:symmetryStrongQG} for weak quantum geometry ($m\gg t$). The hole FM here preserves $C_4$ and breaks $C_4T_{(1,0)}$, opposite to the $m\ll t$ case. }
\label{tab:symmetryWeakQG}
\centering
\renewcommand{\arraystretch}{1.25}
\resizebox{\textwidth}{!}{%
\begin{tabular}{l|ccc|ccccccccccc}
\hline\hline
 & $|\mathbf{S}|/N_{\mathrm{site}}$
 & occuN
 & occuX
 & spin-$\U(1)$
 & $T_{(1,0)}$
 & $C_4$
 & $M_x$
 & $\mathcal{T}$
 & $C_4 T_{(1,0)}$
 & $C_4^2$
 & $T_{(1,0)}\mathcal{T}$
 & $C_4\mathcal{T}$
 & $C_4 T_{(1,0)}\mathcal{T}$
 & $C_4^2\mathcal{T}$ \\
\hline\hline
sym-FM                       & $\nu/2$& 2& 2& $\checkmark$ & $\checkmark$& $\checkmark$& $\checkmark$& $\times$& $\checkmark$& $\checkmark$& $\times$& $\times$& $\times$& $\times$\\
\hline
hole FM                  & $(1-\nu)/2$& 2& 1& $\checkmark$& $\times$& $\checkmark$& $\checkmark$& $\times$& $\times$& $\checkmark$& $\times$& $\times$& $\times$& $\times$\\
\hline\hline
\end{tabular}%
}
\end{table*}

\section{Hartree--Fock stability and time-dependent Hartree--Fock spectra for the AM phase}
\label{app:TDHF}
In this section, we examine the HF stability of the AM I phase and its collective modes with the time-dependent Hartree--Fock (TDHF) method~\suppcite{KwanMoire2025,Ge2026NematicWigner}. The density matrix evolves as $i\partial_t\rho=[h^{\text{HF}}(\rho),\rho]$~\suppcite{NegeleRMP1982}, where $h^{\text{HF}}(\rho)$ is the matrix of $\hat{H}^{\text{HF}}$ in \cref{app:HFhAndChannels}. Linearizing around a stationary point $\rho_0$, $[h^{\text{HF}}(\rho_0),\rho_0]=0$, with $\delta\rho=\rho^{1}e^{-i\omega t}+\rho^{1\dagger}e^{i\omega^* t}$ gives
\begin{align}\label{eq:app:TDHFeq}
i\partial_t\delta\rho ={}& [h^{\text{HF}}(\rho_0), \delta \rho] + [h^{\text{HF}}(\rho_0 + \delta\rho) - h^{\text{HF}}(\rho_0), \rho_0],\nonumber\\
\omega\rho^1 ={}& [h^{\text{HF}}(\rho_0), \rho^1] + [h^{\text{HF}}(\rho_0 + \rho^1) - h^{\text{HF}}(\rho_0), \rho_0],\nonumber\\
-\omega\rho^{1\dagger} ={}& [h^{\text{HF}}(\rho_0), \rho^{1\dagger}] + [h^{\text{HF}}(\rho_0 + \rho^{1\dagger}) - h^{\text{HF}}(\rho_0), \rho_0],
\end{align}
where $\rho^{1}$ is not Hermitian. Denote $d^{\dagger}_{n}(\mbf{k})=\sum_{\tau\alpha s}c^{\dagger}_{\mbf{k}\tau\alpha s}U_{\tau\alpha s,n}(\mbf{k})$ which creates the $n$-th eigenstate of $h^{\text{HF}}(\rho_0)$, and $\rho^{1}_{nn'}(\mbf{k},\mbf{k}')=\langle d^{\dagger}_{n'}(\mbf{k}')d_{n}(\mbf{k})\rangle_1$. Since $\rho^2=\rho$ at zero temperature, only the particle-hole and hole-particle components of $\rho^{1}$ are nonzero, and the superlattice translation symmetry decouples different momentum transfers $\mbf{q}$ in the reduced BZ. For each $\mbf{q}$, the coupled components are $\rho^{1}_{n_pn_h}([\mbf{k}+\mbf{q}],\mbf{k})$ and $\rho^{1}_{n_hn_p}(\mbf{k},[\mbf{k}-\mbf{q}])$ for all $\mbf{k}$, where $n_h$ ($n_p$) labels occupied (empty) states and $[\cdot]$ folds a momentum into the reduced BZ. In this basis, \cref{eq:app:TDHFeq} becomes,
\begin{align}\label{eq:app:TDHFeqEigen}
\omega \rho_{n_p n_h}^{1}([\mbf{k} + \mbf{q}], \mbf{k})={}& \left(E_{n_p}([\mbf{k} + \mbf{q}]) - E_{n_h}(\mbf{k})\right) \rho_{n_p n_h}^{1}([\mbf{k} + \mbf{q}], \mbf{k}) + \sum_{\mbf{k}_2} \sum_{n_{p}' n_{h}'} \rho_{n_{p}' n_{h}'}^{1}\left([\mbf{k}_2 + \mbf{q}], \mbf{k}_2\right) V_{n_p [\mbf{k} + \mbf{q}], n_h \mbf{k}, n_{p}'[\mbf{k}_2 + \mbf{q}], n_{h}' \mbf{k}_2}\nonumber\\
&+ \sum_{\mbf{k}_2} \sum_{n_{p}' n_{h}'} \rho_{n_{h}' n_{p}'}^{1} (\mbf{k}_2, [\mbf{k}_2 - \mbf{q}]) V_{n_p[\mbf{k} + \mbf{q}], n_h \mbf{k}, n_{h}' \mbf{k}_2, n_{p}' [\mbf{k}_2 - \mbf{q}]},\nonumber\\
\omega \rho_{n_h n_p}^{1}(\mbf{k}, [\mbf{k} - \mbf{q}])={}& \left(E_{n_h}(\mbf{k}) - E_{n_p}([\mbf{k} - \mbf{q}])\right) \rho_{n_h n_p}^{1}(\mbf{k}, [\mbf{k} - \mbf{q}]) - \sum_{\mbf{k}_2} \sum_{n_{p}' n_{h}'} \rho_{n_{p}' n_{h}'}^{1}\left([\mbf{k}_2 + \mbf{q}], \mbf{k}_2\right) V_{n_h \mbf{k}, n_p [\mbf{k} - \mbf{q}], n_{p}'[\mbf{k}_2 + \mbf{q}], n_{h}' \mbf{k}_2}\nonumber\\
&- \sum_{\mbf{k}_2} \sum_{n_{p}' n_{h}'} \rho_{n_{h}' n_{p}'}^{1} (\mbf{k}_2, [\mbf{k}_2 - \mbf{q}]) V_{n_h\mbf{k}, n_p [\mbf{k} - \mbf{q}], n_{h}' \mbf{k}_2, n_{p}' [\mbf{k}_2 - \mbf{q}]},
\end{align}
with $E_n(\mbf{k})$ eigenvalues of $h^{\text{HF}}(\rho_0)$, and
$V_{n_p [\mbf{k} + \mbf{q}], n_h \mbf{k}, n_{p}'[\mbf{k}_2 + \mbf{q}], n_{h}' \mbf{k}_2} = \frac{2}{N_{\text{site}}}\sum_{\tau\alpha \beta \mu \nu s s' t t'} f_{\alpha s, \beta s';\mu t, \nu t'} U_{\tau \alpha s,n_p}^{*}([\mbf{k} + \mbf{q}]) U_{\tau \beta s', n_h}(\mbf{k}) U_{\tau \mu t, n_p'}([\mbf{k}_2 + \mbf{q}]) U_{\tau \nu t',n_h'}^{*}(\mbf{k}_2) e^{i \left([\mbf{k}_2 + \mbf{q}] + \mbf{k} - [\mbf{k} + \mbf{q}] - \mbf{k}_2\right)\cdot \boldsymbol{\delta}_{\tau}}$ where $\boldsymbol{\delta}_{\tau} = (0, 0), (1, 0)$ for $\tau=A, B$, respectively [\cref{app:symmetryRho}].
The factor $f_{\alpha s,\beta s';\mu t,\nu t'}$ is the coefficient of $c^\dagger_{\mbf{R}\alpha s}c_{\mbf{R}\beta s'}\langle c^\dagger_{\mbf{R}\nu t'}c_{\mbf{R}\mu t}\rangle$ in the HF Hamiltonian, including both Hartree and Fock contributions and with the direct Hartree term of $n_{\mbf{R}}$ subtracted as in \cref{app:HFhAndChannels}.
In matrix form, \cref{eq:app:TDHFeqEigen} reads $\begin{pmatrix}
A(\mbf{q}) & B(\mbf{q}) \\
-B^{*}(-\mbf{q}) & -A^{*}(-\mbf{q})
\end{pmatrix} \begin{pmatrix}
X(\mbf{q})\\ Y(\mbf{q})
\end{pmatrix} =\omega \begin{pmatrix}
X(\mbf{q})\\ Y(\mbf{q})
\end{pmatrix}$. This matrix is not Hermitian, so $\omega$ can be complex. It is related to the stability (Hessian) matrix  $S(\mbf{q}) = \text{diag}(I, -I)\begin{pmatrix}
A(\mbf{q}) & B(\mbf{q}) \\
-B^{*}(-\mbf{q}) & -A^{*}(-\mbf{q})
\end{pmatrix} = \begin{pmatrix}
A(\mbf{q}) & B(\mbf{q}) \\
B^{*}(-\mbf{q}) & A^{*}(-\mbf{q})
\end{pmatrix}$. If $S(\mbf{q})$ is positive definite for all $\mbf{q}$, $\rho_0$ is a local minimum on the manifold of Slater determinants; if $S(\mbf{q})$ has a negative eigenvalue, $\rho_0$ is unstable and $\omega(\mbf{q})$ can be complex.

\begin{figure}
    \centering
    \includegraphics[width=0.7\linewidth]{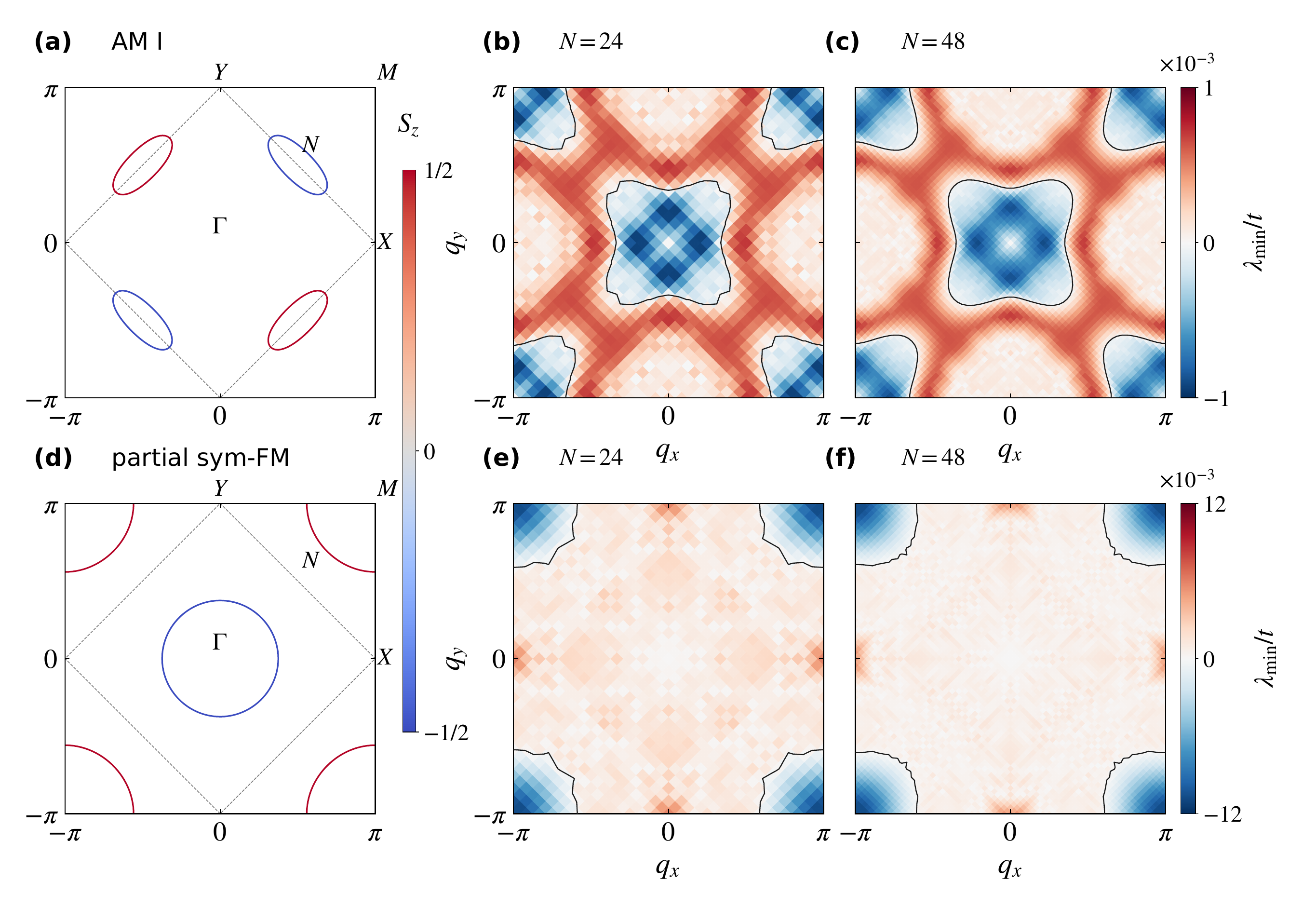}
    \caption{Lowest eigenvalues of TDHF stability matrix $S(\mbf{q})$. Interaction parameters are $U'_{\alpha \beta} = (U' - U_A)\delta_{\langle \alpha\beta\rangle} + (U' + U_A) \delta_{\llangle \alpha\beta\rrangle}, J_{p,\alpha\beta} = 3J_{H, \alpha\beta} =  3U_A\delta_{\llangle \alpha \beta\rrangle}$ where $(U'/U, U_A/U) = (0.25, 0.45)$, same as in \cref{fig:fsAMdiffUA}. (a) Fermi surface of AM I. (b)--(c) the lowest eigenvalue $\lambda_{\min}$ of $S(\mbf{q})$ for (b) $N=24$ and (c) $48$. Within numerical accuracy, $\lambda_{\min}(\mbf{q}=0)=0$. (d)--(f) Same quantities plotted for the partial sym-FM, whose most negative $\lambda_{\min}$ is at $\mbf{q}=(\pi,\pi)$, with a magnitude more than ten times larger than $|\lambda_{\min}|$ in AM I.}
    \label{fig:TDHFstability}
\end{figure}

\begin{figure}
    \centering
    \includegraphics[width=\linewidth]{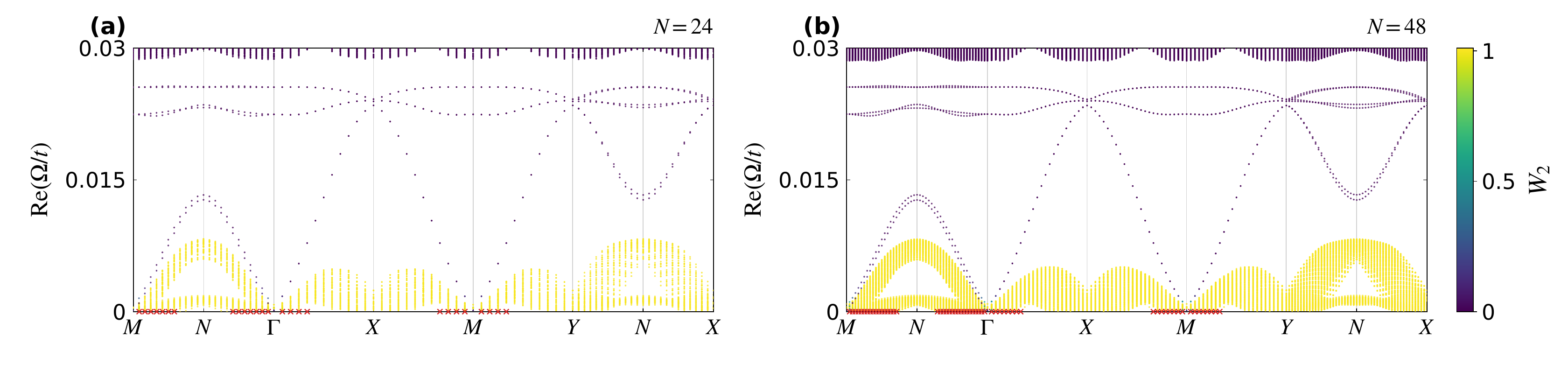}
    \caption{TDHF spectra for the AM I state. Interaction parameters are $U'_{\alpha \beta} = (U' - U_A)\delta_{\langle \alpha\beta\rangle} + (U' + U_A) \delta_{\llangle \alpha\beta\rrangle}, J_{p,\alpha\beta} = 3J_{H, \alpha\beta} =  3U_A\delta_{\llangle \alpha \beta\rrangle}$ where $(U'/U, U_A/U) = (0.25, 0.45)$, same as in \cref{fig:fsAMdiffUA}. Color: Absolute value of the indefinite weight $W_2=|X_2^{\dagger}(\mbf{q})X_2(\mbf{q})-Y_2^{\dagger}(\mbf{q})Y_2(\mbf{q})|/|X^{\dagger}(\mbf{q})X(\mbf{q})-Y^{\dagger}(\mbf{q})Y(\mbf{q})|$ of each real-$\omega$ mode within the lowest two bands. Red crosses at $\omega=0$ mark modes with complex $\omega$.}
    \label{fig:TDHFspectra}
\end{figure}

\Cref{fig:TDHFstability,fig:TDHFspectra} show the lowest eigenvalues of $S(\mbf{q})$ and the TDHF frequencies $\omega(\mbf{q})$, respectively. We keep only the HF bands closest to the Fermi level in the particle-hole basis; the results converge once four bands are kept at each $\mbf{k}$.
[Such truncation leaves $S(\mbf{q})$ square.]
In \cref{fig:TDHFspectra}, we plot $\mathrm{Re}\,\omega$ for real-frequency modes with a positive norm, $X^{\dagger}(\mbf{q})X(\mbf{q})-Y^{\dagger}(\mbf{q})Y(\mbf{q})>0$, which correspond to excitation modes, while complex $\omega$ modes are marked on the horizontal axis. The $N=24$ and $N=48$ meshes give qualitatively similar results. For AM I, $S(\mbf{q})$ has negative eigenvalues at small $\mbf{q}$ [\cref{fig:TDHFstability}(b,c)], so this gapless HF state is unstable against long-wavelength perturbations; the most negative eigenvalues appear near $\mbf{q}=(\pm\pi/4,0)$ and $(0,\pm\pi/4)$.
Similarly, the HF N\'eel state of the Hubbard model is unstable toward a spiral twisting of the spins
upon hole doping away from half-filling~\suppcite{SinghPRB1990,*SinghPRB1990Erratum,SinghPRB1991}.
The competing partial sym-FM, which has the symmetries of the sym-FM (\cref{tab:symmetryStrongQG}) but $|\mathbf{S}|/N_{\text{site}}<(1-\nu)/2$ and preserves translation symmetry, has negative eigenvalues near $\mbf{q}=(\pi,\pi)$ that are an order of magnitude larger in absolute value [\cref{fig:TDHFstability}(e,f)]. This indicates a leading instability toward $\sqrt{2}\times\sqrt{2}$ translation breaking, consistent with the $T_{(1,0)}$-broken AM I having a lower energy than the partial sym-FM (\cref{fig:fsAMdiffUA}).

The TDHF spectrum of AM I in \cref{fig:TDHFspectra} agrees with $S(\mbf{q})$: there are two degenerate modes with $\omega\approx0$ at $\mbf{q}=0$, and at the small $\mbf{q}$ where $S(\mbf{q})$ has negative eigenvalues, at least one mode has a complex frequency.
These modes with complex frequencies lie mostly within the lowest two bands, like the low-lying particle-hole continuum discussed below. They suggest that the HF ground state may carry an additional long-wavelength modulation of the AM I order, driven by particle-hole excitations near the Fermi surface, as in the doped N\'eel state~\suppcite{SinghPRB1991}.
However, the magnitude of the complex frequencies is about $0.0008t$, which is quite small compared to non-interacting bandwidth $4m = 0.04t$.
 Therefore, the existence of the complex frequencies must be due to finite size effect. 
Even if the complex frequencies do persist as the size increases, the largest imaginary parts of the complex frequencies, found at $\mbf{q}\approx(\pm\pi/4,0)$ and $(0,\pm\pi/4)$, correspond to a modulation period of about eight lattice constants, which is much larger than the $\sqrt{2}\times\sqrt{2}$ unit cell, and thus even if it happens, the AM can still exist within each supercell.
The low-lying particle-hole continuum (yellow points in \cref{fig:TDHFspectra}) lies mostly within the lowest two bands, as shown by the absolute value of their indefinite weight in these partially occupied bands, defined by $W_2(\mbf{q})=|X_2^\dagger(\mbf{q})X_2(\mbf{q})-Y_2^\dagger(\mbf{q})Y_2(\mbf{q})|/|X^\dagger(\mbf{q})X(\mbf{q})-Y^\dagger(\mbf{q})Y(\mbf{q})|$, where $X_2(\mbf{q})$ and $Y_2(\mbf{q})$ are the TDHF eigenvector components projected into the lowest two HF bands in the reduced BZ.
Two low-lying TDHF modes are connected to the two $\omega\approx0$ modes at $\mbf{q}=0$, carrying negligible projected indefinite weight $W_2$ onto the two lowest HF bands. We attribute them to the two generators broken by the reduction of spin-$\mathrm{SU}(2)$ to spin-$\mathrm{U}(1)$. These modes are degenerate along $\Gamma$--$X$--$M$--$Y$ but split along $M$--$N$--$\Gamma$ and $Y$--$N$--$X$, similar to the spin splitting of the AM I bands in \cref{fig:bandStrucUA}. The degeneracy along $\Gamma$--$X$--$M$--$Y$ is protected by the preserved $M_x\mathcal{T}$ and $M_y\mathcal{T}$ symmetries.

\FloatBarrier
\renewcommand{\refname}{Appendix References}
\putbib[re]

\end{bibunit}

\end{document}